\documentclass[amssymb,amsmath,twocolumn,superscriptaddress,nofootinbib,floatfix]{revtex4-2}
\usepackage{graphicx}
\usepackage{bm}
\usepackage{float}
\makeatletter
\renewcommand\fs@ruled{%
  \def\@fs@cfont{\bfseries}%
  \let\@fs@capt\floatc@ruled
  \def\@fs@pre{\hrule height.8pt depth0pt \kern6pt}
  \def\@fs@mid{\kern6pt\hrule\kern6pt}
  \def\@fs@post{\kern6pt\hrule\relax}
  \let\@fs@iftopcapt\iftrue}
\makeatother
\usepackage{amssymb,amsmath}
\usepackage{mathrsfs}
\usepackage{latexsym}
\usepackage{color}
\usepackage{comment}
\usepackage[normalem]{ulem}
\usepackage{dcolumn}
\usepackage{xcolor}
\usepackage[colorlinks=true,citecolor=blue,urlcolor=blue]{hyperref}
\usepackage{multirow}
\usepackage{soul}

\usepackage{orcidlink}
\usepackage{enumitem}
\usepackage{algorithm}
\usepackage{algorithmicx}
\usepackage{algpseudocode}
\usepackage{etoolbox}

\AtBeginEnvironment{algorithmic}{\normalsize}

\graphicspath{{./}{figures/}{figures/toy_model}{figures/schematics}}

\allowdisplaybreaks

\newcommand{\CIT}{\affiliation{TAPIR, California Institute of Technology, Pasadena, CA 91125, USA}}
\newcommand{\CITLab}{\affiliation{LIGO Laboratory, California Institute of Technology, Pasadena, CA 91125, USA}}
\newcommand{\UChicago}{\affiliation{Department of Astronomy \& Astrophysics, The University of Chicago, Chicago, IL 60637 USA}}
\newcommand{\ETH}{\affiliation{ETH Zurich, Institute for Particle Physics and Astrophysics, Wolfgang-Pauli-Strasse 27, 8093 Zurich, Switzerland}}

\definecolor{kcmagenta}{rgb}{0.54, 0.17, 0.88}
\definecolor{smgreen}{rgb}{0.26, 0.625, 0.277}

\newcommand{\editaccepted}[1]{{#1}} 

\newcommand{\chieff}{\chi_{\textrm{eff}}}

\newcommand{\maxL}{\mathrm{max.}\,\mathcal{L}}
\newcommand{\dpred}{d^{\rm pred}} 
\newcommand{\dobs}{d^{\rm obs}} 
\newcommand{\lambdapred}{\lambda^{\rm pred}} 
\newcommand{\lambdaobs}{\lambda^{\rm obs}} 
\newcommand{\dpredvec}{\vec{d}^{\,\rm pred}} 
\newcommand{\dobsvec}{\vec{d}^{\,\rm obs}} 
\newcommand{\lambdapredvec}{\vec{\lambda}^{\rm pred}} 
\newcommand{\lambdaobsvec}{\vec{\lambda}^{\rm obs}} 
\newcommand{\pvalue}{\mathbf{p}_T}

\begin{document}

\title{Posterior predictive checks for gravitational-wave populations: Limitations and improvements}

\author{Simona J.~Miller~\orcidlink{0000-0001-5670-7046}} \email{simona.miller@ligo.org} \CITLab \CIT 
\author{Sophia Winney} \UChicago
\author{Katerina Chatziioannou~\orcidlink{0000-0002-5833-413X}}  \CIT \CITLab 
\author{Patrick M. Meyers~\orcidlink{0000-0002-2689-0190}} \ETH \CIT

\date{\today}

\begin{abstract}
When selecting a model to characterize an astrophysical population, it is crucial to assess whether that model fits the data and, if not, how it can be improved. 
To this end, posterior predictive checks (PPCs) are a widely used statistical test of model fit when inferring gravitational-wave source populations.
However, PPCs exhibit limitations when assessing single-event parameters with large measurement uncertainty, like the spin tilt angles of the binary black holes (BBHs) observable with the LIGO-Virgo-KAGRA (LVK) detectors. 
When single-event inference is prior dominated, traditional PPCs fail to flag even very poor model fits.
In this work, we assess the efficacy of various alternative PPCs on poorly constrained parameters.
We compare PPCs conducted on event- vs.~data-level parameters  (e.g. posterior samples vs.~maximum likelihood points), and explore two additional event-level PPCs: partial predictive checks and split predictive checks. 
Independent of measurement uncertainty, we find that PPCs on maximum likelihood parameters are always more discerning of model misspecification than any event-level PPC. 
However, when investigating simulated GWTC-3.0-like catalogs, none of the alternative PPCs show significant improvement over those traditionally used, indicating that at that sensitivity, any limited information in the data about spin tilts is insufficient to diagnose model misspecification.
Finally, we apply our suite of PPCs to the spin magnitude and tilt distributions inferred in the most recent LVK catalog, GWTC-4.0. 
We conclude that the Gaussian Component Spins model used therein under-predicts BBHs with large spin magnitudes and over-predicts those with perfectly anti-aligned tilts. 

\end{abstract}

\maketitle

\section{Introduction}
\label{sec:intro}

The hundreds of gravitational-wave (GW) signals from compact binary mergers~\cite{GWTC2,GWTC2.1,GWTC3,GWTC4_cat} observed over the past decade by the LIGO~\cite{aLIGO}, Virgo~\cite{aVirgo}, and KAGRA~\cite{KAGRA} (LVK) detectors have yielded constraints on the population properties of coalescing binary black holes (BBHs) with masses $3{-}300\,M_\odot$~\cite{GWTC1_pop,GWTC2_pop,GWTC3_pop,GWTC4_pop}. 
The distributions of mass, spin (i.e., intrinsic angular momentum), and redshift across this BBH population are inferred by, first, selecting some model for the underlying population and, second, constraining that model's parameters using hierarchical Bayesian methods~\cite[cf.~Sec.~2 of Ref.][]{GWTC4_pop}. 
Since this process relies on a (sometimes phenomenological) population model, it is important to assess whether that model is indeed a good fit to the observed data, and to understand when conclusions are driven by the data versus the model.

Several model-checking procedures are regularly used in GW population analyses~\cite{Romero-Shaw:2022ctb}.
For instance, one can infer the distribution of mass, spin, and redshift across the same population using different models and compare these models' (dis)agreement~\cite[e.g.,][]{GWTC3_pop,GWTC4_pop,Sadiq:2021fin,Golomb:2022bon,Vitale:2022dpa}.
Of particular utility is contrasting the behavior of strongly parametrized models (like Gaussians or power laws) to more flexible, weakly parametrized models which make minimal assumptions about the shape of the underlying population (like splines, auto-regressive processes, or binned-Gaussian processes)~
\cite[e.g.,][]{Edelman:2022ydv,Golomb:2022bon,Godfrey:2023oxb,Callister:2023tgi,Ray:2024hos,Heinzel:2024jlc,Alvarez-Lopez:2025ltt}.
However, while one can make sure that each is consistent with the maximum likelihood population~\cite{Payne:2022xan,Guttman:2025jkv}, whether or not two inferred distributions are ``similar enough" remains difficult to quantify, especially given the high dimensionality of BBH parameter space. 
Another commonly utilized tool for comparing population models is the Bayes factor---a ratio between the marginal likelihoods of two models. 
However, Bayes factors tend to be highly prior dependent, rely on obscure criteria for quantifying significance, and provide only an aggregate test of model performance without demonstrating which aspects of the data a non-preferred model fails to capture. 
As a result, Bayes factors are most helpful in ranking models, not improving them~\cite{Isi:2022cii}.

In this work, we instead focus on posterior predictive checks (PPCs)~\cite{Bayarri:2008,Gelman:1996,Rubin:1984,Guttman:1967}.
Compared to other model-checking procedures, PPCs have the advantage of both providing interpretable, quantitative results about model fit \textit{and} suggesting sites of model improvement.
Broadly, PPCs evaluate the performance of models by checking the consistency between data predicted by the model and current observations.
They assess whether statistically significant differences exist between the observed GW catalog and catalogs of simulated GW events drawn from the inferred population distribution.
In GW population analyses, PPCs are traditionally conducted on ``event-level" parameters: the true underlying astrophysical black hole (BH) masses, spins, etc.~\cite[e.g.,][]{Fishbach:2019ckx, GWTC3_pop, Romero-Shaw:2022ctb,Thrane:2019,Ray:2026uur}. 
Alternatively, PPCs can be conducted on ``population-level" parameters (hyper-parameters), i.e., features in the population distribution~\cite[e.g.,][]{GWTC1_pop,Farah:2023vsc,Sadiq:2021fin,Mould:2026sww,Vitale:2025lms,Callister:2021fpo}, or ``data-level"  parameters like maximum likelihood points or search pipeline results~\cite{Fishbach:2019ckx,Mould:2026sww}. 
In this work, we \textit{(i)} contrast event and data-level PPCs and \textit{(ii)} implement alternative predictive checks on event-level parameters.

Our study is motivated by previous work in \citet{Miller:2024sui}, which demonstrated that the commonly used event-level PPCs exhibit significant limitations when looking at \textit{poorly constrained} BBH parameters.
When individual event data are weakly informative about a particular parameter, event-level PPCs nearly always indicate that a population model is a ``good fit" to the data, even if we know \textit{a priori} it is not. 
In this regime, the behavior of event-level PPCs becomes dominated by the population model, rather than the limited information contained in the data---the exact problem we wish to diagnose with such tests.  
Specifically, event-level PPCs incorrectly indicate that a poor model is a good fit because they test a \textit{combination of the data and prior} (i.e., population model) against that prior.
In contrast, the data-level PPCs we explore in this work test \textit{just the data} against the prior.
We conclude that data-level PPCs provide a more robust assessment of \textit{(i)} to what extent the data are actually informative and \textit{(ii)} if the data are indeed informative, whether they disagree with the choice of population model.

Developing trustworthy model-checking algorithms for poorly constrained parameters is crucial, as some of the most astrophysically interesting BBH properties---spin magnitudes and directions---are indeed only weakly informative. 
A robust measurement of the BBH spin distribution is central when untangling which BBH formation and evolutionary mechanisms~\cite[e.g.,][]{Mandel:2018hfr,Mapelli:2018uds} contribute to the observed population.  
Spins offer unique insight into angular momentum transport in stars~\cite{Fuller:2019ckz,Fuller:2019sxi}, supernovae kicks~\cite{Kalogera:1999tq,Gerosa:2018wbw, Steinle:2020xej,Wysocki:2017isg,Stevenson:2022hmi,Callister:2020vyz,Baibhav:2024rkn,Tauris:2022ggv}, and two-body interactions hypothesized to impact BBH formation and evolution, such as tides, accretion, and stellar winds~\cite[e.g.,][]{Hut:1981,Packet:1981,Tout:1992,Mandel:2015qlu,Qin:2018sxk,Qin:2018vaa,Bavera:2020uch,Ma:2023nrf}.
Spin-orbit misalignment can also be used to identify dynamically formed BBHs, e.g., in globular clusters~\cite{Rodriguez:2015oxa,Rodriguez:2016kxx,Rodriguez:2017pec,Farr:2017uvj}, the disks of active galactic nuclei~\cite[e.g.,][]{Wang:2021yjf,McKernan:2021nwk,McKernan:2023xio,McKernan:2019beu, Yang:2019okq}, and/or via hierarchical mergers~\citep[e.g.,][]{Rodriguez:2019huv,Zhang:2023fpp,Doctor:2019ruh,Kimball:2020qyd,Payne:2024ywe,Gerosa:2021mno,Fishbach:2017dwv,Gerosa:2017kvu,Mould:2022ccw,Mahapatra:2021hme,Mahapatra:2022ngs,Mahapatra:2024qsy}.

Previous studies have found that the BBH population has preferentially small but non-zero spin magnitudes, and a wide range of angles (``tilts") between the spin and orbital angular momenta \cite{GWTC4_pop}. 
While these broad conclusions remain qualitatively unimpeachable, the exact quantitative results are model dependent, due to the aforementioned large measurement uncertainty~\cite{Miller:2024sui,Vitale:2022dpa,Vitale:2025lms}.
Model dependence is especially prominent when trying to probe narrow population features, such as the exact fraction of non-spinning BBHs, or features informed by the tails of the population, like the fraction of BBHs with spin vectors pointing below the orbital plane~\cite{Callister:2022qwb,Galaudage:2021rkt,Tong:2022iws,Adamcewicz:2023szp,Adamcewicz:2025phm}.
Thus, in this and previous~\cite{Miller:2024sui} work, we choose to focus on population-level measurements of BBH spin tilts as they provide an astrophysically motivated testbed for the behavior of PPCs when measurement uncertainty is large.

In this paper, we \textit{explore a wide range of predictive checks on GW population data}.
\editaccepted{We begin in Sec.~\ref{sec:result_summary} by providing a top-level road map of our methods and results.}
\editaccepted{We then} revisit PPCs in general, explain their mathematical formulation in detail, \editaccepted{and} describe posterior predictive $p$-values   \editaccepted{(Sec.~\ref{sec:methods})}. 
\editaccepted{We} explore various choices one can make when selecting precisely what predicted vs.~observed data are compared\editaccepted{, applying} three alternative PPCs to spin tilts. \editaccepted{We investigate} simulated populations, where we know the true underlying distribution and intentionally measure it with an insufficiently complex model \editaccepted{(Secs.~\ref{sec:event_vs_data_level}, \ref{sec:partial_PPCs}, \& \ref{sec:split_PPCs}), as well as the LVK's Fourth Gravitational-wave Transient Catalog (GWTC-4.0) \cite{GWTC4_intro,GWTC4_cat} (Sec.~\ref{sec:gwtc4_results})}.
\editaccepted{Finally, recommendations and future work are offered in Sec.~\ref{sec:conclusions}.}

\section{Summary of Main Results}
\label{sec:result_summary}

\begin{figure*}[]
    \centering
    \includegraphics[width=\linewidth]{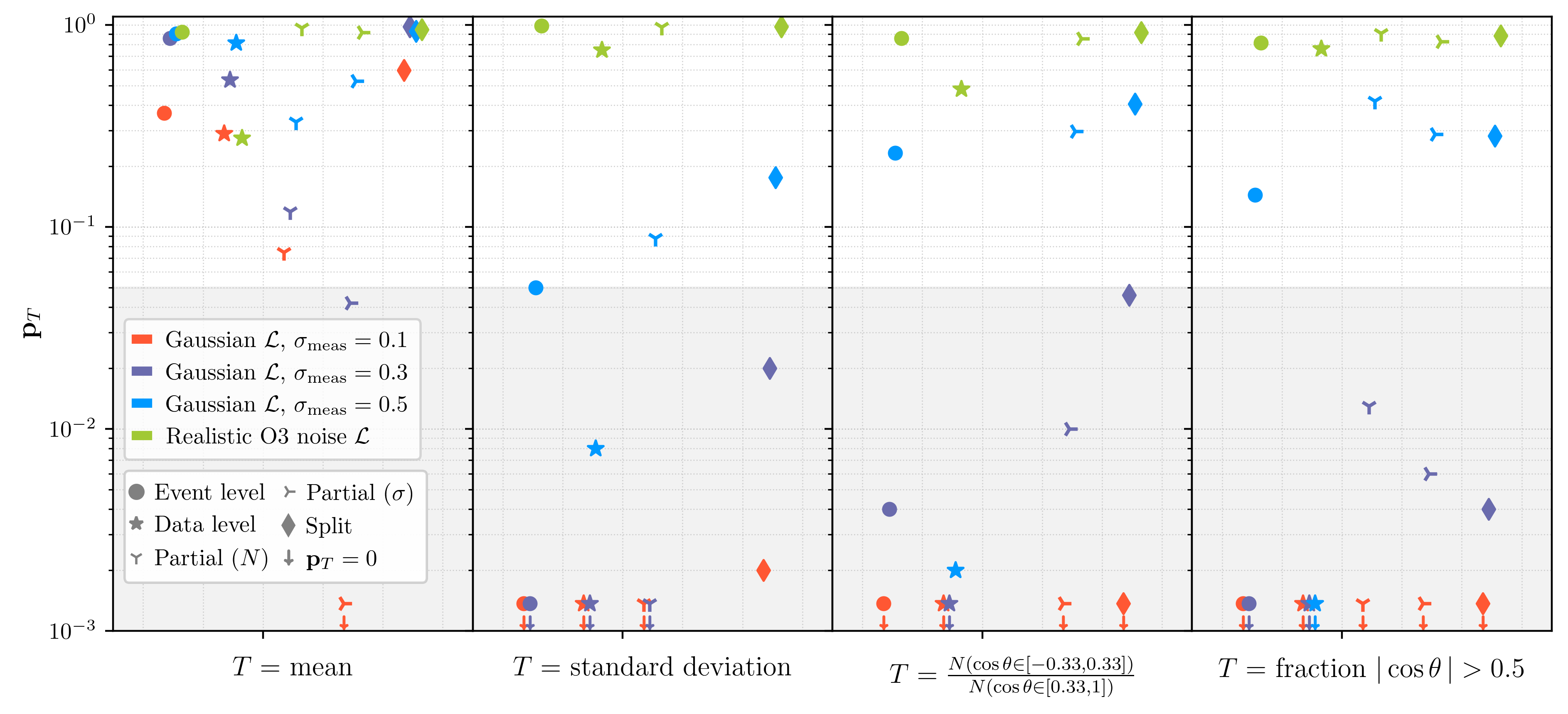}
    \caption{
    Summary of the posterior predictive $p$-values ($\pvalue$) for all PPCs (shapes), test statistics ($T$; subplots), and single-event likelihoods (different colors) that we explore in this work. 
    Here, we are using an intentionally misspecified model to measure the BBH tilt ($\cos\theta$) distribution: the true distribution is bimodal, while our population model is a single Gaussian (Fig.~\ref{fig:data_event_PPCs}).
    The goal is to successfully identify this model misspecification with low $p$-values ($\pvalue<0.05$, gray shaded region).
    Each $\pvalue$ is generated from 1,000 PPC traces.
    Smaller values of $\pvalue$ indicate a more robust test of model fit. 
    Points in the bottom row atop downward arrows have $\pvalue<10^{-3}\approx 0$, the smallest value we can probe with 1,000 PPC traces. 
    The unitless horizontal axis groups the $p$-values by $T$ (each subplot).
    From left to right, $T$ is the mean, standard deviation, ratio of the number of events with $\cos\theta \in [-0.33, 0.33]$ to those with $\cos\theta \in [0.33, 1]$, and the fraction of events with $|\cos\theta|> 0.5$.    
    The horizontal placement within each subplot is arbitrary and selected to aid readability. 
    We show results for different PPC (between each gridline per-subplot) and single-event likelihood ($\mathcal{L}$, slightly offset from each other for readability).
    We investigate both Gaussian single-event likelihood models with increasing measurement uncertainty (red, purple, and blue), as well as \editaccepted{a} realistic O3 noise model (green), as discussed in Sec.~\ref{subsec:simulated_pops}.
    For the definition of $\pvalue$, see Secs.~\ref{subsec:pvalues} and \ref{subsec:event_vs_data_level_pvalues}. For descriptions of the various PPCs, see Secs.~\ref{sec:event_vs_data_level} (event vs.~data level), \ref{sec:partial_PPCs} (partial), and \ref{sec:split_PPCs} (split).
    In the legend, ``Partial ($N$)'' refers to the partial PPC where $T = N(\cos\theta \in [-0.33, 0.33])/N(\cos\theta\in[0.33,1])$ is fixed between predicted and observed catalogs, while  ``Partial ($\sigma$)'' refers to that where the standard deviation is fixed. These cases are thus respectively excluded from the third and second columns. Throughout this work, we refer back to the above figure to explain results in greater detail. 
    }
    \label{fig:pvalue_summary}
\end{figure*}

\editaccepted{In this section, we summarize our findings and indicate where readers can turn for more detail.}
We explore the below predictive checks on GW population data, and conclude the following: 
\begin{enumerate}
    \item \textbf{Event\editaccepted{-} vs.~Data\editaccepted{-}level PPCs} \editaccepted{(Sec.~\ref{sec:event_vs_data_level})}: Event-level PPCs are conducted on true single-event parameters\editaccepted{, and are standard practice in GW population analyses~\cite[e.g.,][]{Fishbach:2019ckx,GWTC3_pop,GWTC4_pop,Romero-Shaw:2022ctb,Thrane:2019}}. They involve taking single-event posterior draws, which by definition depend on the choice of prior. We also conduct PPCs on the data itself. Here we use the maximum likelihood parameters, which do not depend on the prior. 
    \vspace{5px} \\
    Averaged over many realizations, \textit{data-level PPCs are always equally or more discerning of model misspecification than event-level PPCs}, as shown in Figs.~\ref{fig:data_event_PPCs} and \ref{fig:toy_model_repeated}. This behavior stems from the event-level PPCs' dependence on the prior, and is thus most stark for poorly constrained parameters like spins.
    \vspace{5px} \\
    To further elucidate event and data level PPCs, we have included a toy-model implementation of both in our data release~\cite{github_release} and Appendix~\ref{app:toy_model}.
    \item \textbf{Partial PPCs} \editaccepted{(Sec.~\ref{sec:partial_PPCs})} target a feature in the population distribution (like the mean, standard deviation, or fraction above some cutoff) and isolate information orthogonal to it \editaccepted{\cite{Bayarri:2008, Robins:2000}} .
    \vspace{5px} \\
    The efficacy of the partial PPC depends on which feature we probe.
    We find that it is \textit{more discerning when the targeted feature is well predicted by the model}, Fig.~\ref{fig:partial_PPC_std}, rather than for features that the model cannot capture, Fig.~\ref{fig:partial_PPC_frac}.
    \item \textbf{Split PPCs} \editaccepted{(Sec.~\ref{sec:split_PPCs})} divide the observed data into two subsets: one subset is used to infer the population distribution and the other is used to generate predictive catalogs  \editaccepted{\cite{Li:2022,Moran:2024}}. Split PPCs are consistently the least informative PPC that we test,  Fig.~\ref{fig:split_PC}.
\end{enumerate}
We conduct both partial and split PPCs on only event-level parameters.


\editaccepted{To quantify each PPC's ability to identify model misspecification, we select population features to probe---called test statistics ($T$)---and from these calculate posterior predictive $p$-values ($\pvalue$). Values of $\pvalue\lesssim 0.05$ indicate that the chosen population model is a poor fit to the data (Sec.~\ref{subsec:pvalues}). 
Focusing on the angle $\theta$ between the BBH spin and orbital angular momentum, we  present results where $T$ is:}
\begin{enumerate}
    \item The mean of $\cos\theta$ across a catalog.
    \item The standard deviation of $\cos\theta$ across a catalog.
    \item The ratio of number of events in a catalog with $\cos\theta \in [-0.33, 0.33]$ to those with $\cos\theta \in [0.33, 1]$.
    \item The fraction of events in a catalog with $|\cos\theta|> 0.5$. 
\end{enumerate}

\editaccepted{
Figure~\ref{fig:pvalue_summary} shows $\pvalue$ for each PPC and $T$ when applied to an intentionally mis-modeled, simulated $\cos\theta$ population across a range of single-event measurement uncertainties and thus acts as a summary of our main results. Lower values of $\pvalue$ indicate that the data level PPCs (stars) are always equally or more discerning of model misspecification than the traditional event-level PPCs (circle), which are themselves always more discerning than the split PPCs (diamonds). The performance of the partial PPC depends on which degree of freedom is targeted.
Figure~\ref{fig:pvalue_summary} also highlights that, for all PPCs, the choice of $T$ (boxed columns) determines the efficacy of $\pvalue$, depending on both the feature we want to probe and the functional form of the population model itself (Sec.~\ref{subsec:event_vs_data_level_pvalues}).
Throughout this work, we will refer back to Fig.~\ref{fig:pvalue_summary} to explain results in greater detail. }

Investigations on simulated GWTC-3.0-like catalogs (green in Fig.~\ref{fig:pvalue_summary}) indicate that none of the PPCs are able to robustly identify model misspecification for any $T$ that we probe \editaccepted{($\pvalue\gg0.05$)}.
Thus, \editaccepted{at the LVK detectors' third observing run (O3) sensitivity level,} any limited information in the data about spin tilts is likely insufficient to diagnose model misspecification.
However, when applying our suite of PPCs to GWTC-4.0 (Fig.~\ref{fig:O4_PPCs})\editaccepted{, i.e., with improved fourth observing run (O4) detector sensitivity}, we find evidence for model misspecification in the spin magnitude and tilt distributions inferred in Ref.~\cite{GWTC4_pop}\editaccepted{. Their headline model} under-predicts BBHs with large spin magnitudes and over-predicts those with perfectly anti-aligned tilts \editaccepted{(Sec.~\ref{sec:gwtc4_results})}. 


\section{Methods}
\label{sec:methods}

To determine various posterior-predictive checking methods' ability to identify discrepancies between population models and observed data, we apply an intentionally misspecified model to a simulated population where the true, underlying parameter distributions are known.
In Sec.~\ref{subsec:simulated_pops}, we describe the simulated population used in this work. 
More details about hierarchical inference and our methods of population simulation and recovery are given in Appendix \ref{app:sim_pop}.
We then present a high-level overview of PPCs (Sec.~\ref{subsec:PPCs}) and associated $p$-values (Sec.~\ref{subsec:pvalues}), with more technical details given in Appendix \ref{app:pvalues}. 

\subsection{Simulated population and its inference}
\label{subsec:simulated_pops}

The simulated BBH population distribution in this work is taken from \citet{Miller:2024sui}, which explored three simulated populations with the same effective aligned spin ($\chieff$) distributions but different underlying spin magnitude and tilt distributions.
The dimensionless quantity $\chi \in [0,1]$ captures the magnitude of the spin angular momentum of the BH, where $\chi=1$ is the maximal spin a BH can support given its mass.
The tilt parameter $\theta \in [0,\pi]$ is the angle between the spin vector of the BH and the Newtonian orbital angular momentum vector of the BBH system. 
We parametrize the tilt through its cosine, $\cos\theta \in [-1,1]$, where $1$ and $-1$ indicate perfect alignment and anti-alignment with the binary's orbital angular momentum, and $0$ indicates fully in-plane spin.
GW signals are primarily influenced not by spin magnitudes and tilts~\cite[e.g.,][]{Purrer:2015nkh,Vitale:2016avz}, but instead by two ``effective" spin parameters~\cite[e.g.,][]{Ajith:2009bn,Ng:2018neg}: the effective aligned spin ($\chi_{\rm eff}$) containing the spin components that are aligned with the orbital angular momentum~\cite{Racine:2008qv}, and the effective precessing spin ($\chi_{\rm p}$) containing the misaligned components~\cite{Schmidt:2010it,Schmidt:2012rh,Schmidt:2014iyl,Thomas:2020uqj,Gerosa:2020aiw} which induce spin-orbit precession. 

In this work, we focus on the \textsc{LowSpinAligned} population of \citet{Miller:2024sui}, shown in their Fig.~1 and replicated in our Fig.~\ref{fig:sim_pop} in Appendix \ref{app:sim_pop}.
This population has a spin magnitude distribution peaking at $\chi=0.1$, with nearly all $\chi < 0.5$. 
Its tilt distribution is \textit{bimodal} with peaks at $\cos\theta=1$ and $\cos\theta=-1$, or perfect alignment and anti-alignment.
These magnitude and tilt distributions produce a $\chieff$ distribution consistent with that observed in real GW data to date, i.e., through the first part of the LVK's fourth observing run (O4a) ~\cite{Miller:2020zox,GWTC2_pop,GWTC3_pop,GWTC4_pop}: narrow and centered slightly above $\chieff=0$.
The simulated mass and redshift populations follow the median distributions found analyzing GWTC-3.0, containing the data through LVK's third observing run (O3)~\cite{GWTC3_pop}.

We consider catalogs of $70$ BBH events (the approximate number of events in GWTC-3.0) and use the same individual-event likelihood models investigated in \citet{Miller:2024sui}, corresponding to different noise scenarios: 
\begin{enumerate}
    \item \textbf{Realistic O3 noise likelihood}: Individual-event posteriors were sampled stochastically via nested sampling~\cite{dynesty} with the parameter-estimation code \textsc{Bilby}~\cite{bilby}, using realistic O3 sensitivity~\cite{O3-sensitivity} and the waveform model \textsc{IMRPhenomXPHM}~\cite{IMRPhenomXPHM}. 
    \item \textbf{Gaussian likelihood}: Individual-event spin magnitude and cosine-tilt posteriors were constructed using simulated Gaussian likelihoods with standard deviations $\sigma_{\rm meas} = 0.1, 0.3, 0.5$.
    Samples from the Gaussian spin posteriors were randomly paired with those from the full stochastically sampled \textsc{Bilby} posteriors for all other parameters, i.e., we assumed spins were un-correlated with mass and redshift.
\end{enumerate}
For reference, the realistic $\cos\theta$ uncertainty at O3 sensitivity averages $\sigma_{\rm meas} \approx 0.5$, which corresponds to posterior distributions often spanning the full allowed parameter range of $\cos\theta \in [-1,1]$.
However, the realistic noise $\cos\theta$ posteriors are non-Gaussian and are correlated with other parameters. 

\editaccepted{In Secs.~\ref{sec:event_vs_data_level} and \ref{sec:gwtc4_results}, we find maximum likelihood ($\maxL$) parameters for simulated and real data.
For the Gaussian single-event likelihoods, we assume that the $\maxL$ spin magnitudes and tilts are drawn from a truncated Gaussian distribution with width $\sigma_{\rm meas}$ centered at the truth.
For the realistic O3 noise case, we find the $\maxL$ parameters through a least-squares minimization algorithm implemented in the parameter-estimation code \texttt{cogwheel}~\cite{Roulet:2022kot,Islam:2022afg,Roulet:2024hwz}.
We find that likelihood \textit{optimization}, rather than taking the $\maxL$ \textit{sample} from a full posterior, is necessary for robust data-level PPCs.
Not only is sampling the posterior computationally infeasible when dealing with many thousands of simulated events, but parameter estimation is not designed to optimize the likelihood, rendering a $\maxL$ posterior sample not a particularly robust data-level parameter for spin tilts.
Appendix~\ref{app:maxL_finding} provides more detail about likelihood optimization and compares different $\maxL$-finding methods.
}

To measure the astrophysical population, we used a \textit{truncated Gaussian distribution} in $\cos\theta$ and inferred its mean and width.
This model was deliberately \textit{not} able to capture the full structure of the true, underlying bimodal population.
In the remainder of this work, we use the individual-event and hyper-posterior samples obtained in \citet{Miller:2024sui} to test whether various PPC methods can identify the discrepancy between the true and inferred $\cos\theta$ distributions.
We refer readers to \citet{Miller:2024sui} (especially Appendices A, B, and C) as well as our Appendix \ref{app:sim_pop} for more details related to our simulated populations and inference.

\subsection{Posterior predictive checks}
\label{subsec:PPCs}

In this section, we describe the concept and associated mathematics of PPCs.
In full generality, PPCs are a means to assess the accuracy of a model fit to observed data $y^{\rm obs}$ based on its ability to predict future data $y^{\rm pred}$ (also known as \textit{replications} of data)~\cite{Gelman:1996}.
The mathematical underpinning of PPCs is that observed data should be plausible under an inferred model's \textit{posterior predictive distribution} (PPD). 
Assuming that the data can be described by a model with parameters $\Theta$, the PPD is
\begin{align}
\mathrm{PPD}(y^{\rm pred} | y^{\rm obs}) & \equiv p(y^{\rm pred} | y^{\rm obs}) \nonumber \\
    & = \int p( y^{\rm pred}|\Theta) \, p(\Theta | y^{\rm obs}) \, d\Theta \,,
    \label{eqn:ppd} 
\end{align}
where $ p(\Theta | y^{\rm obs})$ is what the observed data tell you about the assumed model's parameters and $ p(y^{\rm pred} |\Theta)$ is how the model generates data given parameters.
Conceptually, the PPD is the distribution of data one would expect to observe in the future ($y^{\rm pred}$) given the data we have already observed ($y^{\rm obs}$) and assuming those data can be described by a model with parameters $\Theta$.

In the case of GW populations, the data $y$ in question are a \textit{catalog} of $N_{\rm obs}$ independent BBH detections: $y=\vec{d}\equiv\{d_i\}_{N_{\rm obs}}$.
Throughout this work, we use vectors to describe catalogs, i.e., collections of single-event data.
We assume that each BBH in the catalog can be described by individual-event parameters $\lambda$ (masses, spins, redshift, etc.) via some waveform model for its GW signal.
Accordingly, the posterior distribution $p(\lambda | d)$ is the probability that strain data $d$ consist of a superposition of Gaussian noise and a waveform with parameters $\lambda$. 
Going up one level in the hierarchy, the distribution of $\lambda$ across the astrophysical BBH population $p(\lambda | \Lambda)$ is then described by a set of \textit{hyperparameters} $\Lambda$; see Eq.~\eqref{eqn:pop_posterior}. 

When dealing with a hierarchical model structure, the PPD is a joint distribution on data $\dpred$ and parameters $\lambdapred$: 
\begin{align}
    \mathrm{PPD}&(\dpred,\lambdapred| \dobsvec)  \equiv p (\dpred,\lambdapred| \dobsvec) \nonumber\\
     &= \int p(\dpred, \lambdapred|\Lambda)\,p(\Lambda | \dobsvec)\, d\Lambda     \label{eqn:ppd_hierarchical} \\
     &= \int \mathcal{L}(\dpred |\lambdapred)\, \pi_{\rm pop}(\lambdapred|\Lambda)\,\,p(\Lambda | \dobsvec)\, d\Lambda\,,  \nonumber
\end{align}
where $\mathcal{L}(\dpred |\lambdapred)$ is the single-event likelihood, $\pi_{\rm pop}(\lambdapred|\Lambda)$ is the population model, and $p(\Lambda | \dobsvec)$ is the hyper-posterior. 
Most commonly in GW population analyses, Eq.~\eqref{eqn:ppd_hierarchical} is integrated over $\dpred$, yielding a PPD on only parameters $\lambdapred$:
\begin{multline}
     \mathrm{PPD}(\lambdapred | \dobsvec)  \equiv p(\lambdapred | \dobsvec)  \\
       = \int \pi_{\rm pop}(\lambdapred|\Lambda)\,\,p(\Lambda | \dobsvec)\, d\Lambda \,\,.
    \label{eqn:ppd_event}
\end{multline}
We can also express the PPD for a full catalog with data $\dpredvec$ and parameters $\lambdapredvec$ (such that $\vec\lambda = \{\lambda_i\}_{N_{\rm obs}}$) rather than a single event:
\begin{multline}
    \mathrm{PPD}(\dpredvec,\lambdapredvec| \dobsvec)  \\ = \int \mathcal{L}(\dpredvec |\lambdapredvec)\, \pi_{\rm pop}(\lambdapredvec|\Lambda)\,\,p(\Lambda | \dobsvec)\, d\Lambda\,, \label{eqn:ppd_hierarchical_catalog}
\end{multline}
where the likelihood in the integrand is a product over single-event likelihoods
\begin{equation}
    \mathcal{L}(\dpredvec |\lambdapredvec) = \prod_i^{N_{\rm obs}} \mathcal{L}(\dpred_i |\lambdapred_i)\,\,, \label{eqn:L_catalog}
\end{equation} 
as is the population distribution,
\begin{equation}
    \pi_{\rm pop}(\lambdapredvec|\Lambda) = \prod_i^{N_{\rm obs}} \pi_{\rm pop}(\lambdapred_i|\Lambda)\,\,. \label{eqn:pi_pop_catalog}
\end{equation}
Equations~\eqref{eqn:ppd_hierarchical_catalog}, \eqref{eqn:L_catalog}, and \eqref{eqn:pi_pop_catalog} are useful when calculating posterior predictive $p$-values, as discussed in the following section. 
We return to discussing PPDs in Sec.~\ref{sec:event_vs_data_level}.

\begin{figure*}
    \centering
    \includegraphics[width=\linewidth]{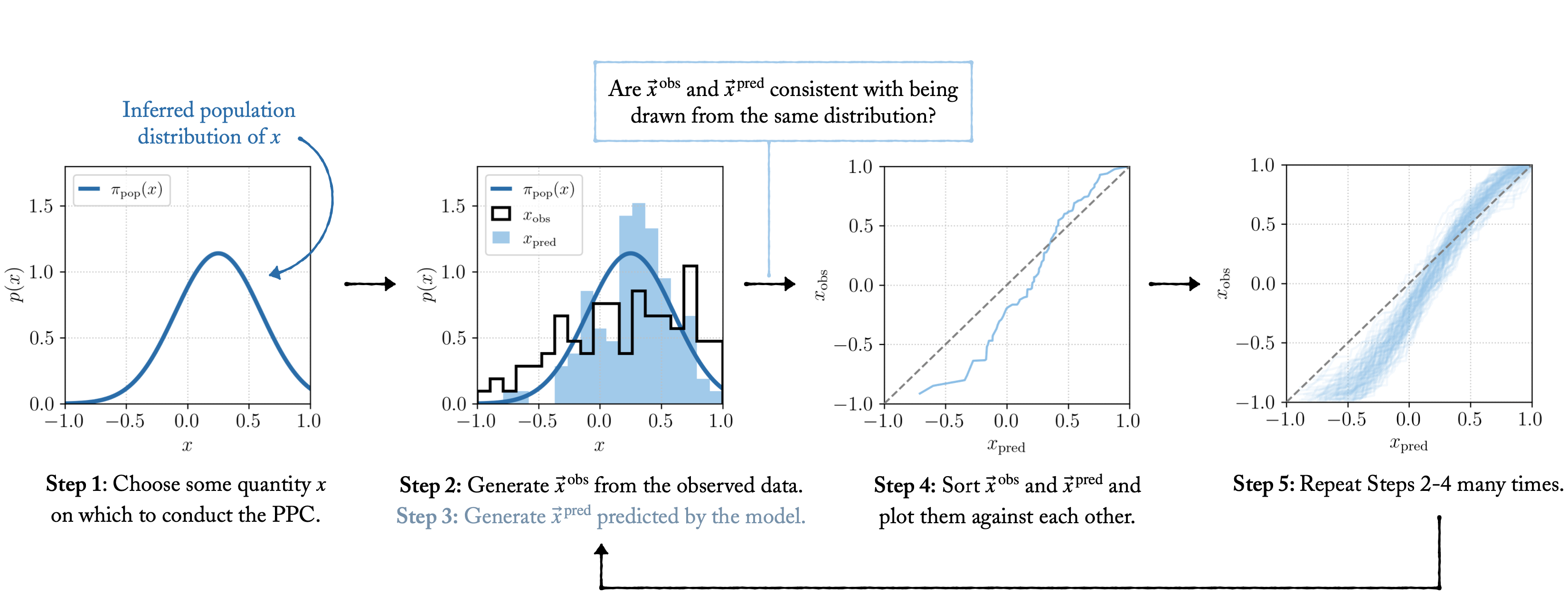}
    \caption{
    Schematic showing the steps for conducting a PPC on an arbitrary parameter $x$ with an inferred distribution $\pi_{\rm pop}(x)$; see also Algorithm \ref{alg:general_PPC}. In this toy example, we pretend that we have perfectly measured $\pi_{\rm pop}(x)$ with some parametrized model (dark blue line in panels 1 and 2). 
    We can tell that the observed data---whose true underlying $x$ distribution we are trying to ascertain---are \textit{inconsistent} with the inferred population distribution because catalogs with $x$-values generated from each (e.g., black vs.~blue histograms in panel 2) produce \textit{non-diagonal PPC traces} (panels 3 and 4).
    Thus, $\pi_{\rm pop}$ is insufficient to capture the shape of the true underlying distribution of $x$. 
    The different types of PPCs explored in this work correspond to exactly what happens between the first two panels i.e., how predicted and observed catalogs are generated.}
    \label{fig:schematic_ppc}
\end{figure*}
\begin{figure*}
    \centering
    \vspace{30px}
    \includegraphics[width=\linewidth]{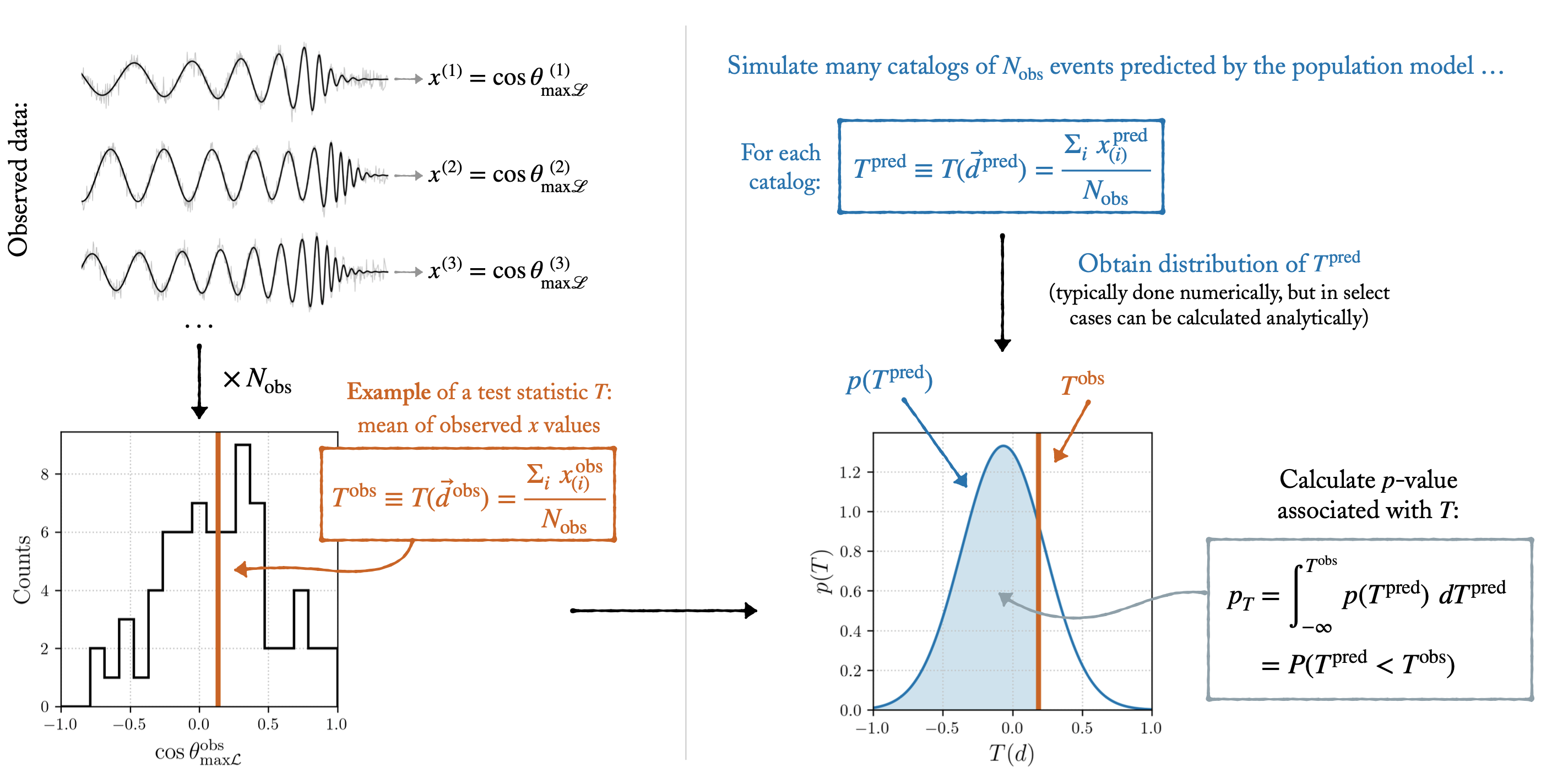}
    \caption{
    Schematic demonstrating how to calculate $p_T$ (Eq.~\ref{eqn:pvalue2}). In this example, we look at the data-level parameter $x=\cos\theta_{\maxL}$ and use the mean of $x$ as our test statistic $T$.
    Refer to Fig.~\ref{fig:schematic_ppc} for how observed and predicted data (catalogs) are generated. 
    }
    \label{fig:schematic_pvalue}
\end{figure*}

In practice, PPCs are conducted numerically by comparing $\dobsvec$ and/or $\lambdaobsvec$ to a series of simulated $\dpredvec$ and/or $\lambdapredvec$ drawn from Eq.~\eqref{eqn:ppd_hierarchical_catalog}.
All PPCs explored in this work can be broken down into the general steps provided in Algorithm \ref{alg:general_PPC} and also visualized in Fig.~\ref{fig:schematic_ppc}.

In the limit of a perfectly measured population distribution with $N_{\rm obs} = \infty$, each of the $\vec x^{\,\rm pred}$ vs.~$\vec x^{\,\rm obs}$ traces will follow an exact diagonal line.
In the realistic scenario where we have finite $N_{\rm obs}$, the \textit{average} of the traces should be consistent with the diagonal if the model is a good fit to the data~\cite[e.g.,][]{Sinharay:2003,Bayarri:2008,Fishbach:2019ckx,Miller:2020zox,Callister:2021}.
Any systematic differences between $\dpredvec$ and $\dobsvec$ will manifest as non-diagonal $\vec x^{\,\rm pred}$ vs.~$\vec x^{\,\rm obs}$ traces\editaccepted{, indicating the population model's inability to correctly infer $\dobsvec$}.
For the above to be a valid technique, $\dpredvec$ must be generated under the exact same set of assumptions as $\dobsvec$, such as detection threshold, single-event likelihood model, etc.

\begin{algorithm}[H]
\caption{General PPC algorithm, also visualized in Fig.~\ref{fig:schematic_ppc}. All PPCs explored in this work can be broken down into these steps. }
\label{alg:general_PPC}
    \begin{algorithmic}[1]
        \State Choose a parameter $x$ on which to conduct the check. 
        In this work we either look at true underlying values of $x\in\lambda$ (denoted $x_{\rm true}$; an example of an \textit{event-level parameter}) or associated maximum likelihood values of $x$ (denoted $x_{\maxL}$; an example of a \textit{data-level parameter}). These quantities are explained in detail in Sec.~\ref{sec:event_vs_data_level}. 
        \vspace{3pt}
        \State Draw $N_{\rm obs}$ values of $x$ that represent the observed data, which we will call $\vec x^{\,\rm obs}$.  This constitutes one ``observed catalog" with values of $x$ consistent with $\dobsvec$. The exact distribution $x$ is drawn from depends on the specific PPC in question, as explained in Secs.~\ref{sec:event_vs_data_level},~\ref{sec:partial_PPCs}, and \ref{sec:split_PPCs}. 
        \vspace{3pt}
        \State Draw $N_{\rm obs}$ values of $x$ predicted by the model, which we will call $\vec x^{\,\rm pred}$. This constitutes a ``predicted catalog" with values of $x$ consistent with one possible instantiation of $\dpredvec$. These are drawn from a distribution analogous to that in Step 2.
        \vspace{3pt}
        \State Sort the elements of $\vec x^{\,\rm pred}$ and $\vec x^{\,\rm obs}$ each from smallest to largest, and plot them against each other. This generates one ``trace" of a PPC. 
        \vspace{3pt}
        \State Repeat steps 2 through 4 many times to generate a collection of PPC traces. 
        \vspace{3pt}
        \State Choose some test statistic $T$ to assess the discrepancy between predicted and observed data. Calculate an associated $p$-value. step is discussed in detail in Sec.~\ref{subsec:pvalues} and visualized in Fig.~\ref{fig:schematic_pvalue}.
    \end{algorithmic}
\end{algorithm}

\subsection{Posterior predictive $p$-values}
\label{subsec:pvalues}

\editaccepted{A classical $p$-value is the probability that a test statistic $T$ computed on replicated data is \textit{at least as extreme} as its observed value, given an assumed distribution for $T$.}
A $p$-value closer to 0 indicates that the data are less likely to have occurred under said model. 
\editaccepted{Expressed through the \textit{posterior predictive $p$-value}, the preceding} idea can be extended to Bayesian statistics, where we move from a single test statistic that summarizes the data, to a distribution of test statistics that summarizes both the data \textit{and} our uncertainty on parameters of the model.
\editaccepted{The posterior predictive $p$-value has previously been used in searching for GWs with Pulsar Timing Arrays, where the GW detection summary statistic depends upon a posterior distribution for individual pulsar noise models~\cite{Vallisneri:2023xay,Meyers:2023nsi,NANOGrav:2024xhc,vanHaasteren:2025jvo}.
In this work, we instead apply it to LVK population analyses.
}

The PPD given in Eq.~\eqref{eqn:ppd} yields a posterior predictive $p$-value~\cite{Gelman:1996} of
\begin{multline}
    p_T(y^{\rm obs}) = \iint I_{[T(y^{\rm pred},\Theta) \geq T(y^{\rm obs},\Theta)]} \\ \times p(y^{\rm pred} |\Theta) ~p(\Theta | y^{\rm obs})~ dy^{\rm pred} d\Theta\,, 
    \label{eqn:pvalue}
\end{multline}
where $I$ is an indicator function,\editaccepted{
\begin{equation}
    I_{[X]} = \begin{cases}
        1 \text{ if $X$ true},\\
        0\text{ if $X$ false}.
    \end{cases}
\end{equation}
Thus, $p_T$ is the probability that replicated data $y^{\rm pred}$, drawn from a model $p(y^{\rm pred} |\Theta)$ with parameters $\Theta$ from the observed posterior $p(\Theta | y^{\rm obs})$, are more extreme than $y^{\rm obs}$, according to some test statistic $T$.
Figure~\ref{fig:schematic_pvalue} illustrates an example $p_T$ computation.
}

Calculating $p_T$ requires \editaccepted{choosing $T$}, a function from data and/or parameter space to the real numbers, compared between \editaccepted{the observed GW catalog and replications predicted by a given model}~\cite{Gelman:1996}. 
\editaccepted{A smart choice for} $T$ is context dependent and tailored to \editaccepted{the population features} one is trying to probe.
For instance, to test whether a model correctly predicts aligned \editaccepted{spin} tilts, a strong $T$ \editaccepted{could} be the fraction of events with $\cos\theta > 0.5$ or the ratio \editaccepted{$p(\cos\theta=1):p(\cos\theta=0)$}.
The efficacy of a given $T$ \editaccepted{also} depends on the \editaccepted{functional form of the} population model itself. 
For example, \editaccepted{a Gaussian distribution will, by definition, correctly infer the population's mean and standard deviation (in the absence of selection effects and assuming the true mean \editaccepted{and standard deviation are} within the prior), making these poor choices for test statistics.
Throughout this work, we use the four $T$ enumerated in Sec.~\ref{sec:result_summary} and first shown in Fig.~\ref{fig:pvalue_summary}.}

\editaccepted{For} hierarchical inference, the posterior predictive $p$-value \editaccepted{uses a} test statistic over the joint posterior predictive distribution of data \textit{and} parameters, requiring a more complicated integral than Eq.~\eqref{eqn:pvalue}. 
Here, we \editaccepted{consider} the probability that $T(\dpredvec,\lambdapredvec)\geq T(\dobsvec,\lambdaobsvec)$ where \editaccepted{$\lambdapredvec$ and $\lambdaobsvec$ are the parameters describing $\dpredvec$ and $\dobsvec$, respectively.} 
\editaccepted{Evaluating the integral in Eq.~\eqref{eqn:pvalue2} yields a single $p_T$ for assessing hierarchical model fit:}
\begin{widetext}
    \begin{equation}
        p_T(\dobsvec) = \iiiint  I_{[T(\dpredvec,\lambdapredvec) \geq T(\dobsvec,\lambdaobsvec)]} ~p(\dpredvec, \lambdapredvec\ | \Lambda)~p(\lambdaobsvec, \Lambda | \dobsvec) \,d\Lambda \,d\lambdapredvec\,d\lambdaobsvec\,d\dpredvec  \,.
        \label{eqn:pvalue2}
    \end{equation}
\end{widetext}
The two probability distributions in the integrand of Eq.~\eqref{eqn:pvalue2} can be understood as follows. 
The observed data give us constraints on our models' single-event parameters and hyper-parameters: $p(\lambdaobsvec, \Lambda | \dobsvec)$. 
The population model then generates single-event parameters, which in turn are used to produce predicted data: $p(\dpredvec, \lambdapredvec\ | \Lambda)$.
Equations \eqref{eqn:pvalue_eventlevel} and \eqref{eqn:pvalue_datalevel} in Sec.~\ref{subsec:event_vs_data_level_pvalues} present specific cases where Eq.~\eqref{eqn:pvalue2} can be simplified: $T$ that depend solely on $\vec d$ (data-level) versus solely on $\vec \lambda$ (event-level). 
\editaccepted{Generically, one could also calculate a \textit{distribution} of $p$-values over any of $\Lambda$, $\lambda$, or $\dpred$, instead of marginalizing over all of these parameters in tandem as we do in this work.}

Using the definitions in Eqs.~\eqref{eqn:pvalue} and \eqref{eqn:pvalue2}, $p_T$ is defined on the range $[0,1]$, with values closer to $0.5$ indicating a good fit. 
For ease of comparison with more widely used frequentist (rather than Bayesian) $p$-values, we decide to scale $p_T$ to obtain a new quantity, 
\begin{equation}
 \pvalue = 1 - 2 \times|\,p_T-0.5\,|\,, \label{eqn:real_pvalue}
\end{equation}
such that $\pvalue$ closer to 1 indicates a better fit, and $\pvalue$ closer to $0$ is obtained when $T^{\rm obs}$ is either greater \textit{or} less than $T^{\rm pred}$. 

\editaccepted{A known feature of Bayesian model checking is that posterior predictive $p$-values are \textit{not uniform under the null hypothesis} (i.e., model matches data), complicating their interpretation~\cite{Bayarri:2000,Bayarri:2008, Gelman:1996, Meng:1994, Robins:2000, Hjort:2006}.
Unlike frequentist $p$-values, where $P(p\leq \alpha)=\alpha$, posterior predictive $p$-values concentrate around $p_T\sim 0.5$ (equivalently, $\pvalue \sim 1$) for traditional PPCs under the null hypothesis because the same data are used to both estimate parameters and evaluate the test statistic~\cite{Bayarri:2008}.
However, \citet{Meng:1994} shows that $P(p_T \leq \alpha) \leq 2\alpha$ under the null, implying via Eq.~\eqref{eqn:real_pvalue} that $\pvalue$ is a \textit{conservative} $p$-value, i.e., $P(\pvalue \leq \alpha) \leq \alpha$. 
We therefore treat the canonical threshold of $\pvalue \lesssim 0.05$ as a conservative indication of discrepancy between model and data: 
if $\pvalue \lesssim 0.05$, the model is strongly disfavored; if $\pvalue$ $\gtrsim 0.05$, the model \textit{may} still be a poor fit, but we cannot say so conclusively. 
}

\section{Event vs.~Data Level PPCs}
\label{sec:event_vs_data_level}

In hierarchical inference, tests for goodness of fit can be conducted on different ``levels" in the hierarchy: population-level, event-level, or data-level parameters~\cite{Fishbach:2019ckx,Essick:2023upv}.
\editaccepted{In this work, we avoid population-level parameters, and we instead focus on comparing event- and data-level PPCs.
For completeness, population-level checks include leave-one-out analyses, where $\Lambda$ posteriors are compared across catalogs with each observed event iteratively removed, or robustness assessments for a particular inferred feature (e.g., an overdensity at some mass or spin). 
By simulating populations \textit{without} a feature of interest, one can quantify how often it is spuriously inferred due to Poisson noise~\cite{Farah:2023vsc}---a form of the posterior predictive $p$-value.}

In GW data analysis, PPCs are most commonly conducted on \textbf{event-level} parameters~\cite[with exceptions, e.g.,][]{Fishbach:2019ckx, Mould:2026sww}\editaccepted{, the ``true"} masses, spins, etc.~that describe the underlying astrophysical system.
The probability of drawing parameters $\lambdapred_{\rm true}$ given our observed data $\dobsvec$, and the population model $\pi_{\rm pop}$ is given in Eq.~\eqref{eqn:ppd_event}, where we average $\pi_{\rm pop}(\lambda^{\rm pred}_{\rm true} |\Lambda)$ over the hyper-posterior $p(\Lambda | \dobsvec)$.
In reality, the observed data have a further restriction: detectability. 
To faithfully compare predicted and observed catalogs, 
we care about a model's power to \editaccepted{predict} true parameters that \editaccepted{yield} a \textit{detectable} GW signal, and so must incorporate selection effects into Eq.~\eqref{eqn:ppd_event}:
\begin{multline}
    p(\lambdapred_{\rm true}\, | \,\dobsvec, \mathrm{det}) =  \mathrm{PPD}(\lambdapred_{\rm true} | \dobsvec)  P_{\rm det}(\lambdapred_{\rm true})
    \\ = \mathrm{PPD}(\lambdapred_{\rm true} | \dobsvec) \, \int I_{[D(d)>D_{\rm thr}]} ~\mathcal{L}(d| \lambdapred_{\rm true}) \,d d   \,\,.
    \label{eqn:ppd_event_with_detection}
\end{multline}
\editaccepted{Here,} $P_{\rm det}$ is the detection probability, equivalent to 0 or 1 \editaccepted{according to} whether \editaccepted{the} \textit{detection statistic} $D$ evaluated on the data \editaccepted{exceeds a known} threshold $D_{\rm thr}$,
and the PPD is given in Eq.~\eqref{eqn:ppd_event}.
\editaccepted{The} integral in Eq.~\eqref{eqn:ppd_event_with_detection} \editaccepted{averages the single-event likelihood} $\mathcal{L}(d| \lambdapred_{\rm true})$ over all replications of data $d$ \editaccepted{ (i.e., noise instantiations) that contain a detectable signal from} a BBH with true parameters $\lambdapred_{\rm true}$.
Equation \eqref{eqn:ppd_event_with_detection} is the basis for the event-level PPCs.\footnote{
    Technically, Eqs.~\eqref{eqn:ppd_event_with_detection} and \eqref{eqn:ppd_data_with_detection} should include a factor of $1/p({\rm det}|\dobs)$. However, $p({\rm det}|\dobs) = 1$ because we have, by definition, detected our observed events.
}

\editaccepted{In contrast,} \textbf{data-level} quantities characterize the actual data recorded in the detectors, such as the source parameters' $\maxL$ values
\begin{equation}
    \lambda_{\maxL}\equiv \{ \lambda \;|\; \mathcal{L}(d| \lambda) = \mathrm{max}[\mathcal{L}(d | \lambda)]\}\,.
    \end{equation}
\editaccepted{
In the cases we consider, event-level parameters are determined \textit{probabilistically} from the data, as any posterior draw is, by definition, an equally valid sample of the true underlying parameters of the BBH, while data-level parameters stem \textit{deterministically} from the data: each posterior distribution only has one $\maxL$ point.}

\editaccepted{The probability of producing} \textit{data} ($\dpred$)\editaccepted{---}rather than the true BBH source parameters that (combined with random noise) generated said data\editaccepted{---is} \cite{Fishbach:2019ckx}
\begin{multline}
    p(\dpred\,|\, \dobsvec, \mathrm{det}) =  \mathrm{PPD}(\dpred | \dobsvec)  P_{\rm det}(\dpred)
    \\= \int I_{[D(\dpred)>D_{\rm thr}]} ~\mathcal{L}(\dpred| \lambda) ~ \mathrm{PPD}(\lambda| \dobsvec)\, d\lambda \,,
    \label{eqn:ppd_data_with_detection}
\end{multline}
where the PPD in the integrand is again \editaccepted{given} in Eq.~\eqref{eqn:ppd_event}.\footnote{\editaccepted{We use the same symbol $P_{\rm det}$ and change the argument between $d$ and $\lambda$ depending on which we are referring to, such that $P_{\rm det}(\lambda) = \int P_{\rm det}(d) \mathcal{L}(d|\lambda) dd$.}}
Equations~\eqref{eqn:ppd_event_with_detection} and  ~\eqref{eqn:ppd_data_with_detection} are similar, but the former marginalizes over noise instantiations ($dd$) while the latter does so over masses, spins, etc.~($d\lambda$).
Equation~\eqref{eqn:ppd_data_with_detection} is the basis for the data-level PPCs: to generate $\lambdapred_{\maxL}$, we draw $\dpred$ from Eq.~\eqref{eqn:ppd_data_with_detection} and then calculate \editaccepted{its $\maxL$ parameters} under the same likelihood model used \editaccepted{in the} original parameter estimation \editaccepted{for} $\dobs$.

\editaccepted{In the following subsections, we provide an \textit{algorithmic} description of how to conduct PPCs using Eqs.~\eqref{eqn:ppd_event_with_detection} and \eqref{eqn:ppd_data_with_detection}, which may be more illuminating than the integral expressions themselves.}
We present results for PPC traces (Sec.~\ref{subsec:event_vs_data_traces}) and $p$-values (Sec.~\ref{subsec:event_vs_data_level_pvalues}) for event-level and data-level parameters, and discuss pros and cons of each.
Additionally, in Appendix~\ref{app:toy_model} we repeat these exercises with a simple, one-dimensional, analytic toy model; we direct readers there to gain intuition.

\subsection{Event vs.~data level PPC traces}
\label{subsec:event_vs_data_traces}

Here, we add specificity to the general PPC algorithm presented in Algorithm \ref{alg:general_PPC}.
One trace of a traditional \textbf{event-level} PPC is generated with the following steps~\cite{Callister:2021} in Algorithm \ref{alg:event_level_PPC}:
\begin{algorithm}[H]
\caption{Generating one trace of a traditional event-level PPC.}
\label{alg:event_level_PPC}
    \begin{algorithmic}[1]
        \State Draw one sample of $\Lambda$ from the hyperposterior $p(\Lambda|\dobsvec)$.
        \vspace{3pt}
        \State \textit{Observed values} ($\vec\lambda^{\rm obs}_{\rm true}$): Draw one sample from each of the $N_{\rm obs}$ individual-event posteriors in the observed catalog, \textit{reweighted}~\cite{Callister:2021} from its parameter-estimation prior to $\pi_{\rm pop}(\lambda_{\rm true}|\Lambda)$.
        \vspace{3pt}
        \State \textit{Predicted values} ($\vec\lambda^{\rm pred}_{\rm true}$): Draw $N_{\rm obs}$ \textit{detectable}\footnote{Specifically, we draw $N_{\rm obs}$ values from the found injections used to determine detection efficiency (Eq.~\ref{eqn:xi}) reweighted from the injected distribution to $\pi_{\rm pop}$.} values of $\lambda_{\rm true}$ from the distribution $\pi_{\rm pop}(\lambda_{\rm true}|\Lambda)$, following Eq.~\eqref{eqn:ppd_event_with_detection}. 
    \end{algorithmic}
\end{algorithm}
\noindent In contrast, for the \textbf{data-level} PPC, there is only one set of observed values, consisting solely of the maximum likelihood parameters from each individual-event posterior ($\lambdaobs_{\maxL}$). \editaccepted{Our approach for finding $\maxL$ parameters is described in Sec.~\ref{subsec:simulated_pops} and Appendix~\ref{app:maxL_finding}.}
To create each set of predicted values, we follow Algorithm \ref{alg:data_level_PPC}:
\begin{algorithm}[H]
\caption{Generating the predicted values for one trace of a data-level PPC.}
\label{alg:data_level_PPC}
    \begin{algorithmic}[1]
        \State Draw one sample of $\Lambda$ from the hyperposterior $p(\Lambda|\dobsvec)$.
        \vspace{3pt}
        \State Draw $N_{\rm obs}$ \textit{detectable} values of $\lambda_{\rm true}$ from the distribution $\pi_{\rm pop}(\lambda_{\rm true}|\Lambda)$ (Eq.~\ref{eqn:ppd_event_with_detection}).
        \vspace{3pt}
        \State Generate a GW signal for each $\lambda_{\rm true}$ and inject it into random Gaussian noise drawn from our power spectral density.\footnote{We self-consistently used a threshold on optimal signal-to-noise ratio (SNR) to quantify detectability when analyzing the simulated populations in \citet{Miller:2024sui} and thus do the same in this work, as discussed in Appendix~\ref{app:sim_pop}. The use of optimal SNR means that detectability can be assessed in Step 3, rather than Step 4 (as would be necessitated for matched-filter SNR).} 
        \vspace{3pt}
        \State For each simulated signal, find the maximum likelihood value of the parameters ($\lambdapred_{\maxL}$) using the same method as was used for $\lambdaobs_{\maxL}$. 
    \end{algorithmic}
\end{algorithm}
\noindent We repeat the steps for the event- and data-level PPCs 1,000 times apiece, stochastically sampling over $\Lambda$ to obtain catalog realizations (``traces'').

\begin{figure*}
\centering
    \includegraphics[width=0.85\textwidth]{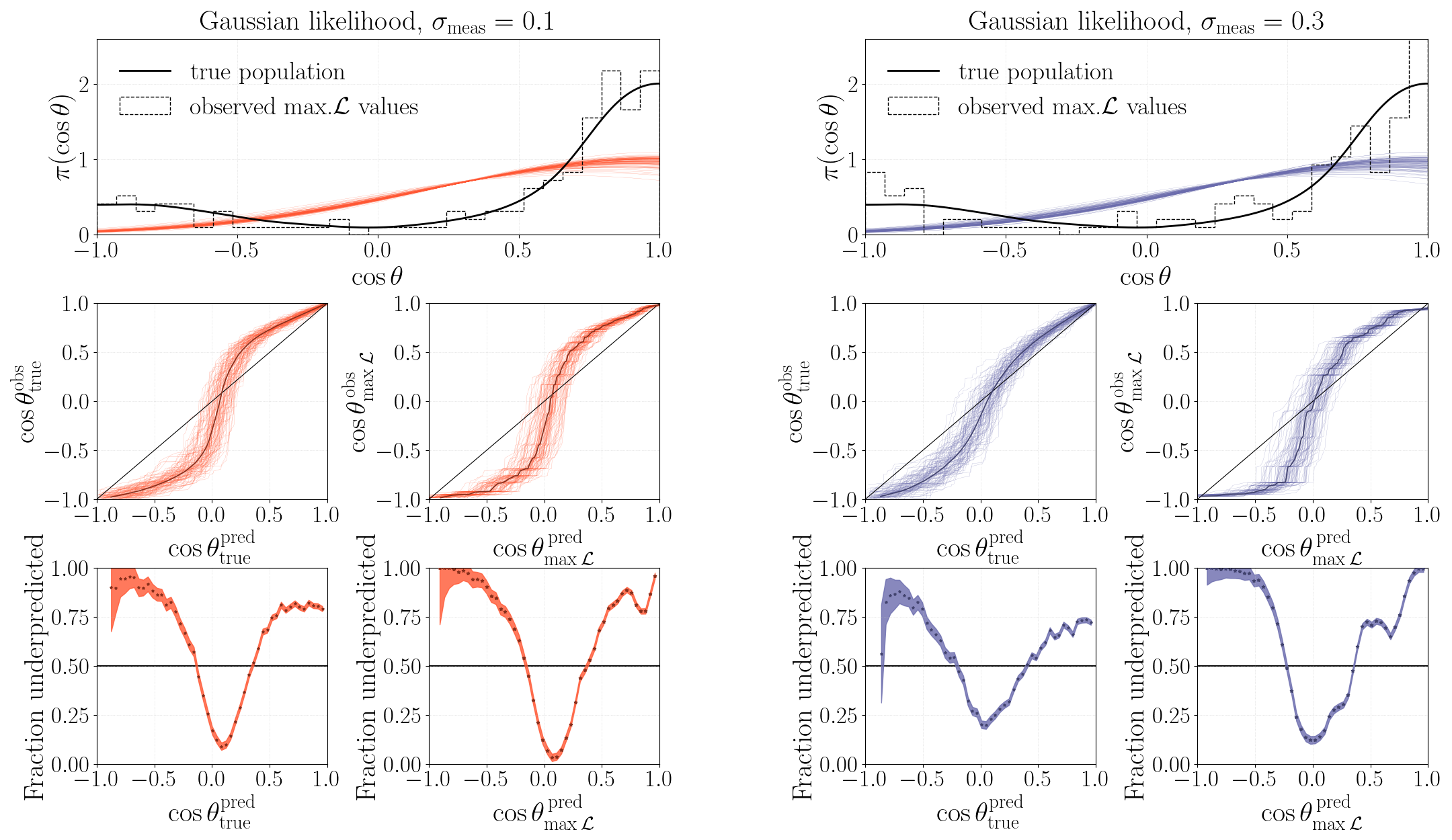}\\
    \vspace{20pt}
    \includegraphics[width=0.85\textwidth]{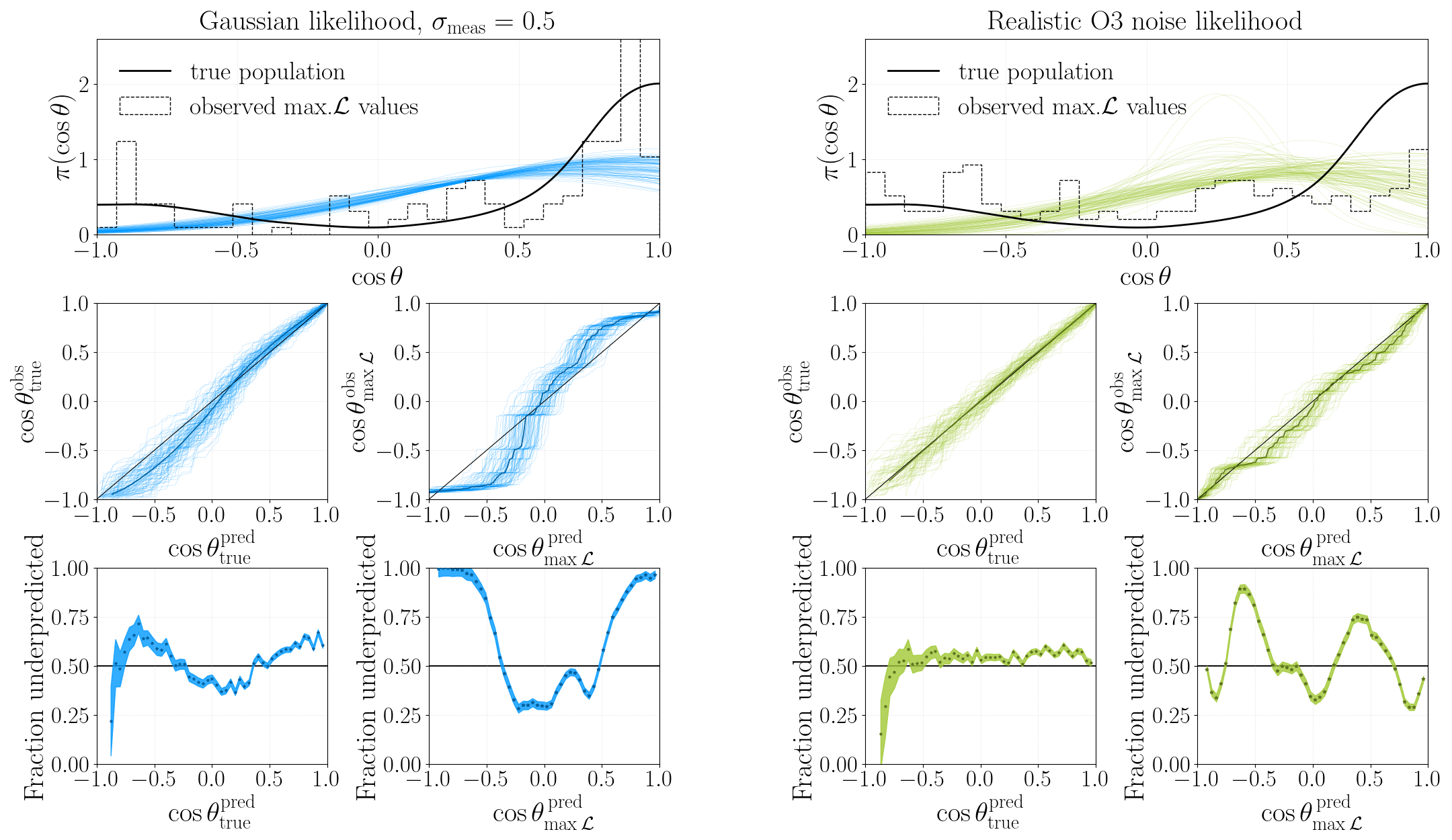}
    \caption{Results of population inference and event-level versus data-level PPCs for $\cos\theta$ using a Gaussian population model on simulated catalogs containing $70$ events which are from a bimodal population.
     Each quadrant/color shows a different individual-event likelihood: Gaussian likelihoods with measurement uncertainty $\sigma_{\rm meas}= 0.1\ \text{(red)},\ 0.3\ \text{(purple), and } 0.5\ \text{(blue)}$ versus the realistic O3 noise likelihood (green).
     The following descriptions correspond to the subplots within each quadrant.
    \textit{(First row)}: The true (black, solid) population distribution for $\cos\theta$ compared to traces obtained from hyper-posterior draws (colors).
    Maximum likelihood values of $\cos\theta$ from the observed histogram are shown in the black-dashed histograms.
    \textit{(Second row)}: Traditional event-level (left) versus data-level (right) PPC traces from 100 catalogs. The bold line is the average of the traces across the catalog instantiations.
    If the population model is a good fit to the data, these traces should follow the diagonal (black) on average.
    \textit{(Third row)}: The fraction of catalogs within a given bin of $\cos\theta$ for which the traces in the middle row have slope $<$ 1, indicating the model underpredicts the true population. Points mark the average over ten 100-trace PPCs; the shaded region indicates $\pm N^{-1/2}$ error where $N$ is the total number of draws in that bin across all 1,000 traces.}
    \label{fig:data_event_PPCs}
\end{figure*}

Figure \ref{fig:data_event_PPCs} \editaccepted{presents event- and data-level PPCs (left and right columns within each color)} for $\cos\theta$ of our simulated population with $N_{\rm obs} = 70$ events \editaccepted{and four different individual-event likelihoods (different colors)}.\footnote{
    \editaccepted{The event-level PPC traces in Fig.~\ref{fig:data_event_PPCs} are replicated from Figs.~3 and 4 of \citet{Miller:2024sui}.}
}
\editaccepted{The top row of Fig.~\ref{fig:data_event_PPCs} demonstrates that }a Gaussian population model \editaccepted{is} a poor fit \editaccepted{(lines of various colors)} to the true bimodal $\cos\theta$ distribution of our simulated population \editaccepted{(black solid line)---a deliberate mismatch that lets us test how well the different PPCs identify model misspecification}.
\editaccepted{We show results for a case where the population \textit{is} well-described by the model in Appendix \ref{subapp:toy_model_good}.}

With low individual-event measurement uncertainties of $\sigma_{\rm meas} = 0.1, 0.3$ (red, purple), the discrepancy between the observed and predicted \editaccepted{catalogs} is visually clear in both the event-level and data-level PPCs: their traces are highly non-diagonal.
However, when $\sigma_{\rm meas}$ increases to $0.5$ (blue) or we use realistic O3 noise (green),
the event-level PPC traces begin to approach the diagonal\editaccepted{, indicating} that the model is a good fit to the data even though \textit{a priori} we know it is not. 
\editaccepted{Figure \ref{fig:data_event_PPCs} thus exhibits how event-level PPCs struggle to diagnose an inaccurate population model when individual-event posterior distributions are prior dominated, as previously shown in \citet{Miller:2024sui}.}
\editaccepted{This failure} stems from Step 2 of the event-level PPC algorithm (Algorithm \ref{alg:event_level_PPC})\editaccepted{: when individual-event measurement uncertainty is high, each re-weighted posterior} becomes near identical to the population distribution, \editaccepted{so the observed and predicted PPC traces closely coincide.}
\editaccepted{Here, the} information contained in the population distribution dominates \editaccepted{over the likelihood}, regardless of whether that information comes from our (perhaps arbitrary) assumptions about the shape of the distribution or the actual limited information in the observed events themselves.
\editaccepted{Implications are concerning: when analyzing real data, we only have access to the inferred populations, not the truth.}

\editaccepted{We find that \textbf{data-level PPCs outperform standard event-level PPCs at diagnosing population model misspecification for poorly constrained parameters} like spin tilts}.
Data-level PPCs avoid re-weighting individual-event posteriors to the population, making them immune to the prior \editaccepted{domination that can plague} event-level PPCs.
\editaccepted{The improvement is most obvious for} the $\sigma_{\rm meas}=0.5$ single-event likelihood \editaccepted{(blue):} the data-level PPC shows significantly more non-diagonality \editaccepted{than the event-level PPC}, properly identifying \editaccepted{model-data mismatch}. 
\editaccepted{Deviations in the data-level PPC traces are also apparent in the realistic noise case (green) but with a different pattern than the Gaussian likelihood cases}.
Such behavior is not unexpected: Gaussian intuition only goes so far, and realistic GW likelihoods have a \editaccepted{less-simplistic, higher-dimensional} structure. 
For reference, the observed $\maxL$ values of $\cos\theta$ \editaccepted{are shown in} black-dashed histograms in the top row of each quadrant of Fig.~\ref{fig:data_event_PPCs}\editaccepted{; these increasingly stray from the true population (solid black line) as $\sigma_{\rm meas}$ increases (top left to bottom right), i.e., noise scatters the $\maxL$ values from their corresponding true values}.
In Appendix~\ref{app:toy_model}, we \editaccepted{report} analogous behavior with a toy model.

\editaccepted{In addition to flagging that a model is a poor fit, PPCs reveal \textit{where} the model fails---guiding model improvement beyond what is accessible with Bayes factors.}
\editaccepted{To this end, the third row in each quadrant of Fig.~\ref{fig:data_event_PPCs} shows the fraction of traces for which the Gaussian model under-predicts the event count across a grid in $\cos\theta$.}
\editaccepted{The fraction under-predicted is calculated from the} \textit{slopes} of the \editaccepted{PPC} traces\editaccepted{: if,} at a given value of $\cos\theta$, the slope of a trace is steeper than the diagonal, the model is \textit{over-predicting} events with that $\cos\theta$; if the slope is shallower than the diagonal, \editaccepted{the model} is \textit{under-predicting}.
\editaccepted{A good model fit yields a fraction consistent with 0.5 (within error bars) for all $\cos\theta$; departures from $ 0.5$ indicate where new model features are needed.}
\editaccepted{In Fig.~\ref{fig:data_event_PPCs}, we average the fraction under-predicted over} ten $100$-trace PPCs \editaccepted{(dots)}; the shaded regions enclose $\pm N^{-1/2}$ error, where $N$ is the total number of draws in \editaccepted{a given} $\cos\theta$ bin across all 1,000 traces.
Near $\cos\theta\sim -1$, \editaccepted{both predicted and observed catalogs contain fewer events, giving} consistently larger uncertainty regardless of single-event likelihood.

\editaccepted{For} $\sigma_{\rm meas} = 0.1,0.3$, both event- and data-level PPCs identify the \editaccepted{Gaussian model's} inability to capture the \editaccepted{bimodality} of the true population distribution, with \editaccepted{the} fraction under-predicted clearly inconsistent with $0.5$ across $\cos\theta$.
\editaccepted{At} $\sigma_{\rm meas} = 0.5$, the event-level PPC shows little scatter around a fraction under-predicted of $0.5$, while the data-level PPC shows large deviation.
The curve mirrors the shape of the true population relative to \editaccepted{that inferred with} the model: for \editaccepted{$|\cos\theta|\gtrsim 0.5$} the model under-predicts the truth while for \editaccepted{$|\cos\theta|\lesssim 0.5$, it} over-predicts. 
\editaccepted{If we did not have prior knowledge of} the true $\cos\theta$ distribution, seeing \editaccepted{this} PPC result would \editaccepted{reveal} unresolved bimodality. 

The fraction under-predicted curve for the realistic-noise case (green) has a similar shape to the Gaussian \editaccepted{likelihood cases when $|\cos\theta|\lesssim 0.5$}, but differs as $\cos\theta \to \pm1$.
We consider several plausible explanations. 
First, data-level PPCs are sensitive to small-number statistics: \editaccepted{since we} compare each predicted trace to the \textit{same} observed trace, a random ``bump" in the observed $\maxL$ values from Poisson uncertainty can have an outsized effect.
\editaccepted{Such fluctuations are responsible for the deviations at $\cos\theta\sim0.7$ ($\sigma_{\rm meas}=0.3$) and $\cos\theta\sim0.3$ ($\sigma_{\rm meas}=0.5$).
The realistic noise case could be explained by similar (albeit larger) deviations as $\cos\theta \to \pm1$.
We also observe seemingly-random over-densities in the realistic-noise $\cos\theta_{\maxL}^{\,\rm obs}$ histogram at $\cos\theta \sim -0.7$ and $0.4$ that coincide to the turning points in the fraction under-predicted curve.} 
Second, \editaccepted{the likelihood may carry} so little information about $\cos\theta$ that \textit{all} we are seeing is the impact of small-number statistics\editaccepted{---an explanation supported by the fact that different $\cos\theta_{\rm true}$ distributions all produce near identical distributions of $\cos\theta_{\maxL}$ in the realistic-noise case (Fig.~\ref{fig:true_vs_maxL_cdfs_3pops})}.
Third, random noise \editaccepted{might impact the} full 15-dimensional BBH likelihood \editaccepted{differently than} the Gaussian likelihood, especially near \editaccepted{parameter space edges}.
\editaccepted{From the observed \editaccepted{realistic-noise} $\maxL$ histogram, we cannot cleanly separate the impacts of the preceding three scenarios.
However, their combined signature is apparent: the realistic-noise $\cos\theta_{\maxL}$ distribution is flatter than those from the Gaussian likelihoods, retaining less information about the $\cos\theta_{\rm true}$ distribution and leading to divergent behavior at  $\cos\theta_{\maxL} \to \pm1$.}
\\
\vspace{-0.5cm}

\subsection{Event vs.~data level $p$-values}
\label{subsec:event_vs_data_level_pvalues}

PPC traces \editaccepted{offer} a qualitative assessment of a model's goodness of fit.
While the fraction \editaccepted{of events} under-predicted across parameter space is one useful \editaccepted{quantitative} metric, we can also look at posterior predictive $p$-values ($\pvalue$), as \editaccepted{introduced} in Sec.~\ref{subsec:pvalues}.
Equations to calculate $p_T$ using event and data-level parameters are given in Eqs.~\eqref{eqn:pvalue_eventlevel} and \eqref{eqn:pvalue_datalevel}, respectively:

\begin{widetext}
\begin{align}
     & \hspace{3in} \mathrm{Event~level:} \nonumber 
     \\&p_T(\dobsvec) =  \iiint I_{[T(\lambdapredvec)\geq T(\lambdaobsvec)]} ~P_{\rm det}(\lambdapredvec,\lambdaobsvec)~\pi_{\rm pop}(\lambdapredvec |\Lambda)~p(\lambdaobsvec| \dobsvec,\Lambda )~p(\Lambda|\dobsvec)~d\Lambda ~d\lambdaobsvec ~d\lambdapredvec\,,
    \label{eqn:pvalue_eventlevel} \\\nonumber\\
    & \hspace{3in} \mathrm{Data~level:} \nonumber 
     \\ & p_T(\dobsvec) = \iiint I_{[T(\dpredvec)\geq T(\dobsvec)]}~P_{\rm det}(\dpredvec)~\mathcal{L}(\dpredvec|\lambdapredvec)~\pi_{\rm pop}(\lambdapredvec | \Lambda)~p(\Lambda|\dobsvec)~d\Lambda ~d\lambdapredvec~d\dpredvec\,.
    \label{eqn:pvalue_datalevel}
\end{align}
\end{widetext}
\editaccepted{Eqs.~\eqref{eqn:pvalue_eventlevel} and \eqref{eqn:pvalue_datalevel}} are derived from Eq.~\eqref{eqn:pvalue2} in Appendix \ref{app:pvalues}. 
To get $\pvalue$, we then use Eq.~\eqref{eqn:real_pvalue}. 
Except for in extremely rare cases, these $p$-values cannot be calculated analytically.

In both Eqs.~\eqref{eqn:pvalue_eventlevel} and \eqref{eqn:pvalue_datalevel}, $p(\Lambda | \dobsvec)$ is the hyperposterior on $\Lambda$ given the observed data $\dobsvec$ (Eq.~\ref{eqn:pop_posterior}) and $\pi_{\rm pop}(\lambdapredvec |\Lambda)$ is the corresponding population distribution (Eq.~\ref{eqn:pi_pop_catalog}).
Algorithmically, we draw $\Lambda$ from $p(\Lambda | \dobsvec)$ and then $\lambdapredvec$ from $\pi_{\rm pop}(\lambdapredvec |\Lambda)$ for each event in the catalog.
\editaccepted{For} the event-level $p_T$, this $\Lambda$ is also used to calculate  $p(\lambdaobsvec| \dobsvec,\Lambda)$, the population-reweighted individual-event posteriors, from which we draw $\lambdaobsvec$.
The test statistic $T$ is calculated on $\lambdaobsvec$ and $\lambdapredvec$ and the indicator function $I$ is evaluated.
Selection effects on the predicted parameters (Eq.~\ref{eqn:ppd_event_with_detection}) are incorporated via the $P_{\rm det}(\lambdapredvec,\lambdaobsvec)$ term.

\editaccepted{For} the data-level $p_T$, there is no individual-event posterior \editaccepted{in the integral}, just the likelihood $\mathcal{L}(\dpredvec|\lambdapredvec)$ (Eq.~\ref{eqn:L_catalog}) which, crucially, does not depend on $\Lambda$.
Algorithmically, the population model generates single-event parameters $\lambdapred$, which in turn are used to produce predicted data $\dpred$ by evaluating a waveform model at $\lambdapred$ and injecting it onto Gaussian noise \editaccepted{under} the same waveform model and noise power-spectral density as the observed data. 
To obtain a catalog ($\dpredvec$), this \editaccepted{process} is repeated $N_{\rm obs}$ times.
The test statistic $T$ is calculated on $\dobsvec$ and $\dpredvec$ and the indicator function $I$ is evaluated. 
Here, \editaccepted{one can think of the $T$ computation as} \textit{including} finding $\maxL$ parameters from each $\vec d$. 
Selection effects on the predicted data are applied via $P_{\rm det}(\dpredvec)$, cf.~Eq.~\eqref{eqn:ppd_data_with_detection}.
Since by definition we have already detected $\dobsvec$, no corresponding selection term need be applied.

\begin{figure*}
    \centering
    \includegraphics[width=\linewidth]{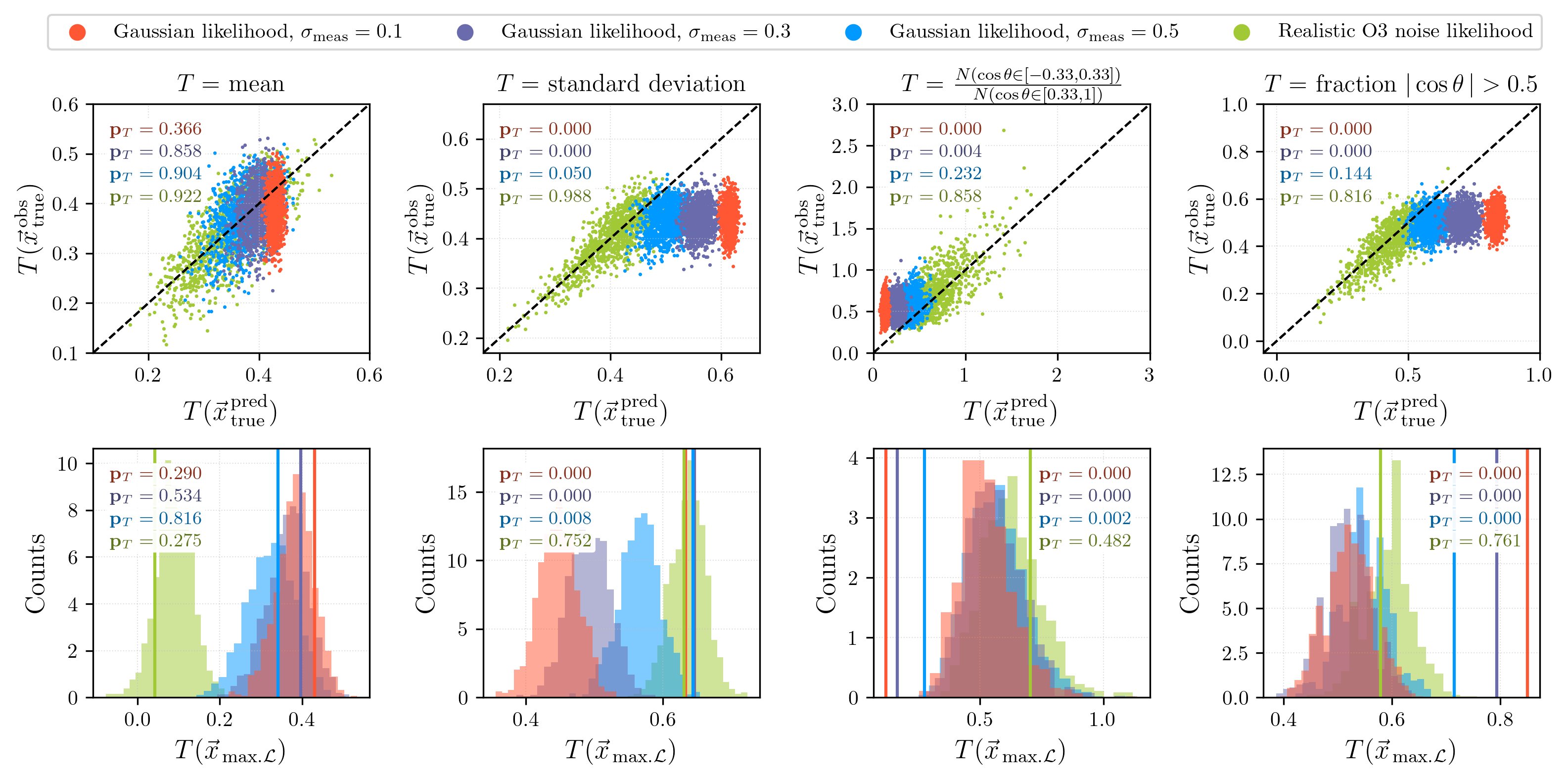}
    \caption{
    Distributions and associated posterior predictive $p$-values ($\pvalue$) for four test statistics $T$ calculated on $x=\cos\theta$ for 500 catalogs, each with 70 events, using the four likelihood variants: Gaussian likelihoods with $\sigma_{\rm meas} = 0.1$ (red), $0.3$ (purple), and $0.5$ (blue), compared to the realistic O3 noise likelihood (green). 
    From left to right, $T$ is the mean of $\cos\theta$ for each catalog, its standard deviation, the ratio of number of events with $\cos\theta\in[-0.33,0.33]$ to $\cos\theta\in[0.33,1]$, and the fraction of events with $|\cos\theta\,|>0.5$.
    The top row shows $T$ on event-level parameters ($\cos\theta_{\rm true}$), with values calculated from $\dpredvec$ on the horizontal axis and $\dobsvec$ on the vertical axis.
    The bottom row shows $T$ on data-level parameters ($\cos\theta_{\maxL}$), with the histogram showing those from $\dpredvec$ and the vertical line showing that from $\dobsvec$. 
    }
    \label{fig:pvalues_event_vs_data}
\end{figure*}

Figure~\ref{fig:pvalues_event_vs_data} shows distributions for \editaccepted{four example} test statistics $T$ and corresponding values of $\pvalue$ \editaccepted{(also collected in Fig.~\ref{fig:pvalue_summary})}.
\editaccepted{From left to right, $T$ is the mean $\cos\theta$ across the catalog}, the standard deviation, the ratio of events with $\cos\theta\in[-0.33,0.33]$ to $\cos\theta\in[0.33,1]$, and the fraction \editaccepted{of events} with $|\cos\theta|>0.5$\editaccepted{; the latter two specifically probe the population's bimodality. 
As before, we examine catalogs of events with Gaussian (red, purple, blue) and realistic-noise (green) single-event likelihoods.}
The top row compares \editaccepted{the test statistics} on the event-level $\cos\theta_{\rm true}$. The horizontal axis \editaccepted{is} $T$ from a predicted catalog generated from one hyper-parameter draw $\Lambda$ \editaccepted{and} the vertical axis \editaccepted{is} $T$ from an observed catalog \editaccepted{under} the same $\Lambda$; one scatterpoint \editaccepted{thus} corresponds to one event-level PPC trace in Fig.~\ref{fig:data_event_PPCs}.
Each event-level \editaccepted{$\pvalue$} is the fraction of scatterpoints below the diagonal, scaled \editaccepted{by} Eq.~\eqref{eqn:real_pvalue}. 
The bottom row shows each \editaccepted{test statistic} on the data-level $\cos\theta_{\maxL}$.
The histogram \editaccepted{is} $T$ from predicted catalogs \editaccepted{and} the vertical line is $T$ from the (single) observed catalog. 
Each data-level \editaccepted{$\pvalue$} is the fraction of points to the right of the vertical line, again scaled \editaccepted{by Eq.~\eqref{eqn:real_pvalue}}.
Model misspecification is successfully identified when $\pvalue \lesssim 0.05$.

For all four likelihoods, the data-level $\pvalue$ are more discerning \editaccepted{(i.e., lower)} than the event-level. 
Most notable is the Gaussian likelihood \editaccepted{case} with $\sigma_{\rm meas}=0.5$ (blue). 
When $T$ is the ratio \editaccepted{of} systems with $\cos\theta\in[-0.33,0.33]$ to $\cos\theta\in[0.33,1]$ (third column)\editaccepted{, the event-level is $\pvalue = 0.232$, indicating a good model fit, while the data-level is $\pvalue=0.002$, well below the threshold for significance}.
The contrast is even more stark \editaccepted{when $T$ is the fraction of systems with $|\cos\theta|>0.5$ (fourth column)}: the event-level $\pvalue=0.144$ and the data-level $\pvalue=0$. 

\editaccepted{For every $T$ we explore,} the realistic O3 noise likelihood case (green) yields an event-level $\pvalue >0.8$, \editaccepted{quantitatively confirming} that the event-level PPC cannot identify model misspecification. 
The data-level $\pvalue$ fares marginally better, but \editaccepted{remains $> 0.05$ throughout, meaning that it is} not considerably more discerning \editaccepted{at O3 sensitivity---at least for this particular observed catalog instantiation}. 
\editaccepted{The case of $T$ as the fraction of events with $|\cos\theta|>0.5$ is especially concerning.
Around $85\%$ of events in the true, underlying population have $|\cos\theta|>0.5$,  while the Gaussian model predicts $\sim 50\%$, yet neither the event- nor data-level PPC identifies this large discrepancy. 
The failure at the data level arises because only $\sim 50\%$ of the corresponding observed $\maxL$ distribution also satisfy $|\cos\theta|>0.5$ (Fig.~\ref{fig:data_event_PPCs}), meaning the likelihood is so uninformative about $\cos\theta$ that nearly all information about the corresponding true distribution is lost; see also the figures in Appendix \ref{app:maxL_finding}.}

\begin{figure*}
    \centering
    \includegraphics[width=0.9\linewidth]{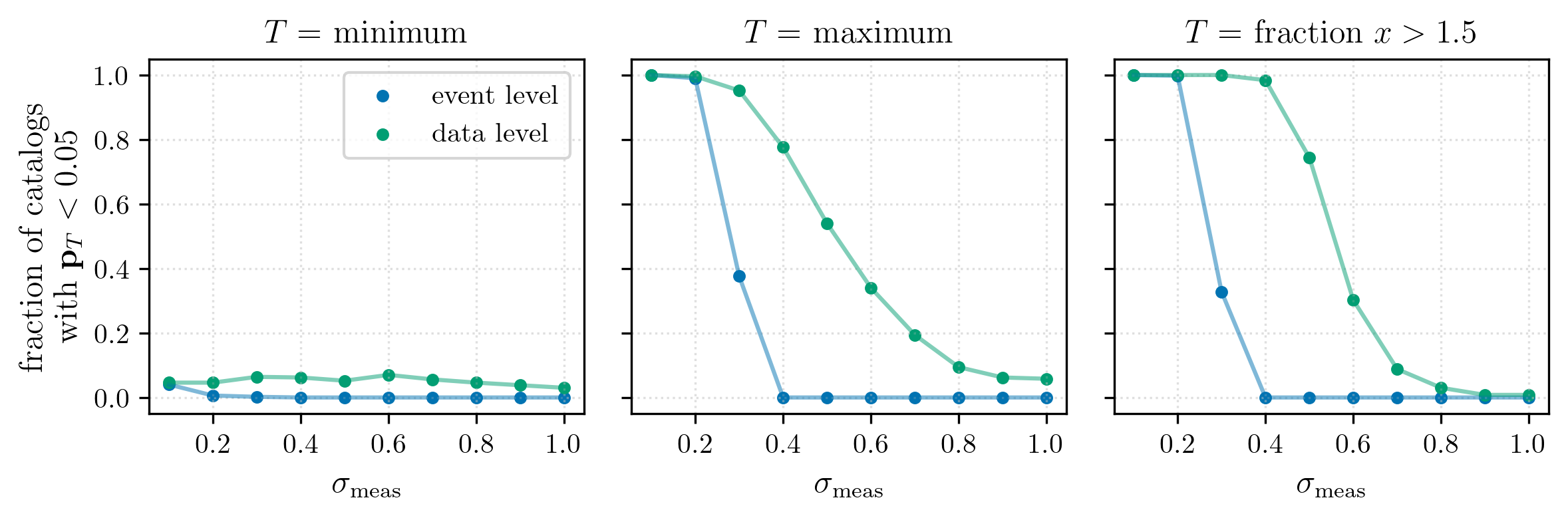}
    \caption{Single-event measurement uncertainty $\sigma_{\rm meas}$ versus the fraction of 500 simulated universes where $\pvalue < 0.05$, using the analytic toy model described in Appendix~\ref{app:toy_model} and three representative test statistics $T$.
    (Note that although the hyper-posterior can be calculated analytically for the toy model, the distribution of $p_T$ cannot for these choices of test statistic, hence sampling over 500 universes.)
    The employed population model is intentionally unable to fully characterize the true underlying distribution; see Fig.~\ref{fig:toy_model_bad}.
    A fraction of $1$ means that the test can always tell that there is model mismatch; a fraction of 0 means it can never tell that there is model mismatch. 
    The data-level $\pvalue$ (green) using $\maxL$ parameters are always more discerning of model misspecification than the event-level $\pvalue$ (blue) using true parameters, although both become uninformative when single-event measurement uncertainty is sufficiently large. 
    The overall degree of informativeness depends on which $T$ is used. Here, e.g., $T=$ minimum (first panel) is substantially less informative than $T=$ maximum (second panel) or $T=$ fraction of $x>1.5$ (third panel).}
    \label{fig:toy_model_repeated}
\end{figure*}

To move beyond a single test case, we compare event- and data-level $\pvalue$ over \textit{many instantiations} of observed catalogs (i.e., many simulated universes)\editaccepted{---a test only} computationally feasible \editaccepted{with} a toy model and analytic likelihood, the details of which are given in Appendix \ref{app:toy_model}.
Again measuring an underlying bimodal \editaccepted{population} with a \editaccepted{single} Gaussian model, we \editaccepted{use} single-event likelihoods with ten values of $\sigma_{\rm meas}$ linearly spaced between $0.1$ and $1$. 
For each $\sigma_{\rm meas}$, we generate 500 instantiations of observed catalogs \editaccepted{with} 1,000 PPC traces \editaccepted{apiece}.
For \editaccepted{several representative} $T$, we calculate the fraction of these 500 catalogs for which $\pvalue < 0.05$\editaccepted{, shown as a function of} $\sigma_{\rm meas}$ in Fig.~\ref{fig:toy_model_repeated}.
A fraction of $1$ \editaccepted{(0)} means that the test can always \editaccepted{(never) identify the known} model mismatch.
\editaccepted{For the range of cases we consider,} \textbf{the data-level $\pvalue$ is \textit{always} equally or more discerning of model misspecification than the event-level $\pvalue$}.
\editaccepted{The data-level $\pvalue$ outperforms the event level regardless of $\sigma_{\rm meas}$ and $T$---}although both \editaccepted{grow} less discerning as $\sigma_{\rm meas}$ \editaccepted{increases} and eventually \editaccepted{become uninformative, a limit we may be hitting with the realistic O3-noise case in Fig.~\ref{fig:pvalues_event_vs_data}}. 
Figure~\ref{fig:toy_model_repeated} also reiterates that \textbf{the most discerning \editaccepted{$T$} depends on the specific population features of interest and the model used to measure \editaccepted{them}.} 
\editaccepted{In this toy example, the Gaussian population model more-or-less always infers the correct population minimum (left) but, at sufficiently low $\sigma_{\rm meas}$, cannot replicate the correct maximum (middle).}

While we can only confirm that data-level PPCs are more discerning than event-level PPCs in the specific cases explored in this work, we expect their superiority to translate across applications, as event-level PPCs will always include the impact of the prior.
\editaccepted{However, a reason one may nonetheless prefer event-level PPCs is computational cost:}
obtaining $\maxL$ parameters for an arbitrarily large number of synthetic GW events is computationally intensive due to the high-dimensional parameter space and non-analytic likelihood model.
Moreover, \textit{consistently} computing $\maxL$ values for predicted vs.~observed catalogs \editaccepted{requires care when} testing real data.
Waveform models, power spectral densities, selection effects, etc.~must be consistent between the (real) observed and (\editaccepted{by necessity,} synthetic) predicted data, which \editaccepted{is} difficult when real GW catalogs \editaccepted{mix} these attributes, some of which are \editaccepted{set by hard-to-replicate} human decisions.
Developing faster, more robust methods \editaccepted{for computing data-level quantities} across observed and predicted GW catalogs is left for future work.

\vspace{-10pt}
\section{Partial PPCs}
\label{sec:partial_PPCs}
\vspace{-10pt}

One shortcoming of the traditional PPCs of Sec.~\ref{sec:event_vs_data_level} is that they use information present in $\dobsvec$ to both estimate parameters \textit{and} assess model incompatibility using $\pvalue$~\cite{Bayarri:2000}.
Partial PPCs (pPPCs) \editaccepted{avoid this} double use of data by fixing a degree of freedom between observed and predicted catalogs \cite{Bayarri:2008, Robins:2000}.
The distribution from which the predicted catalogs are drawn \editaccepted{for} a pPPC changes from the traditional PPCs, \editaccepted{where we use the} full posterior predictive distribution, conditioned on the observed data.
In the pPPC, the distribution from which \editaccepted{we draw predicted catalogs} is \textit{additionally} constrained to match the observed value of a chosen test statistic $T_0$:
\begin{multline}
    \mathrm{PPD}^{\rm pPPC}(\lambdapred_{\rm true} | \dobsvec, T_0) = \\ \int \pi_{\rm pop}(\lambdapred_{\rm true} |\Lambda, T_0(\dobsvec)) \, p(\Lambda | \dobsvec) \,d\Lambda
    \label{eqn:ppd_pPPC}\,, 
\end{multline}
resulting in the $p$-value given in Eq.~\eqref{eqn:pPPC_pvalue}:
\begin{widetext}
\begin{multline}
    \mathrm{Partial}: ~p^{\rm pPPC}_T(\dobsvec, T_0) = \\ \iiint I_{[T(\lambdapredvec)\geq T(\lambdaobsvec)]}~P_{\rm det}(\lambdapredvec,\lambdaobsvec)~\pi_{\rm pop}(\lambdapredvec |\Lambda,  T_0(\dobsvec))~p(\lambdaobsvec| \dobsvec,\Lambda )~p(\Lambda|\dobsvec)~d\Lambda ~d\lambdaobsvec ~d\lambdapredvec\,.
     \label{eqn:pPPC_pvalue}
\end{multline}
\end{widetext}

\vspace{-8pt}
We implement the pPPC following Algorithm \ref{alg:partial_PPC}:
\vspace{-5pt}
\begin{algorithm}[H]
\caption{Partial PPC}
\label{alg:partial_PPC}
    \begin{algorithmic}[1]
        \State Choose a test statistic $T_0$ to keep constant between the predicted and observed catalogs. 
        \vspace{3pt}
        \State Draw one sample of $\Lambda$ from $p(\Lambda|\dobsvec)$.
        \vspace{3pt}
        \State \textit{Observed values} ($\vec\lambda^{\rm obs}_{\rm true}$): Draw one sample from each individual-event posterior in the observed catalog, \textit{reweighted} from its parameter-estimation prior to $\pi_{\rm pop}(\lambda_{\rm true}|\Lambda)$. Calculate $T_0(\dobsvec)$.
        \vspace{3pt}
        \State \textit{Predicted values} ($\vec\lambda^{\rm pred}_{\rm true}$): Draw $N_{\rm obs}$ \textit{detectable} values of $\lambda_{\rm true}$ from the distribution $\pi_{\rm pop}(\lambda_{\rm true}|\Lambda)$. Calculate $T_0(\dpredvec)$. If
        $$T_0(\dobsvec) - \delta T_0 \leq T_0(\dpredvec) \leq T_0(\dobsvec) + \delta T_0,$$ 
        continue to the next step. Otherwise, redraw $N_{\rm obs}$ events until the condition is satisfied, resulting in a predictive distribution following Eq.~\eqref{eqn:ppd_pPPC}. We choose $\delta T_0 = (2 N_{\rm obs})^{-1/2}$.\footnote{
            We cannot set $\delta T_0=0$ because we are dealing with catalogs of finite size.
            Instead, we want $\delta T_0$ to be as close to zero as possible. However, in reality disagreement between the predicted and observed populations can be so extreme that it is practically impossible to satisfy $$T_0(\dobsvec) - \delta T_0 \leq T_0(\dpredvec) \leq T_0(\dobsvec) + \delta T_0$$ via random population draws if $\delta T_0$ is sufficiently small. 
            We consider that standard error when dealing with finite events is $\propto (N_{\rm obs})^{-1/2}$.
            The factor of 2 is included for additional constraining power; if increased we encounter sampling insufficiency. 
            Figures~\ref{fig:partial_PPC_frac} and \ref{fig:partial_PPC_std} show the minor spread of $T_0(\dpredvec)$ vs.~$T_0(\dobsvec)$ about the diagonal due to our choice of $\delta T_0$.
        }
        \vspace{3pt}
        \State Repeat steps 2-4 for each $\Lambda$.
    \end{algorithmic}
\end{algorithm}

\begin{figure*} 
    \centering 
    \includegraphics[width=0.6\linewidth]{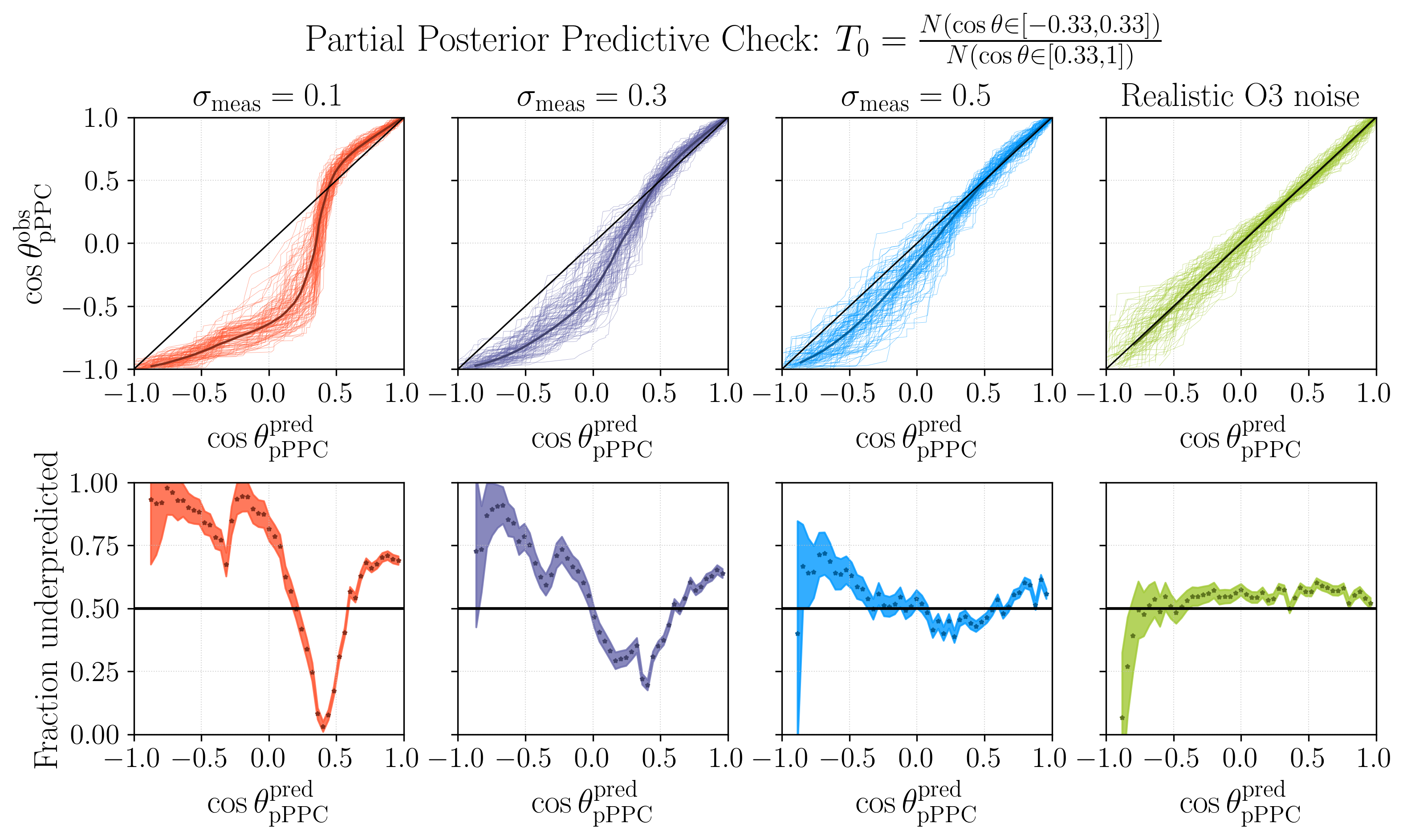} 
    \includegraphics[width=0.83\linewidth]{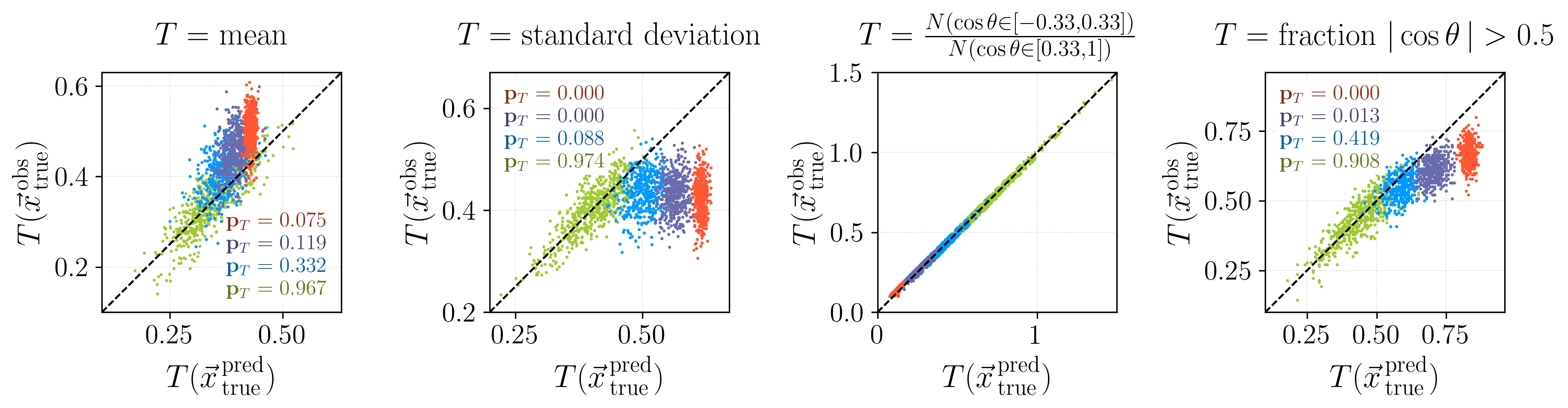}
    \caption{PPC traces (top row), fraction under-predicted (middle row), and $\pvalue$ (bottom row) using a pPPC where the ratio of tilts with $\cos\theta \in [-0.33, 0.33]$ to tilts with $\cos\theta \in [0.33, 1]$ is the fixed degree of freedom, $T_0$. 
    The third column of the bottom row shows predicted vs.~observed $T_0$, which naturally follows the diagonal.
    Colors and labels retain meaning from Figs.~\ref{fig:data_event_PPCs} and \ref{fig:pvalues_event_vs_data}.
    } 
    \label{fig:partial_PPC_frac} 
\end{figure*}

\begin{figure*} 
    \centering 
    \includegraphics[width=0.6\linewidth]{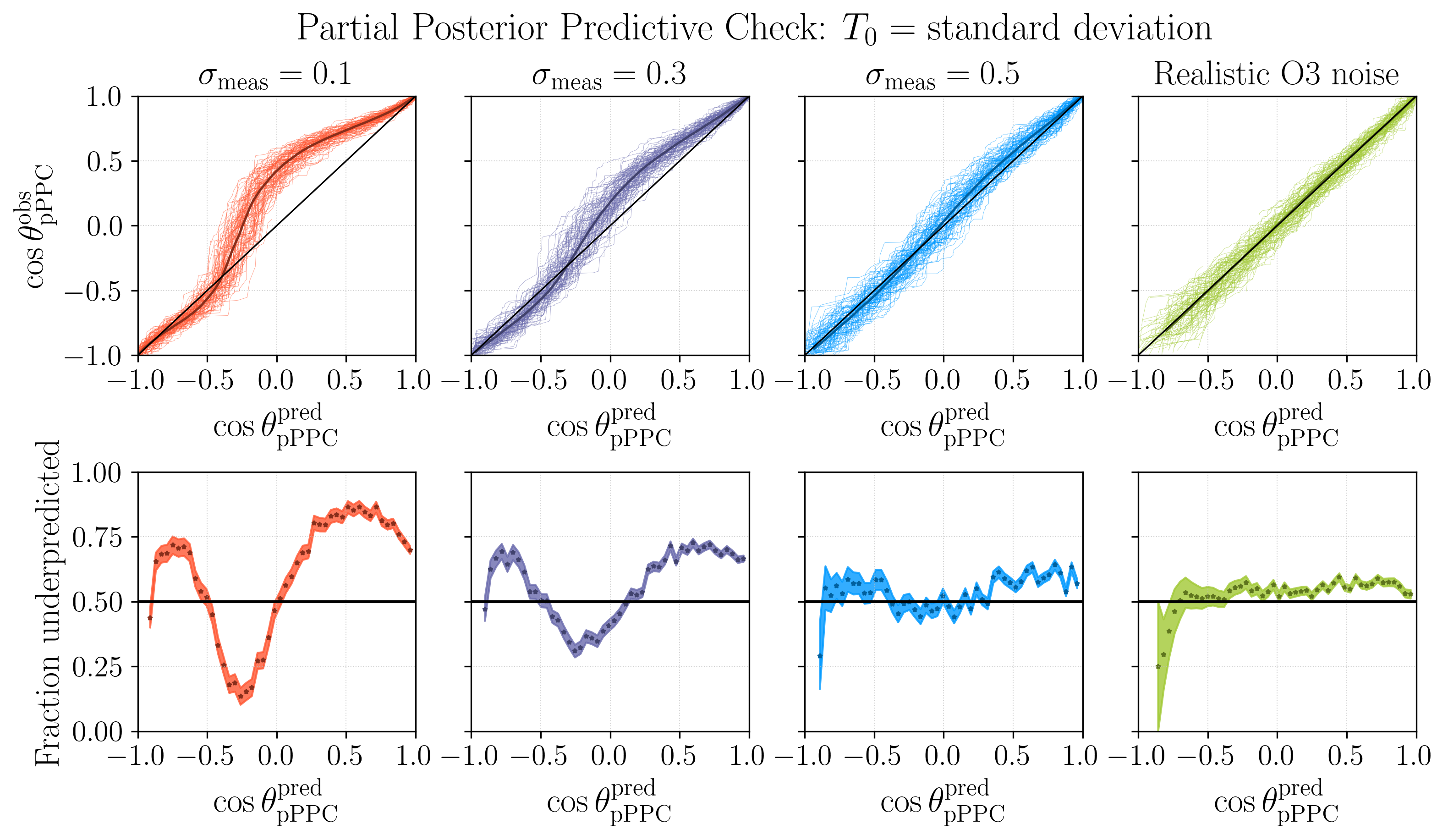} 
    \includegraphics[width=0.83\linewidth]{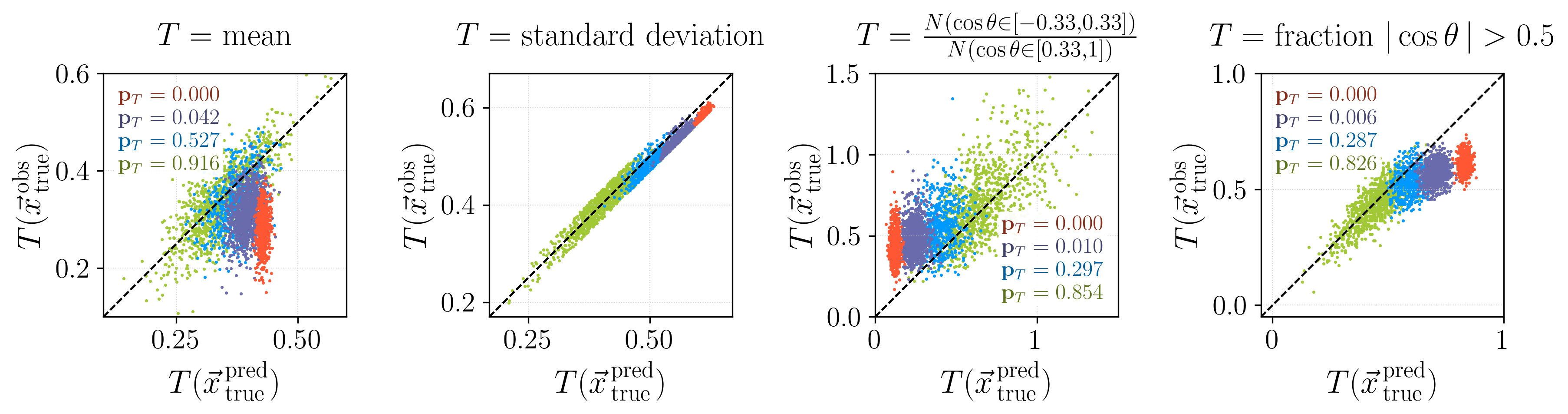} 
    \caption{Same as Fig.~\ref{fig:partial_PPC_frac} but where the fixed degree of freedom $T_0$ is the standard deviation of a catalog's $\cos\theta$ draws.} 
    \label{fig:partial_PPC_std} 
\end{figure*}

\editaccepted{While our pPPC method still occurs at the event level and therefore uses reweighted posteriors, it only uses information not present in $T_0(\dobsvec)$ when drawing from the posterior predictive distribution.
By fixing one degree of freedom between observed and predicted catalogs, we can assess whether the remaining degrees of freedom become more informative; ideally, a pPPC leads to a more discerning $\pvalue$ for the non-fixed $T$.}
We focus on two cases. 
First, we choose a $T_0$ which is highly mismatched between the inferred vs.~true underlying population distributions: the ratio of tilts with $\cos\theta \in [-0.33, 0.33]$ to tilts with $\cos\theta \in [0.33, 1]$ \editaccepted{(Fig.~\ref{fig:partial_PPC_frac})}. 
Second, we define $T_0$ as the standard deviation of $\cos\theta$ draws per catalog---a quantity which is theoretically well measured by the Gaussian population model \editaccepted{(Fig.~\ref{fig:partial_PPC_std})}.

\editaccepted{For small single-event measurement uncertainty (red, purple), our two choices of $T_0$ yield predicted vs.~observed traces that drastically differ in shape from both each other and the traditional event-level PPCs in Fig.~\ref{fig:data_event_PPCs}.}
In these cases, fixing one degree of freedom does indeed yield more informative predictive checks and corresponding $\pvalue$ for (some of) the non-fixed test statistics. 
The most pronounced improvement is when \editaccepted{$T_0=$ standard deviation and $T=$ mean, where the pPPC} is actually \textit{more} informative than even the data-level PPC \editaccepted{(see first column of Fig.~\ref{fig:pvalue_summary})}. 
However, for other test statistics, fixing $T_0$ makes $\pvalue$ less discerning (higher $\pvalue$) than the traditional event-level PPC; e.g., when $T=$ fraction $|\cos\theta|>0.5$.
\editaccepted{In the cases with higher single-event measurement uncertainty (blue, green)}, neither pPPC shows improvement over the traditional, event-level PPC.
The reweighting process already results in predicted values that overly reflect the model,  to the extent that fixing $T_0(\dobsvec)=T_0(\dpredvec)$ has a sub-dominant effect and the data-level PPCs still reign supreme. 

In conclusion, the efficacy of the  pPPC depends on single-event measurement uncertainty and which specific population features we decide to probe.
We find that, generally, \textbf{pPPCs are more discerning when the targeted feature is well predicted by the \editaccepted{population} model} (Fig.~\ref{fig:partial_PPC_std}), rather than one that the model cannot capture (Fig.~\ref{fig:partial_PPC_frac}).
However, when single-event uncertainty is large, reweighting has already created biased individual event posteriors, and there is no discernible advantage in using the pPPC to diagnose model misspecification over the traditional event-level PPC.

\vspace{-5pt}

\section{Split Predictive Checks}
\label{sec:split_PPCs}

Lastly, we implement the split predictive check (SPC)~\cite{Li:2022} (also \editaccepted{called} the holdout predictive check~\cite{Moran:2024}) as an alternative event-level PPC. 
\editaccepted{In an SPC, the data are divided} into two subsets: one to infer the population distribution and another to generate predictive catalogs.
The process is \editaccepted{similar} to a leave-one-out analysis, in which population inference is performed on all but one event to see how the resultant population prior impacts the posterior of the left-out event~\cite[e.g.,][]{Fishbach:2019ckx,Essick:2021vlx,GWTC1_pop,GWTC2_pop}. 
The SPC again changes the distribution from which predicted catalogs are drawn:
\begin{multline}
    p^{\rm SPC}_{\rm pop}(\lambdapred_{\rm true} | \{\dobsvec_1, \dobsvec_2\}) = \\ \int \pi_{\rm pop}(\lambdapred_{\rm true} |\Lambda, \dobsvec_1) \, p(\Lambda | \dobsvec_2) \,d\Lambda\,,
    \label{eqn:ppd_SPC}
\end{multline}
where the full observed data are $\dobsvec = \{\dobsvec_1, \dobsvec_2\}$.
This PPD results in the $\pvalue$ given in Eq.~\eqref{eqn:SPC_pvalue}, where the held out data $\dobsvec_2$ are used to generate the hyperposterior $p(\Lambda|\dobsvec_2)$, which is then used to draw $\lambdapred$ and reweight $p(\lambdaobs|\dobsvec_1)$, from which we draw $\lambdaobs$:
\begin{widetext}
\begin{multline}
     \mathrm{Split}:~ p^{\rm SPC}_T(\dobsvec_1, \dobsvec_2) = \\ \iiint I_{[T(\lambdapredvec)\geq T(\lambdaobsvec)]}~P_{\rm det}(\lambdapredvec,\lambdaobsvec)~\pi_{\rm pop}(\lambdapredvec |\Lambda)~p(\lambdaobsvec| \dobsvec_1,\Lambda )~p(\Lambda|\dobsvec_2)~d\Lambda ~d\lambdaobsvec ~d\lambdapredvec\,.
     \label{eqn:SPC_pvalue}
\end{multline}
\end{widetext}

\begin{figure*}
\centering
    \includegraphics[width=0.6\textwidth]{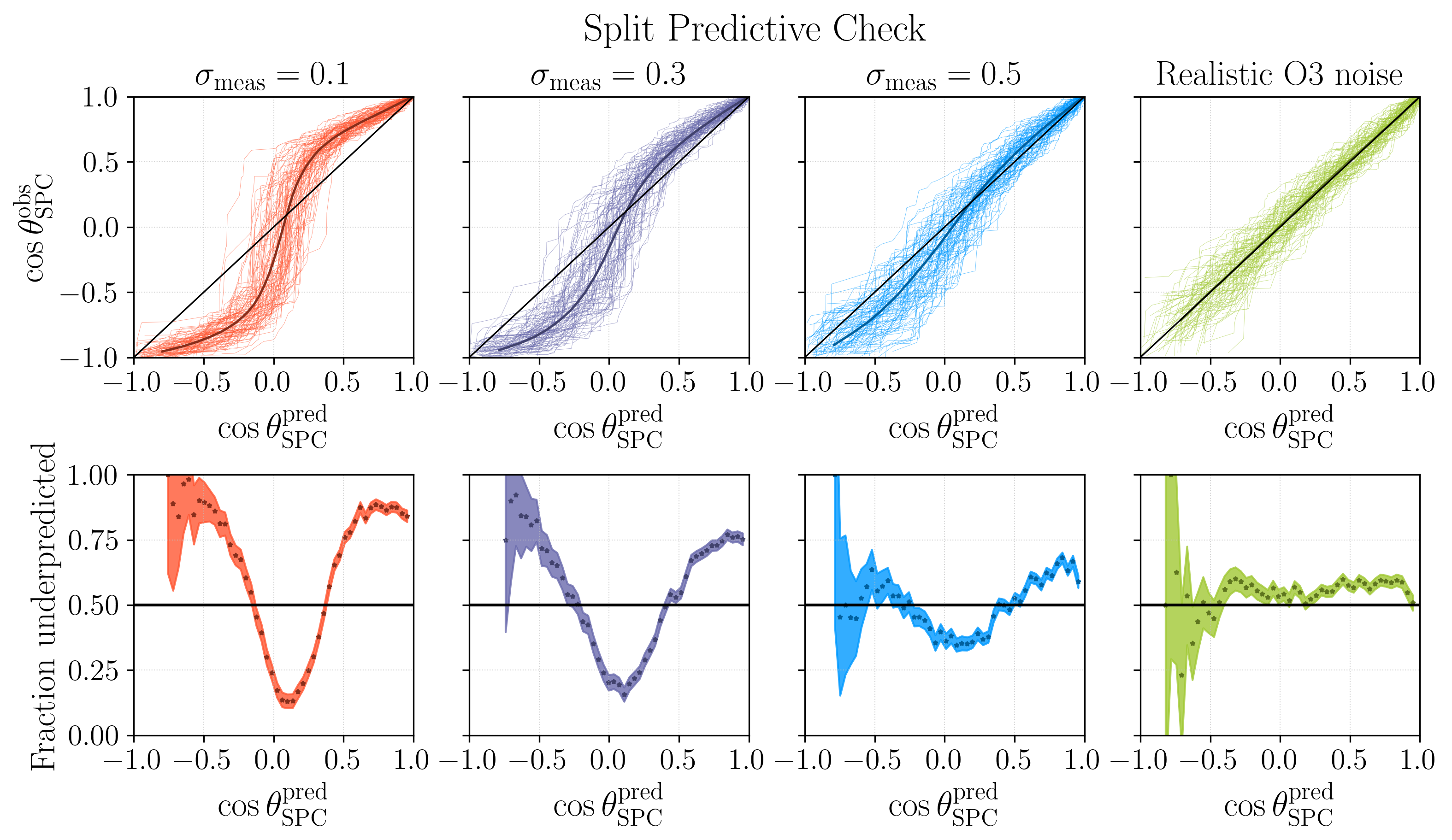}
    \includegraphics[width=0.82\textwidth]{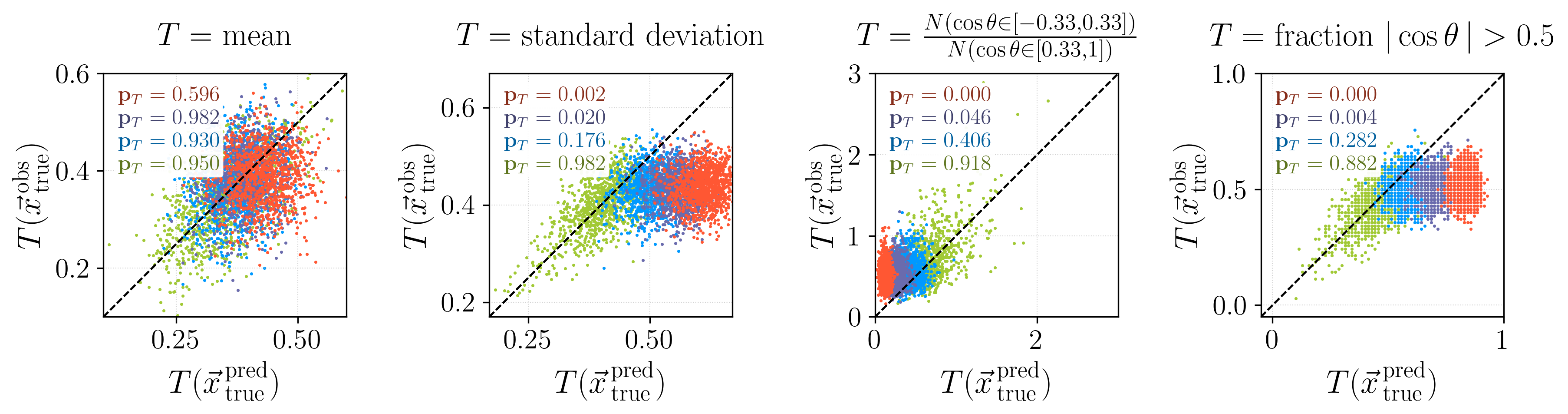}
    \caption{Same as Fig.~\ref{fig:partial_PPC_frac} but for the SPC, rather than \editaccepted{the} pPPC.}
    \label{fig:split_PC}
\end{figure*}

\editaccepted{In theory, an SPC requires rerunning hierarchical inference on many combinations of events in the observed catalog. 
To avoid this, we use a hyperposterior reweighting scheme, hinging on the fact that the} likelihood of a catalog  $\dobsvec$ is the product of the likelihoods of each event in that catalog \editaccepted{(Eq.~\ref{eqn:pop_likelihood})}.
Therefore, 
\begin{align}
    \mathcal{L}(\dobsvec|\Lambda)= \mathcal{L}(\dobsvec_1|\Lambda)\,\mathcal{L}(\dobsvec_2 | \Lambda) \,\,.
\end{align}
To generate $\Lambda$ draws from $p(\Lambda|\dobsvec_2)$ in Eq.~\eqref{eqn:SPC_pvalue}, we can \editaccepted{thus draw from the existing $p(\Lambda|\dobsvec)$ according to the} weights 
\begin{align}
    w_\Lambda  & = \frac{\mathcal{L}(\dobsvec_2 |\Lambda)}{\mathcal{L}(\dobsvec|\Lambda)}= \frac{1}{\mathcal{L}(\dobsvec_1|\Lambda)}\,,
    \label{eqn:SPC_Lambda_weights}
\end{align}
where 
\begin{equation}
    \mathcal{L}(\dobsvec_1|\Lambda)
    = \prod_{\rm i=0}^{n_1} \mathcal{L}(d^{\rm obs, i}|\Lambda) 
\end{equation}
and $n_1$ is the number of events \editaccepted{in the subset} $\dobsvec_1$. 
\editaccepted{As the weights $w_\Lambda$ can be entirely calculated in post-processing, we need not re-run hierarchical inference on any catalog subsets}.


To conduct an SPC, we follow Algorithm \ref{alg:split_PPC}:
\vspace{-10pt}
\begin{algorithm}[H]
\caption{Split PPC}
\label{alg:split_PPC}
    \begin{algorithmic}[1]
        \State Randomly partition the full observed catalog into disjoint subsets $\dobsvec_1$ and $\dobsvec_2$.
        \vspace{3pt}
        \State Draw one sample of $\Lambda$ from the hyperposterior $p(\Lambda|\dobsvec_2)$. 
        In practice, to do this we perform a \textit{weighted} draw from the already available $p(\Lambda|\dobsvec)$ with the weights given in Eq.~\eqref{eqn:SPC_Lambda_weights}.
        \vspace{3pt}
        \State \textit{Observed values} ($\vec\lambda^{\rm obs}_{\rm true}$): Draw one sample from each of the \editaccepted{$n_1$} individual-event posteriors in $\dobsvec_1$, reweighted from its parameter-estimation prior to $\pi_{\rm pop}(\lambda_{\rm true}|\Lambda)$~\cite{Callister:2021}.
        \vspace{3pt}
        \State \textit{Predicted values} ($\vec\lambda^{\rm pred}_{\rm true}$): Draw \editaccepted{$n_1$} detectable values of $\lambda_{\rm true}$ from the distribution $\pi_{\rm pop}(\lambda_{\rm true}|\Lambda)$.
        \vspace{3pt}
        \State Repeat steps 1-4 many times.
    \end{algorithmic}
\end{algorithm}
\noindent We choose to split $\dobsvec$ into subsets of equal size, such that $n_1=n_2=N_{\rm obs}/2$.

\editaccepted{Figure~\ref{fig:split_PC} shows that the SPC presents} no advantage over the traditional event-level PPC \editaccepted{(as does Fig.~\ref{fig:pvalue_summary})}, and in fact is the least discerning of all the PPCs we test.
Because we \editaccepted{use} event-level parameters, SPCs still \editaccepted{trigger the failure mode from} reweighting individual events to the population model. 
We attribute the SPC's heightened \editaccepted{deficiency} to the fact that it utilizes smaller predicted \editaccepted{and} observed catalogs (here, by a factor of two).
\editaccepted{The decreased catalog size causes} a wider spread in the traces \editaccepted{and $T$}, effectively ``blurring out" some of the information from Fig.~\ref{fig:data_event_PPCs}.
\editaccepted{Exploring SPCs with larger catalog sizes and different ratios of $n_1:n_2$ is left to future work.}

\section{PPCs applied to GWTC-4.0}
\label{sec:gwtc4_results}

\begin{figure*}
\centering
    \includegraphics[width=\textwidth]{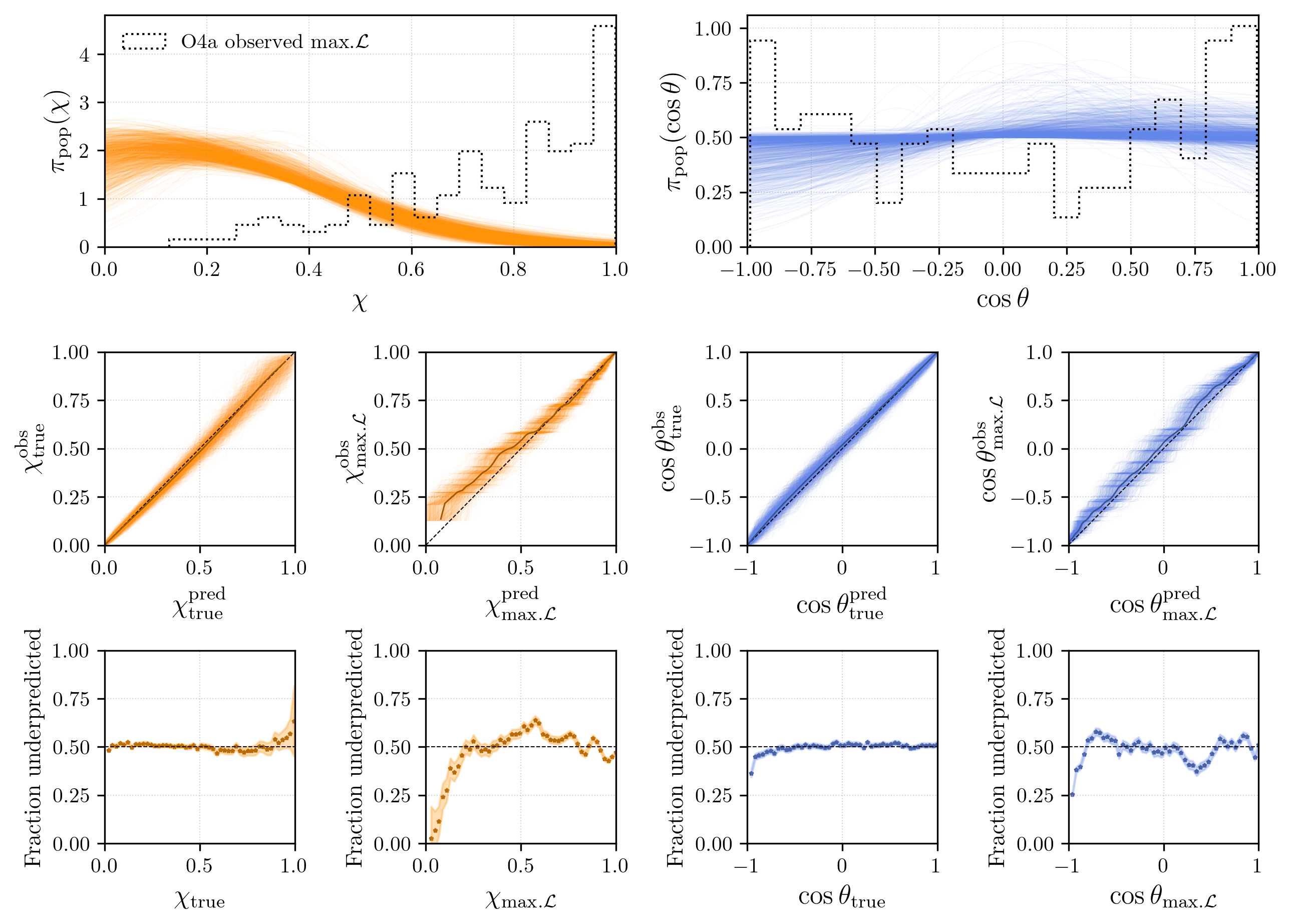}
    \caption{Results of standard event-level (left within each color) and data-level (right within each color) PPCs for GWTC-4.0 spin magnitudes (orange) and tilt (blue). 
    \textit{(First row)}: Recovered population distribution for $\chi$ and $\cos\theta$ (traces) given in Ref.~\cite{GWTC4_pop}, reproduced from their Fig.~7~\cite{GWTC4_pop_data_release} and compared to $\maxL$ values from the observed O4a population (histogram).
    Note that direct comparison of the traces versus histograms does not indicate model misspecification: the former showcases true parameters (used for event-level PPCs) while the latter shows $\maxL$ parameters (used for data-level PPCs); see also Fig.~\ref{fig:GWTC4_hists} in Appendix~\ref{app:maxL_finding}.
    \textit{(Second row)}: PPC traces from 100 catalogs. \textit{(Third row)}: Fraction of traces in each binned $\chi$ or $\cos\theta$ with a slope $<$ 1, corresponding to the fraction of events in that parameter range under-predicted by population model.  The points and shaded region retain their meaning from Fig.~\ref{fig:data_event_PPCs}.
    }
    \label{fig:O4_PPCs}
\end{figure*}

We have shown that among our tested suite of PPCs, data-level PPCs are holistically the most capable of identifying model mismatch.
Thus, we apply and compare the traditional event-level and data-level PPCs of Sec.~\ref{sec:event_vs_data_level} to the BBH mergers in the LVK's Fourth Gravitational-wave Transient Catalog (GWTC-4.0)~\cite{GWTC4_intro,GWTC4_cat}, using the \editaccepted{\textsc{Gaussian Component Spins}} model of Ref.~\cite{GWTC4_pop}\editaccepted{, cf. their Eqs.~(B26) and (B27)}.
Under this model, the spin magnitude ($\chi$) \editaccepted{distribution is a truncated} Gaussian on $[0,1]$ and $\cos\theta$ is a mixture of a \editaccepted{truncated} Gaussian distribution on $[-1, 1]$ and an isotropic distribution.

For the event-level PPCs, we use all 153 BBHs from GWTC-4.0 with a false-alarm rate of less than one per year~\cite{GWTC4_cat}; for the data-level case, we only use the 84 new BBHs from O4a when generating PPC traces. 
The latter restriction allows the use of a single injection set---that for O4a sensitivity estimates~\cite{Essick:2025zed,O4a_injections}\footnote{
    The sensitivity estimates from the LVK's first two observing runs are semi-analytic~\cite{Essick:2023toz} and do not contain all extrinsic parameters necessary to re-produce the GW signals~\cite{GWTC3_injections, GWTC4_cumulative_injections}. 
    Sensitivity estimates from the LVK's third observing run also exclude the merger phase~\cite{O3b_injections}.
}---and a single representative power spectral density~\cite{GWTC4_intro,GWTC4_psds,LIGO:2024kkz,Capote:2024rmo} for the predicted catalogs.
To find $\maxL$ values for the data-level PPCs, we use the parameter-estimation code {\tt cogwheel}~\cite{Roulet:2022kot,Islam:2022afg,Roulet:2024hwz} and follow the procedure described in Appendix~\ref{app:maxL_finding}. 

Figure~\ref{fig:O4_PPCs} shows results for the event-level and data-level PPCs on GWTC-4.0 spin magnitudes (orange) and tilt angles (blue).
Focusing first on the event-level PPCs, our results indicate that \editaccepted{\textbf{when applied to GWTC-4.0, the} \textsc{Gaussian Component Spins} \textbf{population model under-predicts the number of BBHs with large spin magnitudes and over-predicts those with perfectly anti-aligned tilts} ($\cos\theta=-1$)}.
Similar conclusions were offered in the LVK analysis~\cite{GWTC4_pop} using \editaccepted{both} a weakly parametrized \editaccepted{\textsc{B-Spline}} model~\cite{Edelman:2022ydv,Godfrey:2023oxb} \editaccepted{and} a strongly parametrized  \editaccepted{$\cos\theta$ model with a hard minimum cutoff}, inferred to be between $-0.90$ and $-0.56$ at 90\% credibility. 
Comparing our results to Fig.~7 in Ref.~\cite{GWTC4_pop} indicates that \editaccepted{their \textsc{B-Spline}}  model might be a better fit to the data than the \editaccepted{\textsc{Gaussian Component Spins}} model.
\editaccepted{Given its simple functional form, the latter may be insufficiently flexible to capture the shape of the $\chi$ and/or $\cos\theta$ distributions near their boundaries, necessarily extrapolating constraints from the more observation-dense regions of parameter space.}
The requirement of high spin magnitudes and a dearth of perfectly anti-aligned tilts are also found by \citet{Guttman:2025jkv} in the maximum likelihood population framework~\cite{Payne:2022xan}.
\editaccepted{The spin magnitude feature may originate from a broad, rapidly spinning subpopulation that becomes prominent at high masses~\cite{Hussain:2026pfm}.}

Turning to the data-level PPC results, we see \editaccepted{some} deviation from the diagonal, potentially indicative of model misspecification.
\editaccepted{However, as} with the simulated populations (Fig.~\ref{fig:data_event_PPCs}), the traces for the $\maxL$ PPCs\footnote{
    \editaccepted{We remind readers that disagreement between the observed $\maxL$ parameters for O4a (histograms in the top row of Fig.~\ref{fig:O4_PPCs}) and the population traces is not \editaccepted{itself} an indicator of model misspecification, as the former is a data-level quantity while the latter is an event-level quantity. 
    Rather, we use the event-level traces to then generate predicted $\maxL$ values (Appendix~\ref{app:maxL_finding}, Fig.~\ref{fig:GWTC4_hists}) which are compared to the observed $\maxL$ histogram.}
} are highly sensitive to the local fluctuations resulting from the particular noise realizations of each observed event, \editaccepted{making it} difficult to ascertain which features are real versus spurious (Sec \ref{subsec:event_vs_data_traces}).
\editaccepted{By eye, the most substantial deviation appears to be at $\chi_{\maxL} \lesssim 0.1$, with the fraction under-predicted sharply approaching $0$ as $\chi\to0$.
Using the minimum $\chi$ as a test statistic, we find a data-level $\pvalue = 0.37$, indicating that the observed minimum $\chi_{\maxL} \approx 0.1$ is consistent with a random noise fluctuation; thus we cannot claim that the Gaussian $\chi$ population model is over-predicting small spin magnitudes.}
\editaccepted{For spin tilts, the data-level PPC finds a more pronounced version of the event-level behavior as $\cos\theta\to-1$.
Using the minimum of the $\cos\theta$ distribution as a test statistic yields $\pvalue=0.005$, passing our threshold for significance and thus robustly indicating that the GWTC-4.0 spin tilt model indeed under-predicts perfectly anti-aligned spins.}
The remaining fluctuations in the $\chi$ and $\cos\theta$ $\maxL$ traces are consistent with the random Poisson variation \editaccepted{observed with even} the lowest single-event uncertainty in Fig.~\ref{fig:data_event_PPCs}.


\section{Recommendations and Future Work}
\label{sec:conclusions}

\editaccepted{We conclude by reminding the reader that our main results are summarized in Sec.~\ref{sec:result_summary} and that Fig.~\ref{fig:pvalue_summary} collects the $p$-values for all PPCs and test statistics considered in this work.}
For \editaccepted{those} looking to use PPCs to probe model misspecification in GW population analyses, we make the following recommendations: 
\begin{enumerate}
    \item For moderate- to poorly constrained parameters, we recommend the use of data-level PPCs alongside event-level PPCs. 
    While any model misspecification uncovered by event-level PPCs is robust, we \textbf{caution against over-interpreting event-level PPC results that indicate a \textit{good} model fit}.
    \editaccepted{If negligible information exists about a parameter in the single-event likelihood, neither class of PPC will be informative.}
   \item Additionally, we advise against assigning significance to every fluctuation in data-level PPC traces that might stem from Poisson noise rather than a feature in the underlying astrophysical population. Small-number effects naturally cause  fluctuations in \editaccepted{the fraction of events over/under-predicted by the model}. Assessing the consistency of cumulative density functions~\cite{Mould:2026sww} or comparing $p$-values are complementary diagnostics to ensure a mismodeled feature is robustly identified.
    \item For well constrained parameters, we recommend the use of the traditional, event-level PPCs: these are more computationally efficient, and in the low measurement-uncertainty limit, perform equally as well as data-level PPCs.
    \item When calculating posterior predictive $p$-values, we recommend always using a variety of test statistics, rather than just one. 
    \item If trying to probe a specific feature in the data, partial PPCs can be a useful tool, but we only recommend their use for moderately to well constrained parameters. 
    \item At current GW catalog size, we discourage the use of split-PPCs.
\end{enumerate}

Moving forward, we \editaccepted{plan} to explore PPCs on alternative data-level quantities beyond just the maximum likelihood parameters, such as search statistics. 
Likelihood optimization for poorly constrained parameters is difficult (Appendix \ref{app:maxL_finding}), warranting both turning to other data-level properties as well as developing faster, more accurate methods for GW likelihood optimization.
Additionally, given that decreased measurement uncertainty drastically improves PPC performance, we hope to investigate whether only conducting PPCs on subsets of events that pass an SNR threshold~\cite{Wolfe:2025yxu} yields any improvement. 
Finally, all of the methods explored in this work can also be applied to non-parametric models; we leave comparing PPCs on parametric to non-parametric approaches to future work.

Robust model assessment for GW source populations is crucial for downstream astrophysical implications.
There is only so much information our data can provide us with, and knowing those limits is preferable to drawing false conclusions.
Model checking will continue to be important as the catalog of observed GW signals grows in size, and our conclusions become increasingly influential. 


\section*{Data Availability}

Notebooks to generate all the figures appearing in the text (excluding Figs.~\ref{fig:schematic_ppc} and \ref{fig:schematic_pvalue}) are available on Github at Ref.~\cite{github_release}, as is a toy model elucidating the difference between event and data-level PPCs. 
The individual-event posterior samples and hyper-posterior samples used as the basis of this work are from \citet{Miller:2024sui} and can be made available upon request.
All PPC traces generated from those data and GWTC-4.0 are included in Ref.~\cite{github_release}.


\acknowledgements

We thank Maya Fishbach, Thomas Callister, and Amanda Farah for the initial discussions that inspired this work in 2023; Javier Roulet and Colm Talbot for assistance with likelihood optimization; and Noah Wolfe, Matthew Mould, Sofia Alvarez-Lopez, Jack Heinzel, and Salvatore Vitale for many insightful discussions about PPCs throughout this project.
In particular, we extend immense gratitude to Sofia Alvarez-Lopez for serving as our internal LVK reviewer, and thank Derek Davis and Eric Thrane for helpful comments during review. 

S.J.M.~and K.C.~were supported by NSF Grants No.~PHY-2308770 and No.~PHY-2409001.
P.M.M.~acknowledges support through the NSF Physics Frontiers Center Awards No.~1430284 and No.~2020265, and through the Gravitational Physics Professorship at ETH Zurich.
This work was also supported by NSF Grant No.~PHY-2150027 as part of the LIGO Caltech REU Program, which funded S.W. 

The authors are thankful for LIGO Laboratory computing resources, funded by the National Science Foundation Grants No.~PHY-0757058 and No.~PHY-0823459.
This research has made use of data or software obtained from the Gravitational Wave Open Science Center (gwosc.org), a service of the LIGO Scientific Collaboration, the Virgo Collaboration, and KAGRA. 
This material is based upon work supported by NSF's LIGO Laboratory which is a major facility fully funded by the National Science Foundation, as well as the Science and Technology Facilities Council (STFC) of the United Kingdom, the Max-Planck-Society (MPS), and the State of Niedersachsen/Germany for support of the construction of Advanced LIGO and construction and operation of the GEO600 detector. Additional support for Advanced LIGO was provided by the Australian Research Council. Virgo is funded, through the European Gravitational Observatory (EGO), by the French Centre National de Recherche Scientifique (CNRS), the Italian Istituto Nazionale di Fisica Nucleare (INFN) and the Dutch Nikhef, with contributions by institutions from Belgium, Germany, Greece, Hungary, Ireland, Japan, Monaco, Poland, Portugal, Spain. KAGRA is supported by Ministry of Education, Culture, Sports, Science and Technology (MEXT), Japan Society for the Promotion of Science (JSPS) in Japan; National Research Foundation (NRF) and Ministry of Science and ICT (MSIT) in Korea; Academia Sinica (AS) and National Science and Technology Council (NSTC) in Taiwan.

This work made use of the following software: \textsc{emcee}~\cite{emcee}, \textsc{bilby}~\cite{bilby}, \textsc{dynesty}~\cite{dynesty}, \textsc{cogwheel}~\cite{Roulet:2022kot,Islam:2022afg,Roulet:2024hwz}, \textsc{LALSuite}~\cite{lalsuite}, \textsc{numpy}~\cite{numpy}, \textsc{scipy}~\cite{scipy}, \textsc{h5py}~\cite{h5py}, \textsc{h5ify}~\cite{h5ify}, \textsc{matplotlib}~\cite{matplotlib}, \textsc{seaborn}~\cite{seaborn}, \textsc{pandas}~\cite{pandas1,pandas2}, \textsc{astropy}~\cite{astropy1,astropy2}.

\onecolumngrid

\appendix

\section{Simulated Population and Inference}
\label{app:sim_pop}

\begin{figure*}
    \centering
    \includegraphics[width=\linewidth]{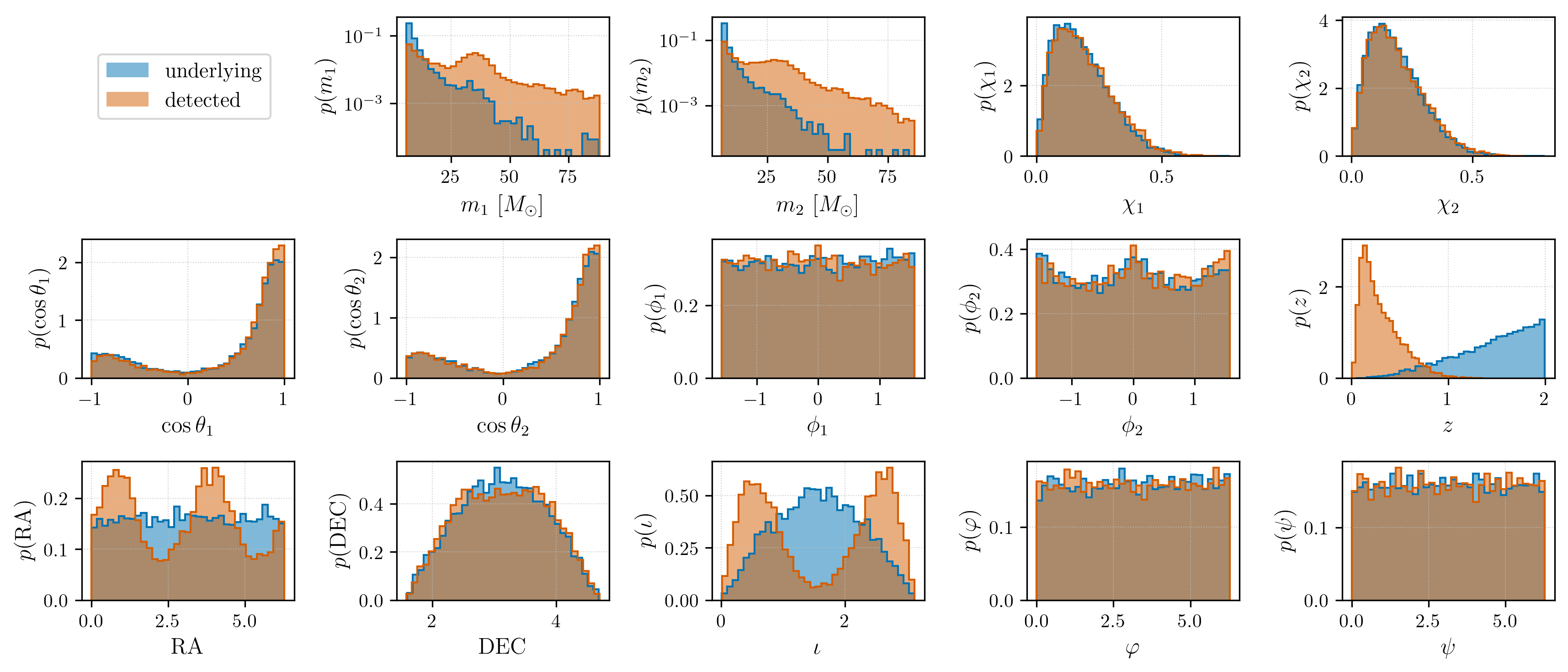}
    \caption{Underlying (blue) and detected (red-orange) distributions of all 15 BBH parameters for the \textsc{LowSpinAligned} population from \citet{Miller:2024sui}, reproduced here for reference. In order, these are: masses $m_i$, spin magnitudes $\chi_i$, spin tilt angles $\cos\theta_i$, spin azimuthal angles $\phi_i$, redshift $z$, right ascension (RA), declination (DEC), inclination $\iota$, phase $\varphi$, and polarization angle $\psi$.}
    \label{fig:sim_pop}
\end{figure*}

We use the simulated \textsc{LowSpinAligned} population from \citet{Miller:2024sui} as the basis for probing the efficacy of various posterior predictive checking methods throughout this work. 
The population's bimodal $\cos\theta$ distribution allows for probing model mismatch with a (unimodal) Gaussian tilt model. 
For reference, we reproduce the \textsc{LowSpinAligned} population's underlying and detectable parameter distributions in Fig.~\ref{fig:sim_pop}. 
A detailed description of the simulated populations and their individual-event and population inference can be found in Appendices A, B, and C of \citet{Miller:2024sui}.
Essential details are summarized as follows. 
For each underlying population we found the corresponding detectable population by applying a network optimal SNR cut of 10 in the LIGO Livingston, LIGO Hanford, and Virgo detector network using O3 power spectral densities~\cite{O3-sensitivity}. 
GW data were simulated using the \textsc{IMRPhenomXPHM} waveform model~\cite{IMRPhenomXPHM} with Gaussian noise. 
Samples were drawn from the 15-dimensional posterior of binary parameters using the \textsc{IMRPhenomXPHM} waveform model, with standard priors for all parameters. 
Posteriors were sampled stochastically using the implementation of the nested sampler \textsc{Dynesty}~\cite{dynesty} in the parameter-estimation code \textsc{Bilby}~\cite{bilby}.
A network optimal SNR cut of 10 was performed on the samples to ensure consistency with the selection process~\cite{Essick:2023upv}.

We establish the following notation for hierarchical/population inference. 
The posterior on population-level parameters $\Lambda$ given a GW catalog $\vec d$ (also called the ``hyper-posterior") is
\begin{equation}
    p(\Lambda | \vec d) = \frac{\mathcal{L}(\vec d|\Lambda) \pi(\Lambda)}{p(\vec d)}\,,
    \label{eqn:pop_posterior}
\end{equation}
where the data $\vec d=\{d_i\}$ are from $N_{\rm obs}$ independent events, and the population likelihood (or ``hierarchical likelihood") is 
\begin{equation}
    \mathcal{L}(\vec d|\Lambda) \propto \prod_i^{N_{\rm obs}} \frac{\int \mathcal{L}(d_i|\lambda) \,\pi_{\rm pop}(\lambda | \Lambda) \, d\lambda}{\xi(\Lambda)} 
    = \frac{1}{\xi(\Lambda)^{N_{\rm obs}}} \prod_i^{N_{\rm obs}} \int \mathcal{L}(d_i|\lambda) \,\pi_{\rm pop}(\lambda | \Lambda) \, d\lambda \,,
    \label{eqn:pop_likelihood}
\end{equation}
with individual-event likelihoods $\mathcal{L}(d_i|\lambda)$ and a population distribution $\pi_{\rm pop}(\lambda | \Lambda)$. 
The detection efficiency $\xi(\Lambda)$---equivalent to the expected fraction of the population described by $\Lambda$---is calculated by 
\begin{equation}
    \xi(\Lambda) = \int \pi_{\rm pop}(\lambda|\Lambda) \, P_{\rm det} (\lambda) \, d\lambda \,,
    \label{eqn:xi}
\end{equation}
where $P_{\rm det}(\lambda)$ is the probability 
that an individual event with parameters $\lambda$ is detected.
In Ref.~\cite{Miller:2024sui}, we conducted hierarchical inference using Markov Chain Monte Carlo sampling with the code \textsc{emcee}~\cite{emcee}.
We used a Beta distribution for $\pi_{\rm pop}(\chi_i)$, Gaussian distribution for $\pi_{\rm pop}(\cos\theta_i)$, and mass/redshift model from Ref.~\cite{GWTC3_pop}.
Monte Carlo uncertainty was accounted for by excluding regions not meeting the effective sample criterion proposed by \citet{Farr:2019}.
To estimate $P_{\rm det}$, we simulated many signals with $\lambda$ drawn from a reference distribution and stored which are detectable~\cite{Tiwari:2017ndi,Mandel:2018mve,Farr:2019,Essick:2022ojx,Essick:2023upv}.
We direct readers to Sec.~2 of Ref.~\cite{GWTC4_pop} or Appendix C of Ref.~\cite{Miller:2024sui} for more details about hierarchical Bayesian inference.

\section{Derivation of posterior predictive $p$-values}
\label{app:pvalues}

In this Appendix, we present the derivation of the expressions for the event- and data-level posterior predictive $p$-values given in Eqs.~\eqref{eqn:pvalue_eventlevel} and \eqref{eqn:pvalue_datalevel}.
We start from Eq.~\eqref{eqn:pvalue2}.
For notational ease, define $y = \dobsvec$, $x=\dpredvec$, and correspondingly $\lambda_y = \lambdaobsvec$, $\lambda_x = \lambdapredvec$. 
In this notation, Eq.~\eqref{eqn:pvalue2} becomes
\begin{align}
    p_T(y) & = \iiiint I_{[T(x,\lambda_x) \geq T(y,\lambda_y)]} \, p(x, \lambda_x | \Lambda)\, p(\lambda_y, \Lambda | y) \,  \,d\Lambda \,d\lambda_x\,d\lambda_y\,dx \,,
    \label{eqn:pvalue_v1}
\end{align}
where  $p(\lambda_y, \Lambda | y)$ indicates how observed data $y$ give us constraints on  $\{\lambda_y, \Lambda\}$, and $p(x, \lambda_x | \Lambda)$ is how the model with hyper-parameters $\Lambda$ generates $\{x, \lambda_x\}$. 
We can expand each of these terms using conditional probabilities: 
\begin{equation}
    p(\lambda_y, \Lambda | y) = p(\lambda_y | y, \Lambda) \, p(\Lambda | y)\,, 
\end{equation}
and
\begin{equation}
    p(x, \lambda_x | \Lambda) = p(x|\lambda_x, \Lambda) \, p(\lambda_x|\Lambda) = \mathcal{L}(x|\lambda_x)\, \pi_{\rm pop}(\lambda_x|\Lambda) \,\,. 
\end{equation}
In the above two expansions, $p(\lambda_y| y, \Lambda)$ is the population-informed posterior, $p(\Lambda | y)$ is the hyperposterior, $\mathcal{L}(x|\lambda_x)$ is the (normalized) product of single-event likelihoods over each event (Eq.~\ref{eqn:L_catalog}), and $\pi_{\rm pop}(\lambda_x|\Lambda)$ is the population distribution (Eq.~\ref{eqn:pi_pop_catalog}).
Substituting into Eq.~\eqref{eqn:pvalue_v1} yields
\begin{equation}
    p_T(y) = \iiiint I_{[T(x,\lambda_x) \geq T(y,\lambda_y)]} \,
    \mathcal{L}(x|\lambda_x) \,
    \pi_{\rm pop}(\lambda_x|\Lambda)\,
     p(\lambda_y | y, \Lambda) \,
     p(\Lambda | y)\,
    \,d\Lambda \,d\lambda_x\,d\lambda_y\,dx \,\,. \label{eqn:pvalue_twolevels}
\end{equation}
Going from Eq.~\eqref{eqn:pvalue_twolevels} to Eqs.~\eqref{eqn:pvalue_eventlevel} and \eqref{eqn:pvalue_datalevel} comes down to how we define our choice of test statistic $T$.

\vspace{10pt}

\noindent\textbf{Derivation of the event-level $p$-value }(Eq.~\ref{eqn:pvalue_eventlevel}): For event-level PPCs, we look at $T$ that only depend on parameters $\{\lambda_x,\lambda_y\}$, not the data itself. Thus, we can integrate over $x$ in Eq.~\eqref{eqn:pvalue_twolevels} because the indicator function $I$ no longer depends on $x$:
\begin{align*}
    &p^{\rm event}_T (y)= \int \mathcal{L}(x|\lambda_x) \, dx
 \iiint I_{[T(\lambda_x) \geq T(\lambda_y)]} \,
    \pi_{\rm pop}(\lambda_x|\Lambda)\,
     p(\lambda_y | y, \Lambda) \,
     p(\Lambda | y)\,
    \,d\Lambda \,d\lambda_x\,d\lambda_y \\
    & \implies p^{\rm event}_T(y) = \iiint I_{[T(\lambda_x) \geq T(\lambda_y)]} \,
    \pi_{\rm pop}(\lambda_x|\Lambda)\,
     p(\lambda_y | y, \Lambda) \,
     p(\Lambda | y)\,
    \,d\Lambda \,d\lambda_x\,d\lambda_y\,, 
\end{align*}
because the likelihood $\mathcal{L}(x|\lambda_x)$ is normalized in $x$ for any $\lambda_x$. 
Finally, we must incorporate selection effects (cf.~Eq.~\ref{eqn:ppd_event_with_detection} in the main text): 
\begin{equation}
    p^{\rm event}_T (y) = \iiint I_{[T(\lambda_x) \geq T(\lambda_y)]} \,
    P_{\rm det}(\lambda_x,\lambda_y)\,
    \pi_{\rm pop}(\lambda_x|\Lambda)\,
     p(\lambda_y | y, \Lambda) \,
     p(\Lambda | y)\,
    \,d\Lambda \,d\lambda_x\,d\lambda_y~. 
    \label{eqn:pvalue_app_event}
\end{equation}
Equation~\eqref{eqn:pvalue_app_event} is equivalent to Eq.~\eqref{eqn:pvalue_eventlevel}.

\vspace{10pt}

\noindent \textbf{Derivation of the data-level $p$-value }(Eq.~\ref{eqn:pvalue_datalevel}):  
For data-level PPCs, we look at $T$ that only depend on the data. We are comparing the observed data $y$ (in the absence of any assumptions about $\Lambda$) to the data generated by the model described by $\Lambda$. 
Thus, we can integrate over $\lambda_y$ in Eq.~\eqref{eqn:pvalue_twolevels} because the indicator function $I$ no longer depends on $\lambda_y$:
\begin{align*}
    &p^{\rm data}_T(y) = \int p(\lambda_y | y, \Lambda)  \, d\lambda_y\iiint I_{[T(x) \geq T(y)]} \, \mathcal{L}(x|\lambda_x) 
    \, \pi_{\rm pop}(\lambda_x | \Lambda)\, p(\Lambda | y) \,d\Lambda\,  d\lambda_x\, dx \\
    & \implies  p^{\rm data}_T(y) = \iiint I_{[T(x) \geq T(y)]} \, \mathcal{L}(x|\lambda_x) 
    \, \pi_{\rm pop}(\lambda_x | \Lambda)\, p(\Lambda | y) \,d\Lambda\,  d\lambda_x\, dx\,, 
\end{align*}
because the posterior $p(\lambda_y | y, \Lambda)$ is normalized in $\lambda_y$ for any $y,\Lambda$. Again, we must incorporate selection effects, yielding
\begin{equation}
    p^{\rm data}_T (y) = \iiint I_{[T(x) \geq T(y)]} \, P_{\rm det}(x)\,\mathcal{L}(x|\lambda_x) 
    \, \pi_{\rm pop}(\lambda_x | \Lambda)\, p(\Lambda | y) \,d\Lambda\,  d\lambda_x\, dx ~.
    \label{eqn:pvalue_app_data}
\end{equation}
Equation~\eqref{eqn:pvalue_app_data} is equivalent to Eq.~\eqref{eqn:pvalue_datalevel}.

\section{Optimizing the Full 15-dimensional Binary Black Hole Gravitational-wave Likelihood}
\label{app:maxL_finding}

\begin{figure}
    \centering
    \includegraphics[width=0.36\linewidth]{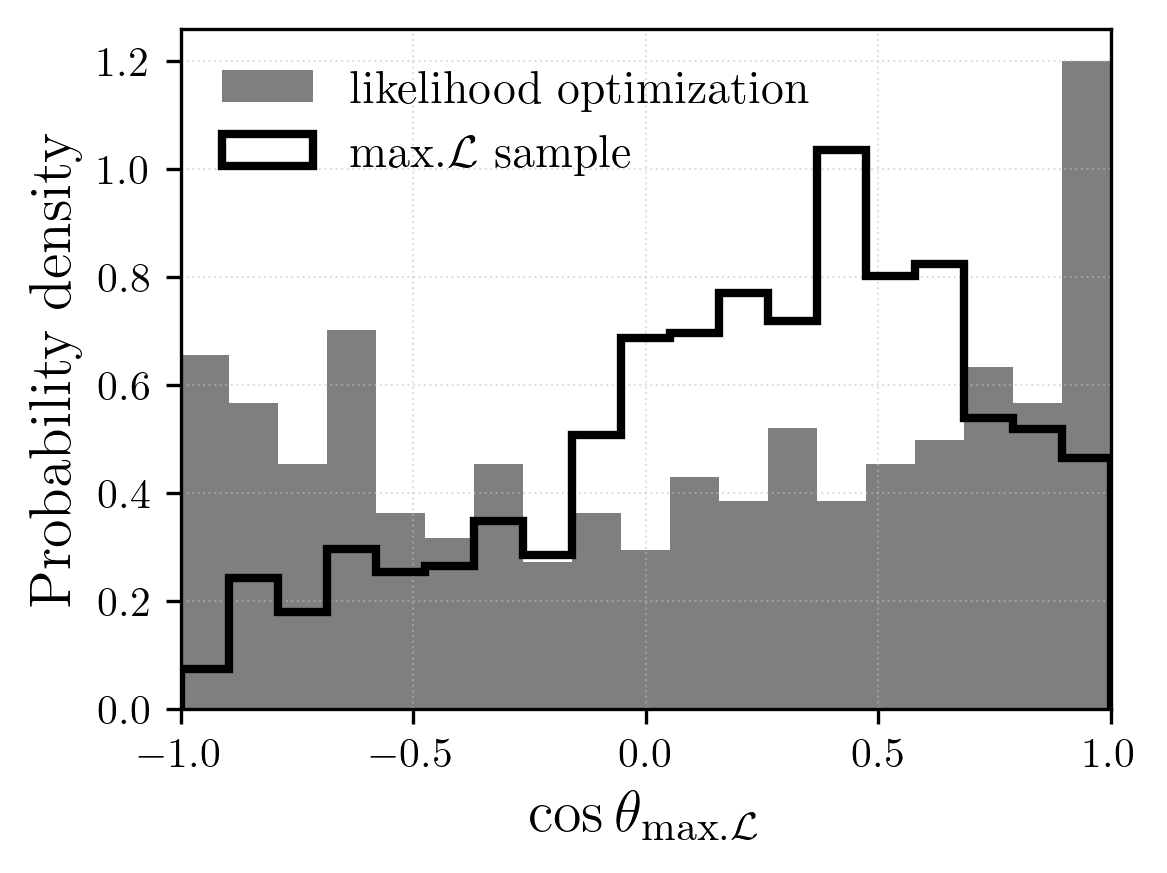}
    \includegraphics[width=0.54\linewidth]{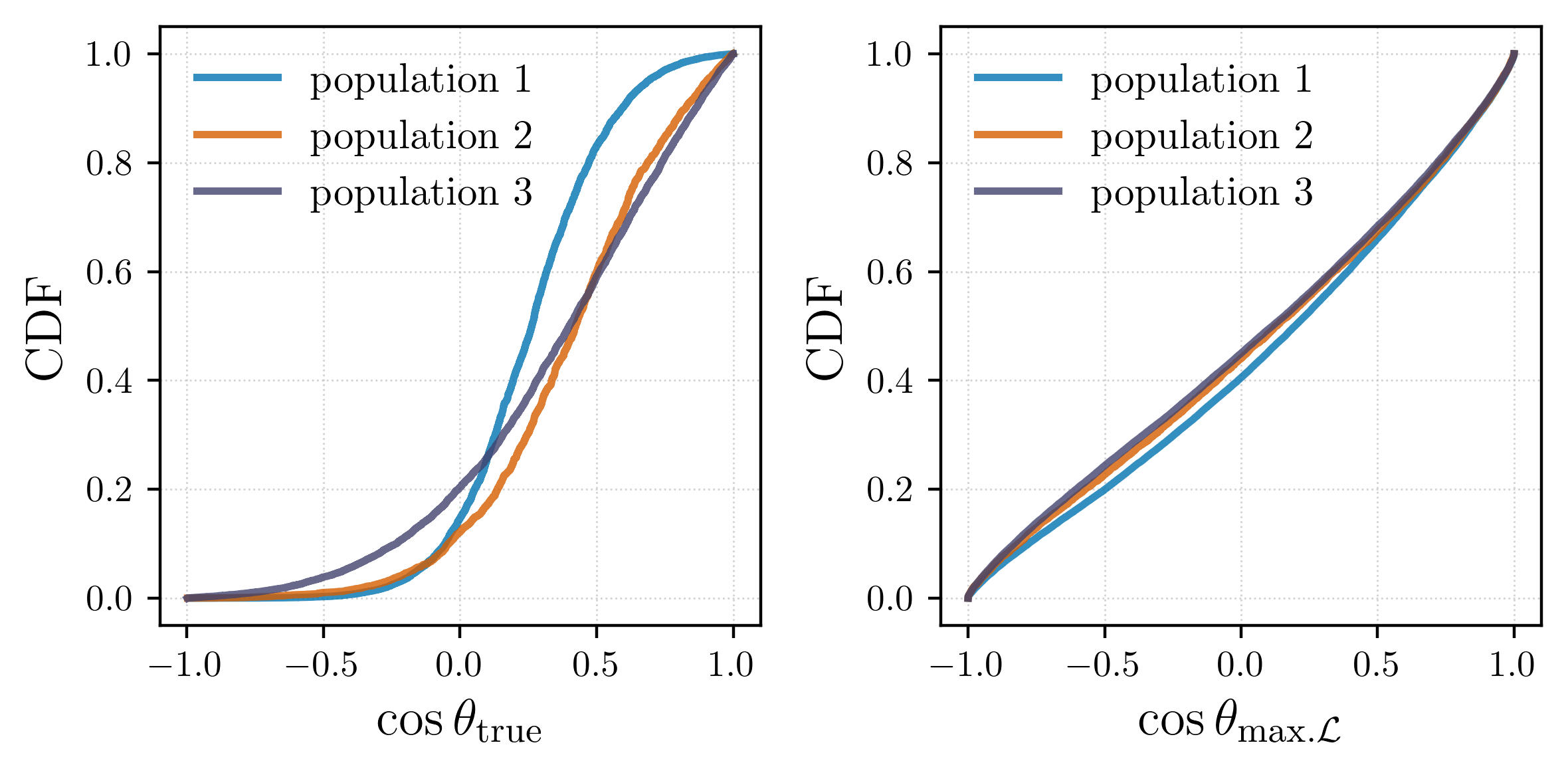}
    \caption{
    (\textit{Left}): Histogram of $\maxL$ samples from stochastically sampled posteriors (empty) versus $\maxL$ values obtained via full likelihood optimization (filled) for the same underlying population.
    (\textit{Middle, right}): The cumulative density functions (CDFs) for three example populations' true and  corresponding $\maxL$ $\cos\theta$ distributions, for the case of realistic O3 noise. 
    While the three true $\cos\theta$ distributions are very different, the $\maxL$ distributions are nearly indistinguishable.}
    \label{fig:true_vs_maxL_cdfs_3pops}
\end{figure}

Many computational tools exist to obtain robust estimates of the full 15-dimensional\footnote{(17-dimensional if including orbital eccentricity)} BBH GW posterior distribution~\cite[e.g.,][]{bilby,dingo,RIFT,Roulet:2024hwz}.
However, perhaps counterintuitively, calculating the maximum likelihood parameters for such a distribution remains more difficult.
Naively, one could simply approximate the posterior distribution via stochastic sampling, and then take the sample that happens to have the highest likelihood value.
For leading-order parameters (like chirp mass), this approximate $\maxL$ will be fairly close to the true $\maxL$. 
For hard-to-constrain parameters like spin tilts that only weakly influence the likelihood, the same is not guaranteed, as shown in the left-hand panel of Fig.~\ref{fig:true_vs_maxL_cdfs_3pops}.
\editaccepted{The difference lies in the fact that parameter estimation is not designed to optimize the likelihood but rather to well-approximate the full posterior distribution.}

Both the $\maxL$ sample and the true $\maxL$ point are valid data-level parameters, as long as they are compared in an apples-to-apples way between predicted and observed data. 
However, running full parameter estimation on tens of thousands of simulated GW signals is computationally infeasible. 
Thus, while $\maxL$ samples are readily available for the observed catalog, they remain difficult to obtain for predicted catalogs.
As such, we turn to likelihood optimization for both the predicted and observed catalogs to obtain true $\maxL$ values.

\begin{figure}
\centering
    \includegraphics[width=0.92\textwidth]{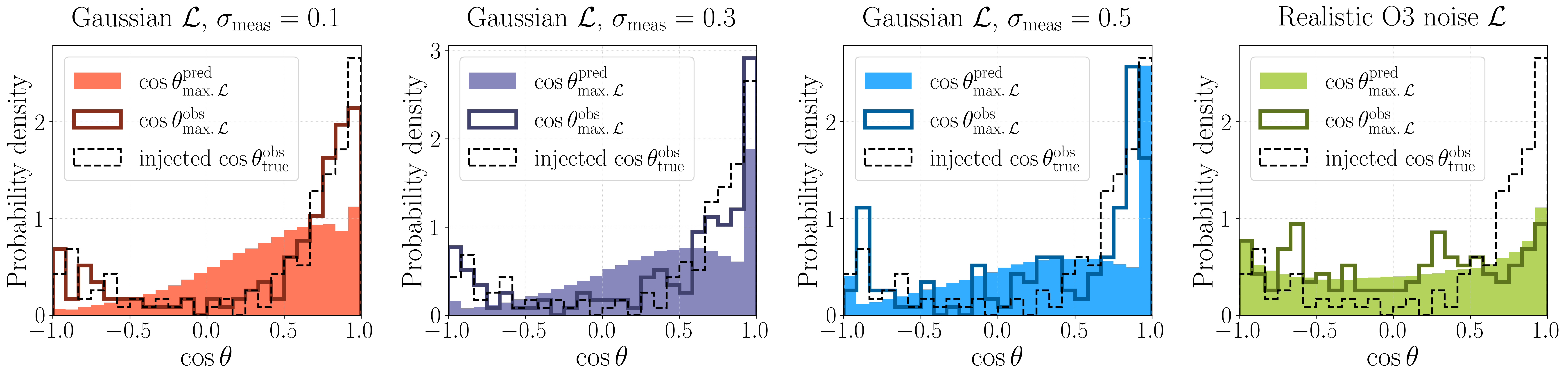}
    \caption{Predicted (filled) vs.~observed (open) $\maxL$ $\cos\theta$ distributions for Gaussian likelihoods with $\sigma_{\rm meas}= 0.1$ (red), $0.3$ (purple), and $0.5$ (blue), and the realistic O3 noise likelihood (green). 
    The histograms show the concatenation of all PPC traces in Fig.~\ref{fig:data_event_PPCs}.
    In black-dashed, we additionally plot the injected $\cos\theta$ distribution, which illuminates that as $\sigma_{\rm meas}$ increases (left to right), the $\maxL$ distribution retains very little information from the true underlying distribution.}
    \label{fig:costheta_hist}
\end{figure}

To optimize the GW likelihood, we use the parameter estimation code {\tt cogwheel}~\cite{Roulet:2022kot,Islam:2022afg,Roulet:2024hwz} for both the simulated GWTC-3.0-like  case (Sec.~\ref{sec:event_vs_data_level}) and the real GWTC-4.0 case (Sec.~\ref{sec:gwtc4_results}). 
The built-in likelihood optimizer in {\tt cogwheel} uses the differential evolution algorithm encoded in {\tt scipy}~\cite{scipy}.
A comparison of injected spin tilts vs. their recovered~$\maxL$ values for three example populations \editaccepted{is} shown in the middle and right-hand panels of Fig.~\ref{fig:true_vs_maxL_cdfs_3pops}.
These three underlying populations are distinct in tilt (middle panel).
However, their $\maxL$ distributions are nearly identical (right panel). 
If we trust that the likelihood optimizer is indeed finding the true $\maxL$ tilts, then Fig.~\ref{fig:true_vs_maxL_cdfs_3pops} indicates that the likelihood (at O3 sensitivity) contains very little information about $\cos\theta$.
This is further reflected in Fig.~\ref{fig:costheta_hist}, where we compare the predicted vs.~observed $\cos\theta$ distributions for the four single-event likelihoods explored in this work. 
With Gaussian likelihoods, the predicted vs.~observed $\maxL$ distributions are visibly distinct, indicating that these likelihoods indeed contain information about $\cos\theta$.
For the realistic O3-noise likelihood, the predicted vs. observed $\maxL$ distributions become visibly similar, echoing what is showcased in Figs.~\ref{fig:data_event_PPCs} and \ref{fig:true_vs_maxL_cdfs_3pops}.
Additionally, as single-event measurement uncertainty increases (left to right), the $\maxL$ distributions look increasingly dissimilar to the injected, true distribution (black-dashed), further exemplifying the likelihood's loss of information as $\sigma_{\rm meas}$ increases.  

\begin{figure}
    \centering
    \includegraphics[width=0.72\linewidth]{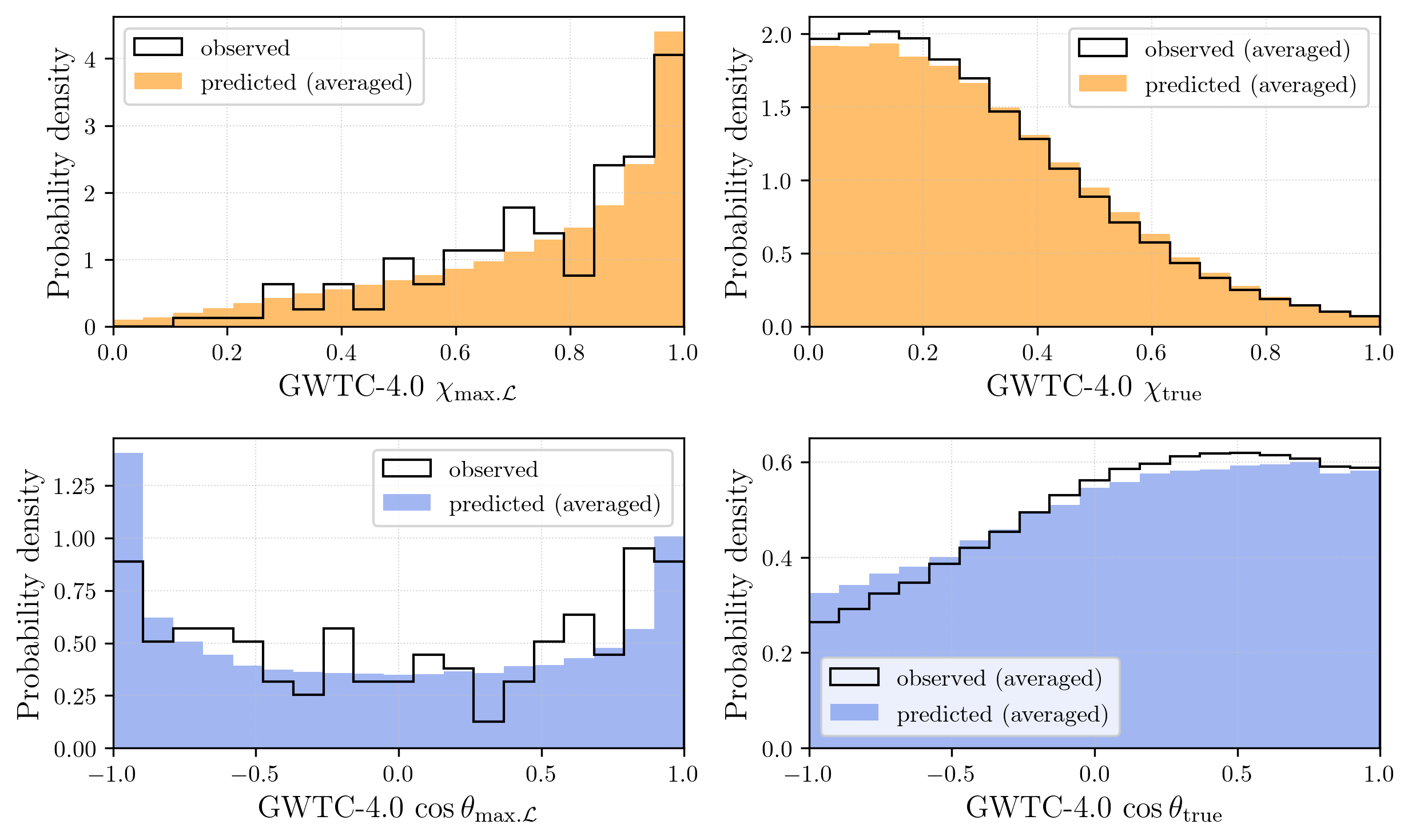}
    \caption{Observed (empty) versus predicted (filled) parameter distributions for $\maxL$ (left column) and true (right column) values of $\chi$ (top row) and $\cos\theta$ (bottom row) for GWTC-4.0, averaged over all of the traces in Fig.~\ref{fig:O4_PPCs}.}
    \label{fig:GWTC4_hists}
\end{figure}

As hinted at in Fig.~\ref{fig:costheta_hist}, the distribution of true vs.~$\maxL$ parameters will generally not be identical. We show this explicitly for the case of GWTC-4.0 results in Fig.~\ref{fig:GWTC4_hists} for spin magnitudes $\chi$ (orange) and tilts $\cos\theta$ (blue), cf.~Sec.~\ref{sec:gwtc4_results} in the main text.
Even if the left ($\maxL$ parameters) vs. right (true parameters) subplots house visibly different distributions, the observed (empty) vs.~predicted (filled) histograms within each subplot look similar, corresponding to PPCs which indicate good model fit. 
Figure~\ref{fig:GWTC4_hists} also helps visualize our conclusion that the GWTC-4.0 Gaussian Component Spins model~\cite{GWTC4_pop} over-estimates BBHs with $\cos\theta\sim-1$.

\section{Simple Toy Model}
\label{app:toy_model}

To help gain intuition, we present a simple toy model where both the individual-event likelihoods and population distribution are Gaussian. 
In this specific case, the hierarchical likelihood of Eqs.~\eqref{eqn:pop_posterior} and \eqref{eqn:pop_likelihood} can be calculated analytically. 
We look at a single individual event-level variable called $x$, which can take on any real value between $-\infty$ and $\infty$, and assume no selection effects, i.e., every value of $x$ is equally likely to be  detected.
The Gaussian population distribution for $\pi_{\rm pop}(x)$ is described by two parameters: the mean $\mu$ and standard deviation $\sigma$.
We implement uniform priors on both. 
Thus, in this toy model, Eq.~\eqref{eqn:pop_posterior} (with $\Lambda = \{\mu,\sigma\}$) becomes
\begin{equation}
    p(\mu, \sigma | \vec d) \propto  \mathcal{L}(\vec d|\mu, \sigma )  \propto \prod_{i}^{N} \int_{-\infty}^{\infty} \mathcal{L}(d_i|x) \, \pi_{\rm pop}(x|\mu, \sigma)\, dx\,,\label{eqn:toymodel_popdist}
\end{equation}
where the population distribution model is
\begin{equation}
    \pi_{\rm pop}(x|\mu, \sigma) = \mathcal{N}(x|\mu, \sigma) = \\ \frac{1}{\sqrt{2\pi \sigma^2}} {\rm Exp} \Big[-\frac{1}{2} \Big(\frac{x-\mu}{\sigma}\Big)^2\Big] 
\end{equation}
and the individual-event likelihood is
\begin{equation}
    \mathcal{L}(d_i|x) = \mathcal{N}(x|x_i^{\mathrm{max}.\mathcal{L}}, \sigma_{\rm meas}) = \\ \frac{1}{\sqrt{2\pi \sigma_{\rm meas}^2}} {\rm Exp} \Big[-\frac{1}{2} \Big(\frac{x-x_i^{\mathrm{max}.\mathcal{L}}}{\sigma_{\rm meas}}\Big)^2\Big]\,,
\end{equation}
where $\sigma_{\rm meas}$ is the measurement uncertainty and $x_i^{\mathrm{max}.\mathcal{L}}$ is the maximum likelihood point for $d_i$.
Under this Gaussian individual-event likelihood, 
\begin{equation}
    x_i^{\mathrm{max}.\mathcal{L}}\sim \mathcal{N}(x_{i}^{\rm true}, \sigma_{\rm meas})\,.
    \label{eqn:x_maxL_draq}
\end{equation}
Solving Eq.~\eqref{eqn:toymodel_popdist}---the integral of the product of two Gaussian distributions---yields
\begin{equation}
     p(\mu, \sigma | \vec d) \propto \prod_{i}^{N} \mathcal{N}\big(x_i^{\mathrm{max}.\mathcal{L}}|\mu, \sqrt{\sigma^2+\sigma_{\rm meas}^2}\,\big)\,\,,
\end{equation}
or equivalently,
\begin{equation}
     p(\mu, \sigma | \vec d) \propto  [2\pi (\sigma^2+\sigma_{\rm meas}^2)]^{-N/2} \\ \times {\rm Exp} \Big[-\frac{1}{2(\sigma^2+\sigma_{\rm meas}^2)} \sum_i^N (x_i^{\mathrm{max}.\mathcal{L}}-\mu)^2\Big]  \,.
     \label{eqn:toy_model_LL}
\end{equation}
The likelihood in Eq.~\eqref{eqn:toy_model_LL} is analytically tractable for a given set of $\{x_i^{\mathrm{max}.\mathcal{L}}\}$.
In this toy model, the maximum likelihood values of $x$ are randomly drawn from a set of $\{x_i^{\rm true}\}$ using Eq.~\eqref{eqn:x_maxL_draq}.
The true values of $x$ can come from any underlying population distribution. 

We first present an example in Appendix~\ref{subapp:toy_model_good} where the true underlying population is also Gaussian, and thus can be accurately recovered by the population model. 
Then, in Appendix~\ref{subapp:toy_model_bad}, we look at a case where the true underlying distribution is highly non-Gaussian, and therefore cannot be well fit by the population model. 
In each, we perform event and data-level PPCs and calculate posterior predictive $p$-values for a series of test statistics. 

\subsection{In which the Gaussian population model is a good fit to the data} \label{subapp:toy_model_good}

\begin{figure*}
    \centering
    \includegraphics[width=0.32\linewidth]{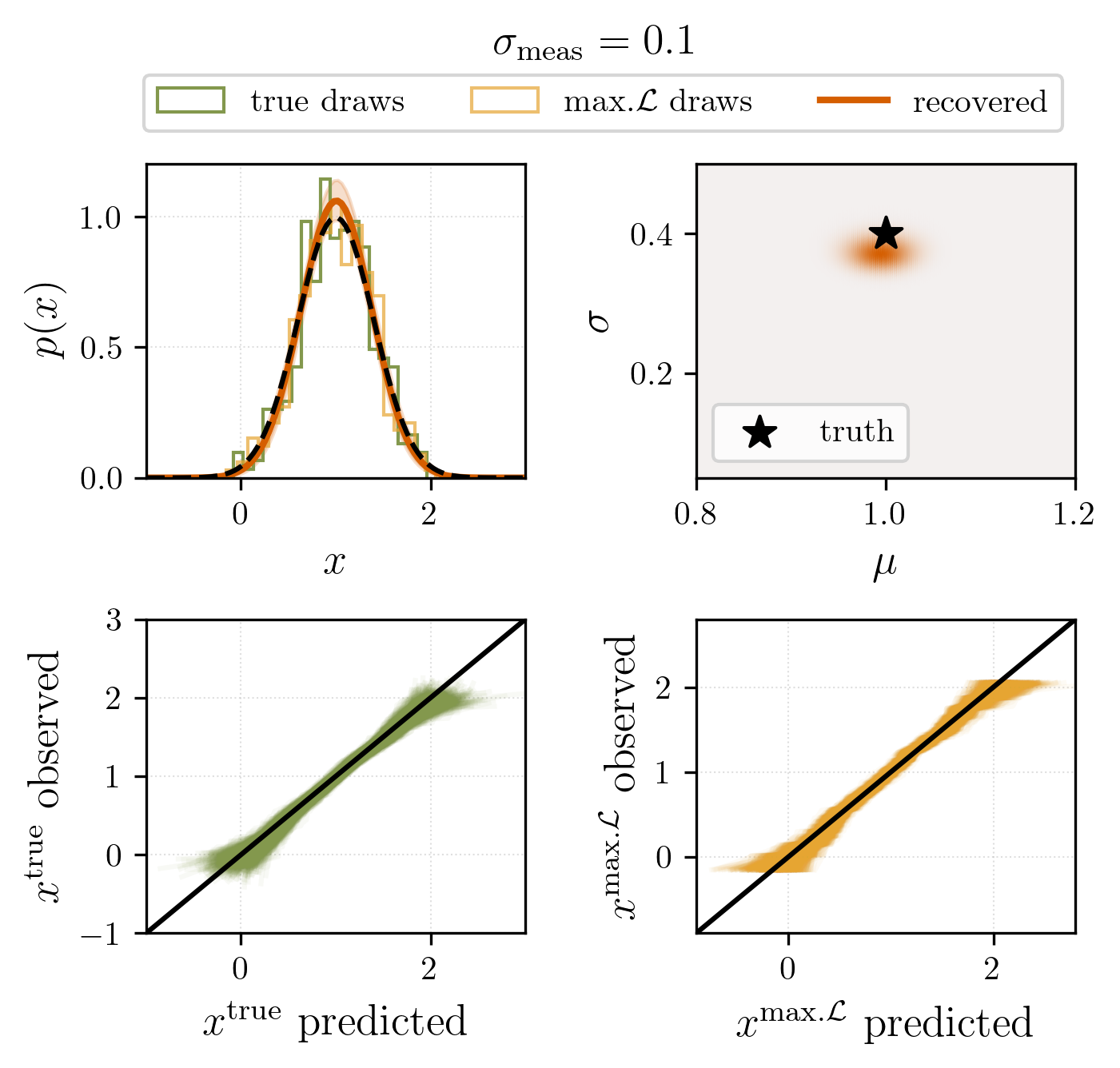} \vline \hfill
    \includegraphics[width=0.32\linewidth]{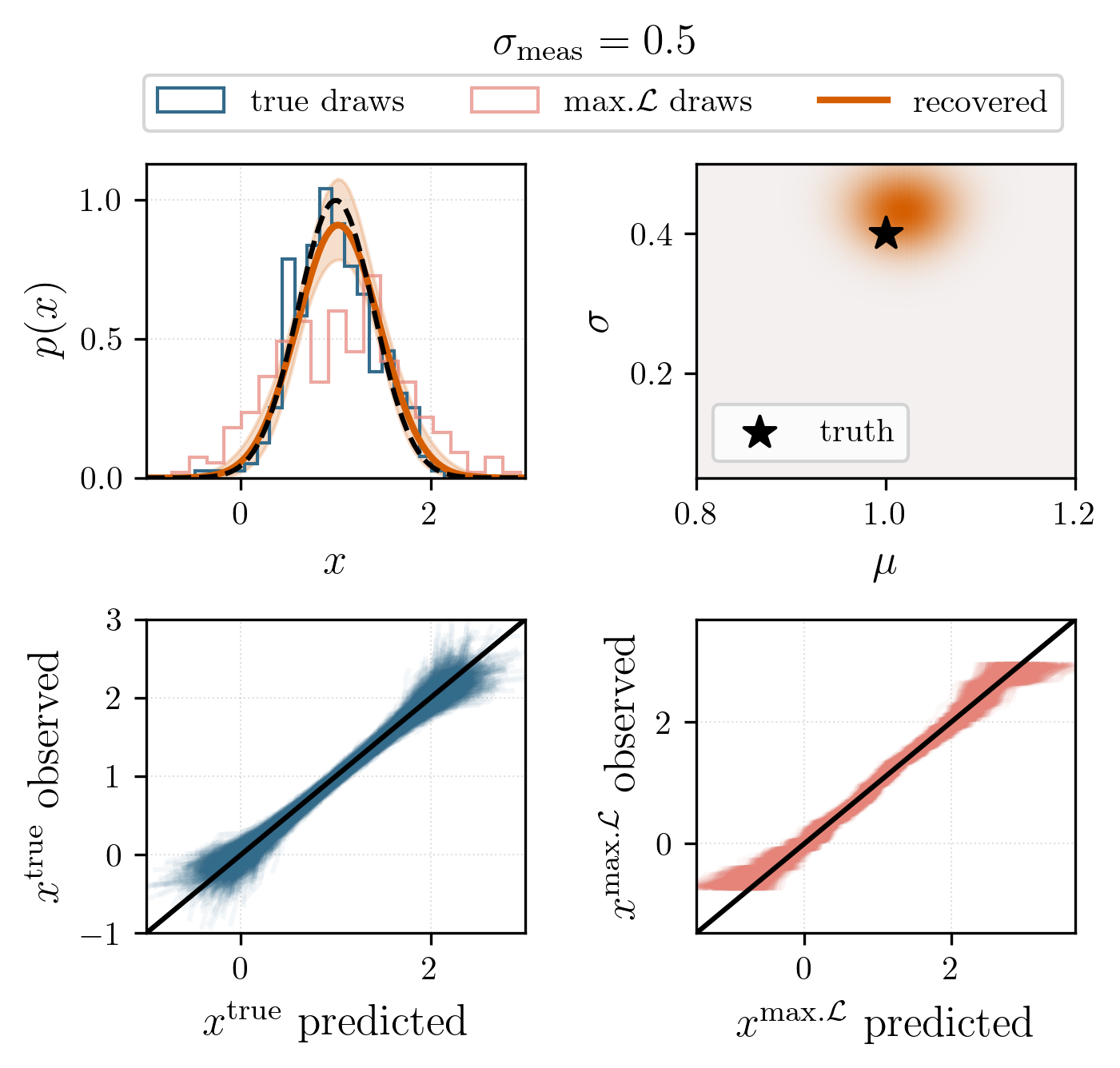} \vline \hfill
    \includegraphics[width=0.32\linewidth]{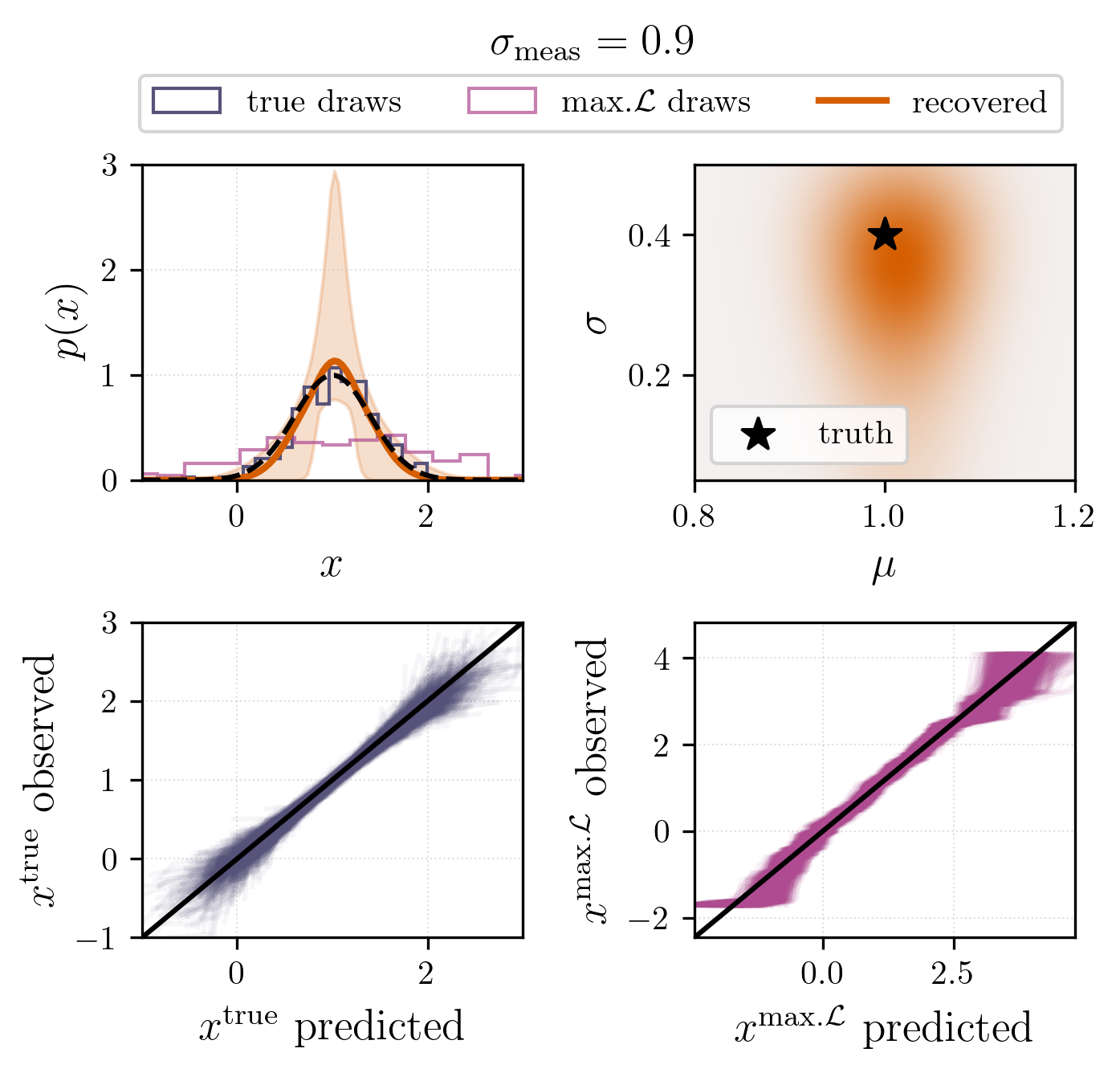}
    \caption{Results from hierarchical inference performed analytically with a Gaussian population model and Gaussian individual-event likelihoods. From the left to right sub-figures divided by vertical lines, the per-event measurement uncertainty increases from $\sigma_{\rm meas} = 0.1, 0.5, 0.9$.
    Within each sub-figure:
    \textit{(Top left)} The true underlying population distribution (black dashed) compared to a 300 event catalog drawn from that distribution, with true values of $x$ and corresponding $\maxL$ values of $x$. The median and 90\% credible region of the reconstructed distribution (from $\mu,\sigma$ in the top right panel) are shown in orange. 
    \textit{(Top right)} Hierarchical likelihood (Eq.~\ref{eqn:toy_model_LL}) on $\mu,\sigma$ given the set $\vec x_{\maxL}$ in the histogram. Darker orange indicates a higher likelihood, and the true underlying value is marked with a black star.
    \textit{(Bottom left)} Event-level PPC traces, consisting of $x^{\rm true}$ for 1,000 catalogs each with 300 events. 
    \textit{(Bottom right)} Data-level PPC traces, consisting of $x_{\maxL}$ for the same number of catalogs and events.}
    \label{fig:toy_model_good}
\end{figure*}

\begin{figure*}
    \centering
    \includegraphics[width=\linewidth]{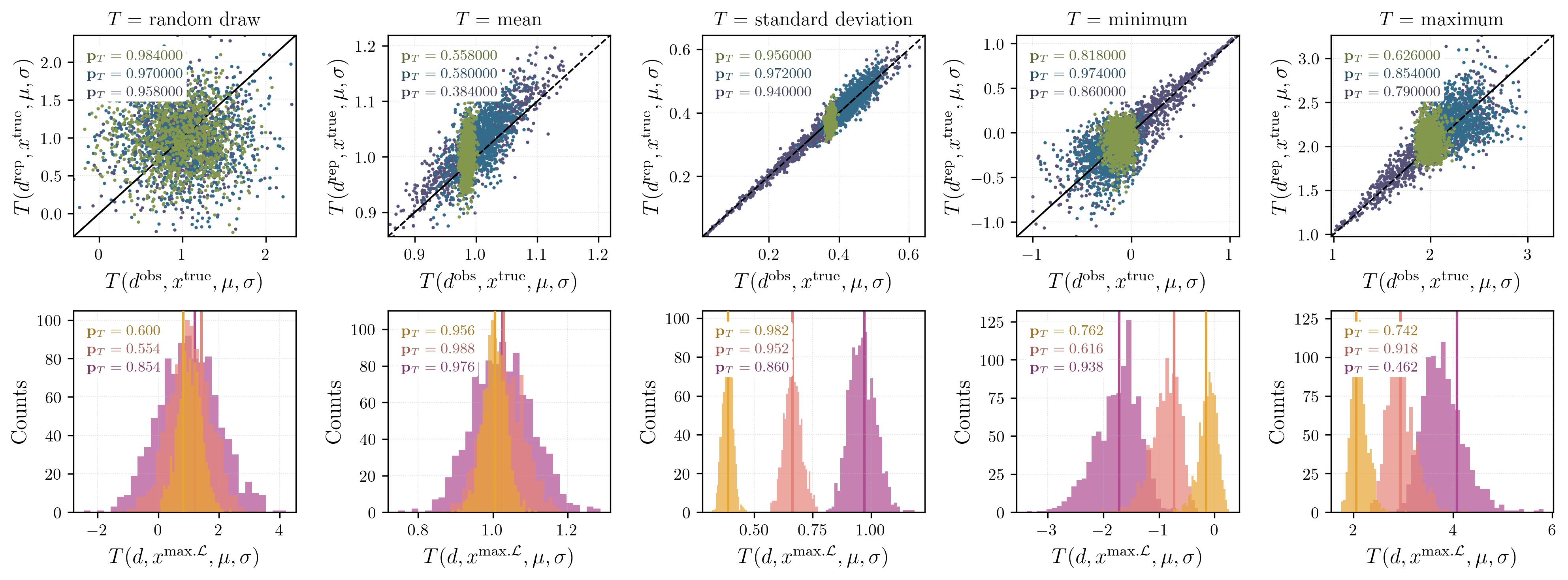}
    \caption{Distributions and associated posterior predictive $p$-values ($\pvalue$) for five test statistics $T$ from the PPCs shown in Fig.~\ref{fig:toy_model_good}, with per-event measurement uncertainties $\sigma_{\rm meas} = 0.1$ (lightest shade of each color), $0.5$ (middle shade), and $0.9$ (darkest shade). 
    From left to right, $T$ is a random draw of $x$ from the catalog, the catalog's mean, its standard deviation, minimum, and maximum.
    The top row show $T$ on event-level parameters ($x^{\rm true}$), with values calculated from $\dobsvec$ on the horizontal axis and $\dpredvec$ on the vertical axis.
    The bottom row shows $T$ on data-level parameters ($x_{\maxL}$), with the histogram showing those from $\dpredvec$ and the vertical line showing that from $\dobsvec$.}
    \label{fig:toy_model_good_test_stats}
\end{figure*}

We begin by examining hierarchical inference, PPCs, and posterior predictive $p$-values for a case where the Gaussian population model is a good fit to the observed data, with results shown in Figs.~\ref{fig:toy_model_good} and \ref{fig:toy_model_good_test_stats}. 
For the true underlying population, we use a Gaussian distribution with $\mu^{\rm true}=0.4$ and $\sigma^{\rm true}=0.4$, shown in the upper left panel of each sub-figure of Fig.~\ref{fig:toy_model_good} by a black-dashed line.
We simulate $300$ events from this population, whose true values $\{x_i^{\rm true}\}$ and corresponding $\{x_i^{\mathrm{max}.\mathcal{L}}\}$ are shown in each histogram, where we look at three possible measurement uncertainties: $\sigma_{\rm meas} = 0.1, 0.5, 0.9$, from left to right sub-figures.
The upper right subplots show the posterior probability (orange gradient) on $\mu,\sigma$---as calculated analytically with Eq.~\eqref{eqn:toy_model_LL}---compared to the true value (black star); deeper orange indicates a higher likelihood.
From the posterior on $\mu, \sigma$ we can reconstruct the population distribution over $x$: the orange lines and bands in each upper left subplot shows the median and 90\% credible interval of the reconstructed distribution.
As expected, the Gaussian population model accurately recovers the true distribution, with uncertainty increasing with $\sigma_{\rm meas}$.

The second row of Fig.~\ref{fig:toy_model_good} shows event-level (lower left, cool-toned colors) and data-level (lower-right, warm-toned colors) PPC traces. 
Each of the 1,000 traces corresponds to a 300-event catalog. 
These are calculated using the procedures outlined in Sec.~\ref{sec:event_vs_data_level} of the main text.
As in the main text, the event-level PPCs are calculated on the values of $x^{\rm true}$, while the data-level are calculated on values of $x_{\maxL}$.  
For all $\sigma_{\rm meas}$, both the event- and data-level PPCs are, by eye, consistent with the diagonal, indicating good model fit. 

We next calculate posterior-predictive $p$-values ($\pvalue$) using five example test statistics, $T$, with results shown in Fig.~\ref{fig:toy_model_good_test_stats}.
From left to right, $T$ is defined as a random draw from a catalog, the catalog's mean, its standard deviation, its minimum, and its maximum---where one catalog corresponds to one trace from the second row of Fig.~\ref{fig:toy_model_good}.
The first row (scatter-plots) shows test statistics and associated $p$-values from the event-level PPCs, while the second (histograms) shows data-level results.
Colors correspond to  Fig.~\ref{fig:toy_model_good}.
All $p$-values, for both the event- and data-level tests, are well above the acceptable threshold of $\pvalue=0.05$, quantifying what we already know: the model is a good fit to the data.

\vspace{-10pt}

\subsection{In which the Gaussian population model is a poor fit to the data} \label{subapp:toy_model_bad}

\begin{figure*}
    \centering
    \includegraphics[width=0.32\linewidth]{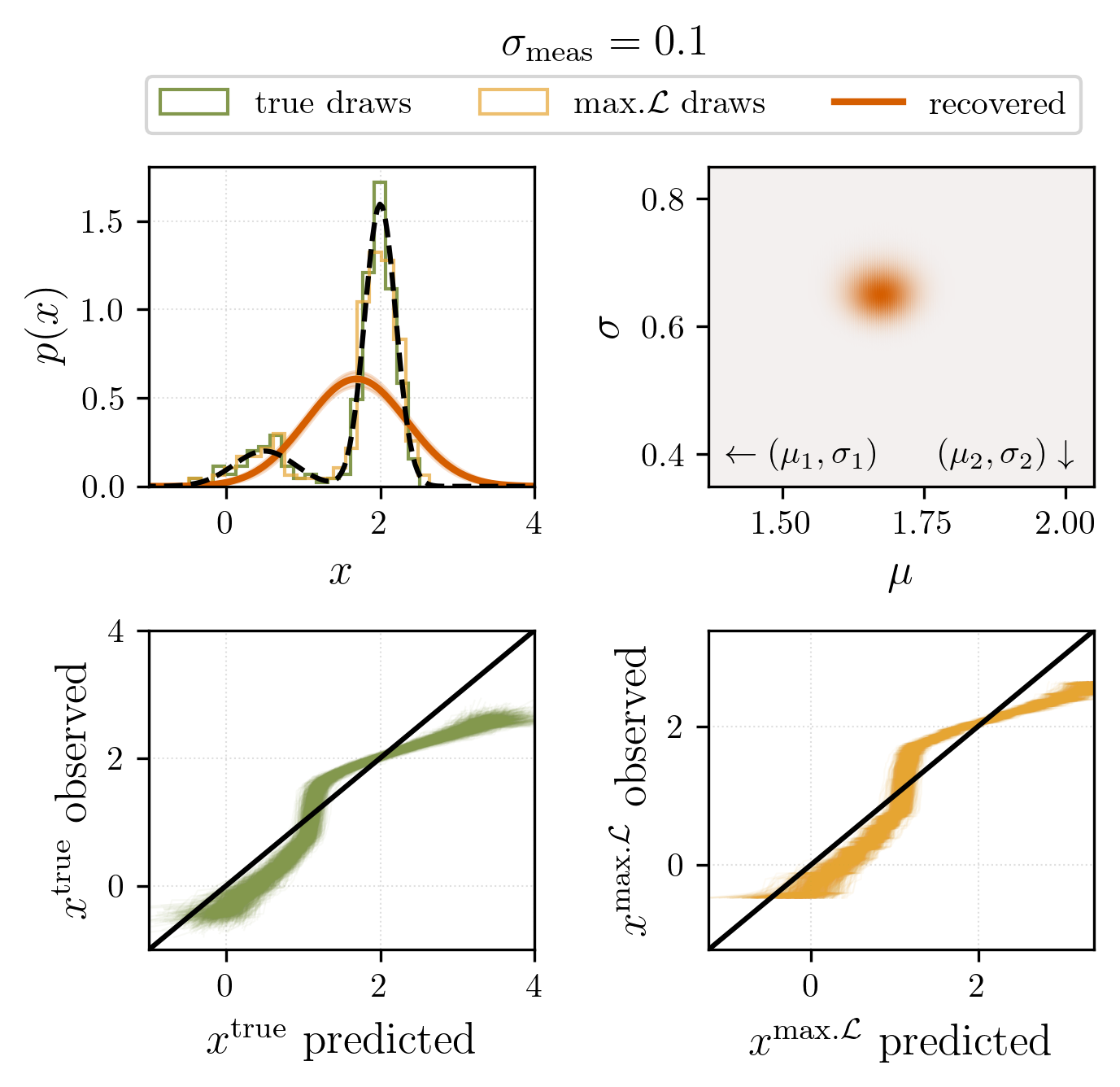} \vline \hfill
    \includegraphics[width=0.32\linewidth]{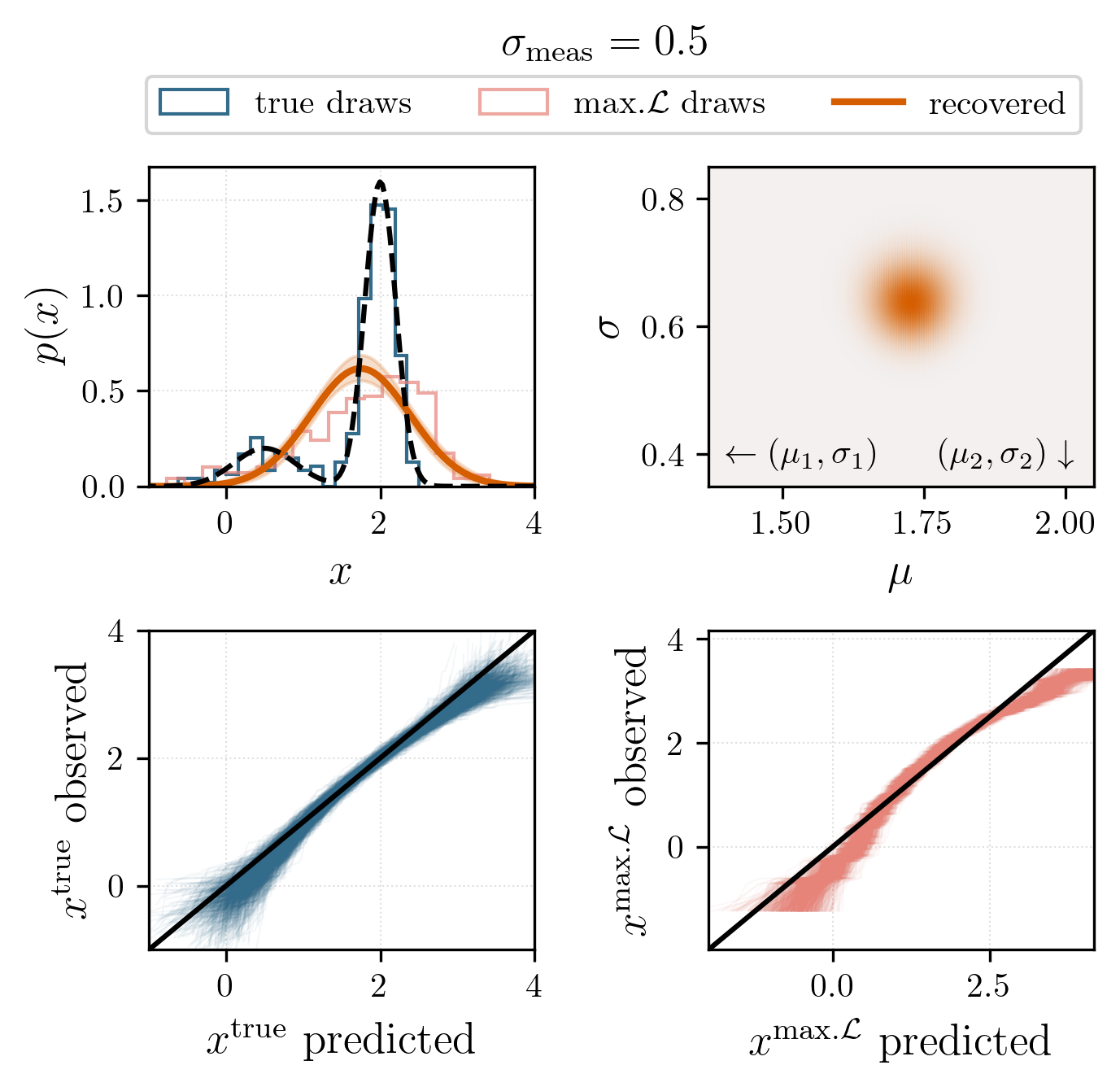} \vline \hfill
    \includegraphics[width=0.32\linewidth]{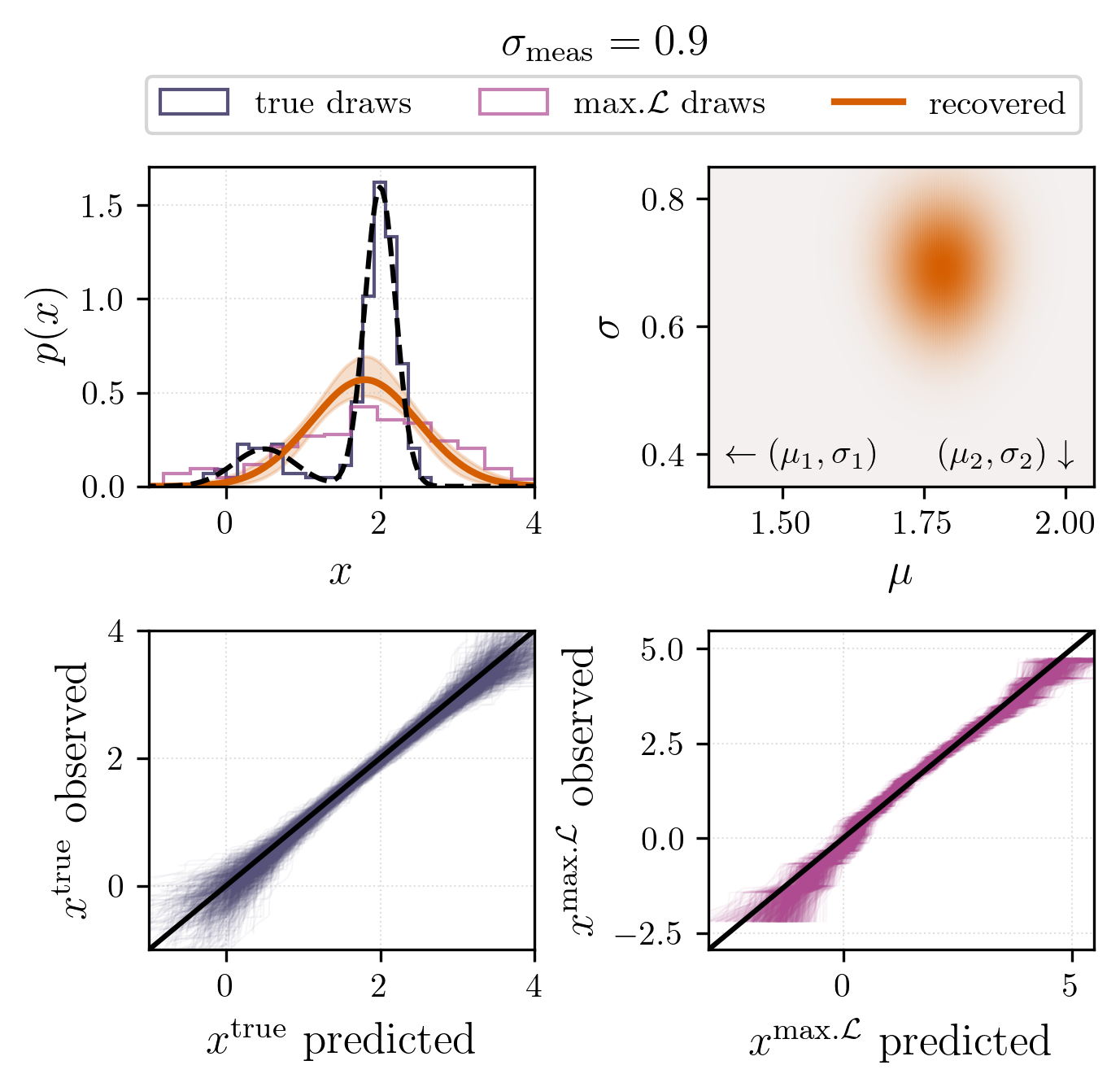}
    \caption{Same as Fig.~\ref{fig:toy_model_good} but with a true underlying  distribution that is bimodal, and therefore cannot be accurately recovered by the Gaussian population model. For the $\sigma_{\rm meas} = 0.1$ case, both the event and data-level PPCs show deviations from the diagonal. For the $\sigma_{\rm meas} = 0.5$ case, only the data-level PPC is informative.}
    \label{fig:toy_model_bad}
\end{figure*}

\begin{figure*}
    \centering
    \includegraphics[width=\linewidth]{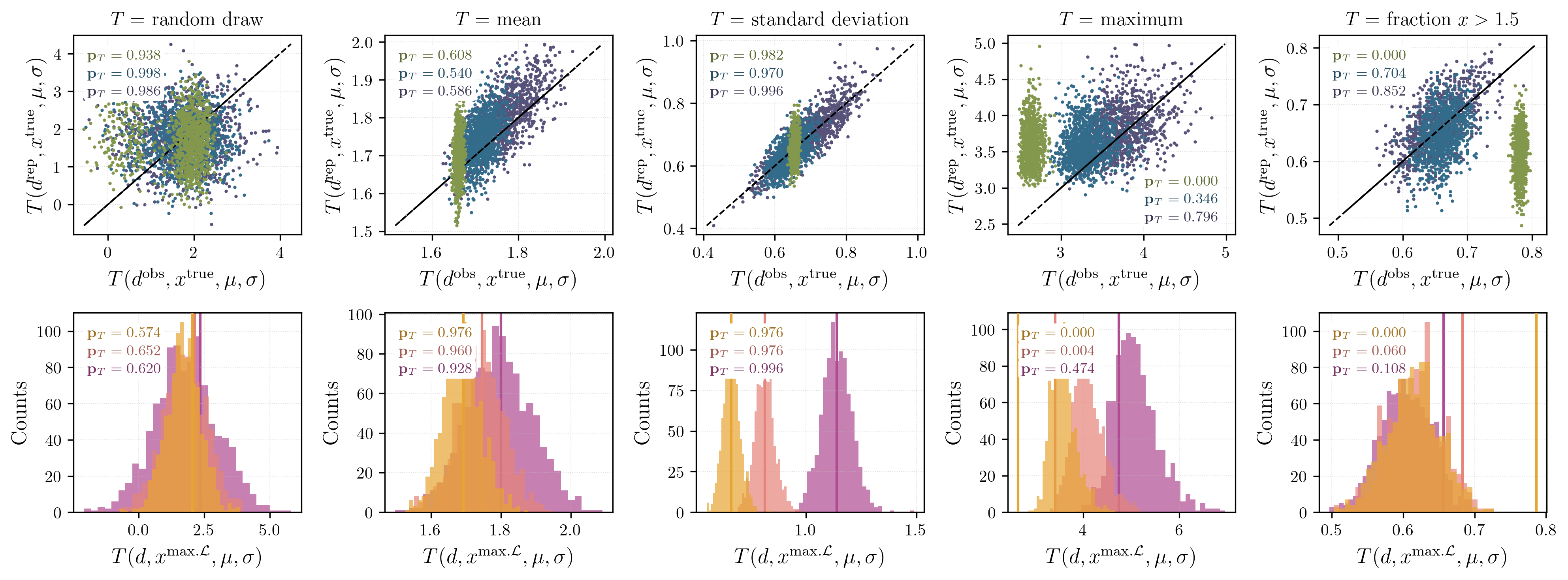}
    \caption{Similar to Fig.~\ref{fig:toy_model_good_test_stats} but for the bimodal underlying population shown in Fig.~\ref{fig:toy_model_bad}, and (from left to right) test statistics $T$ defined as a random draw from a catalog, the catalog's mean, its standard deviation, its maximum, and the fraction greater than $1.5$ (the inflection point of the bimodal distribution). Colors correspond to Fig.~\ref{fig:toy_model_bad}.}
    \label{fig:toy_model_bad_test_stats}
\end{figure*}

We repeat the above exercise with a simulated population that \textit{cannot} be accurately recovered by a Gaussian model, with results shown in Figs.~\ref{fig:toy_model_bad} and \ref{fig:toy_model_bad_test_stats}.
Here, the true underlying population is a bimodal distribution---the mixture of two Gaussian distributions, with the sub-dominant component having $\mu_1, \sigma_1 = (0.5, 0.4)$, the dominant component having $\mu_2, \sigma_2 = (2, 0.2)$, and a mixing fraction of $0.2$ between them. 
Within each sub-figure of Fig.~\ref{fig:toy_model_bad}, the true population distribution is shown in the upper left panel with a black-dashed line, again compared to our simulated observed catalog of 300 values of $x_{\rm true}$ and corresponding $x_{\maxL}$.
The analytically calculated hierarchical likelihood on $\mu, \sigma$ is shown in each upper right panel, with darker orange indicating a higher likelihood.

Clearly, the Gaussian population model does not find the true underlying population. 
Its 90\% recovered credible interval (orange, upper left subplot of Fig.~\ref{fig:toy_model_bad}) does not encompass the truth, for any of the three values of $\sigma_{\rm meas}$.
The true values of the mean and standard deviation of either of the true distribution's subpopulations are far away from the most probable $\mu,\sigma$ for the single Gaussian. 
This makes sense: the Gaussian distribution is indeed finding the mean and width of the underlying population; these two degrees of freedom just do not fully characterize the true, more complicated distribution. 

For the $\sigma_{\rm meas} = 0.1$ case, both the event and data-level PPCs show deviations from the diagonal, while neither do when $\sigma_{\rm meas} = 0.9$. 
We thus focus on the middle case of $\sigma_{\rm meas} = 0.5$, which is coincidentally the closest analog to BBH spins. 
The discrepancy between model and truth can be clearly seen by eye in the data-level PPC traces, shown in pink in the bottom right panel of Fig.~\ref{fig:toy_model_bad}'s middle sub-figure---but \textit{not} so much in the event-level traces (lower left panel, blue). 

The strength of the data-level PPC is also reflected in the posterior predictive $p$-values in Fig.~\ref{fig:toy_model_bad_test_stats}.
For $\sigma_{\rm meas} = 0.5$, the event-level $p$-values are all well above the threshold of $0.05$, even though we know \textit{a priori} that the model is not a good fit to the data. 
For some choices of $T$, the data-level $p$-values are much more informative. 
Although the random draw (first column), mean (second column), and standard deviation (third column) test statistics would indicate that our model is a very good fit (which makes sense, given that we are using a Gaussian model which \textit{does} accurately constrain the mean and standard deviation degrees of freedom), the others are either more inconclusive or return a negative result. 
Most notable is when $T$ is the distribution's maximum (fourth column): In the event-level case, $\pvalue = 0.34$, well above the threshold. In the data-level case, however, $\pvalue=0.004$, an order of magnitude below the threshold.
The histogram of $T$ further tells us that the model nearly always over-predicts the maximum $x_{\maxL}$ of a catalog.
In the upper-left panel of Fig.~\ref{fig:toy_model_bad}, we see that the distributions' upper tails are where the true and inferred populations are most discrepant, consistent with what is reported by $\pvalue$.
The fraction of events with $x>1.5$ (the inflection point in the true underlying bimodal distribution) is also flagged by the data-level PPC, with $\pvalue=0.06$ in the data level case---right on the threshold for passing---as compared to $\pvalue = 0.7$ in the event-level PPC. 
Even the $\sigma_{\rm meas} = 0.9$ case has a relatively low $\pvalue = 0.108$ for this final choice of test statistic.

The Gaussian toy model has trivial computational cost, making it easy to repeat population inference and calculations of $\pvalue$ for many instantiations of $\dobsvec$, a process which remains computationally infeasible for the results presented in the main text. 
We repeat the results shown in Fig.~\ref{fig:toy_model_bad_test_stats} over ten values of $\sigma_{\rm meas}$ linearly spaced between $0.1$ and $1$. For each $\sigma_{\rm meas}$ we generate 500 instantiations of observed catalogs (thus repeating the exercise in Fig.~\ref{fig:toy_model_bad_test_stats} 5000 times total).
For each $T$ and $\sigma_{\rm meas}$, we calculate the fraction of these 500 catalogs for which $\pvalue < 0.05$, indicating strong evidence for model mismatch (which we know exists). 
A fraction of $1$ means that the test can always tell that there is model mismatch; a fraction of 0 means it can never tell that there is model mismatch. 
Results are presented in Fig.~\ref{fig:toy_model_repeated} in the main text, and show that the data-level $\pvalue$ are always more discerning of model misspecification than the event-level $\pvalue$, although both become uninformative when single-event measurement uncertainty is sufficiently large. 
For comparison, when repeating this analysis with the good-fit population of Appendix~\ref{subapp:toy_model_good}, the fraction of $\pvalue < 0.05$ is clustered at $0$ for all test statistics at both the event and data level parameters. 

%

\bibliographystyle{apsrev4-2} 
\twocolumngrid
\bibliography{OurRefs}

\end{document}